\documentclass[aps,pra,twocolumn,groupedaddress,floatfix]{revtex4-1}

\usepackage{graphicx}
\usepackage{dcolumn}
\usepackage{bm}
\usepackage{amsmath}
\usepackage{xcolor}
\usepackage{braket}

\usepackage{amsfonts}
\usepackage{blkarray}
\usepackage{amssymb}
\usepackage{multirow}
\usepackage{mathrsfs}

\newcommand{\bhzz}{\ket{300}}
\newcommand{\bzhz}{\ket{030}}
\newcommand{\bzzh}{\ket{003}}

\newcommand{\btoz}{\ket{210}}
\newcommand{\btzo}{\ket{201}}
\newcommand{\botz}{\ket{120}}
\newcommand{\bzto}{\ket{021}}
\newcommand{\bozt}{\ket{102}}
\newcommand{\bzot}{\ket{012}}

\newcommand{\booo}{\ket{111}}

\newcommand{\cu}{{\cal U}}
\newcommand{\ca}{{\cal A}}
\newcommand{\cb}{{\cal B}}
\newcommand{\cc}{{\cal C}}
\newcommand{\bc}{{\bf c}}

\begin{document}

\title{All-order momentum correlations of three ultracold bosonic atoms confined in triple-well traps:
Signatures of emergent many-body quantum phase transitions and analogies with three-photon quantum-optics 
interference}

\author{Constantine Yannouleas}
\email{Constantine.Yannouleas@physics.gatech.edu}
\author{Uzi Landman}
\email{Uzi.Landman@physics.gatech.edu}

\affiliation{School of Physics, Georgia Institute of Technology,
             Atlanta, Georgia 30332-0430}

\date{11 December 2019}

\begin{abstract}
All-order momentum correlation functions associated with the 
time-of-flight spectroscopy of three spinless ultracold bosonic interacting neutral atoms 
confined in a linear three-well optical trap are presented. The underlying Hamiltonian employed for
the interacting atoms is an augmented three-site Hubbard model. 
Our investigations target matter-wave interference of massive particles, aiming at the establishment of 
experimental protocols for characterizing the quantum states of trapped attractively or repulsively 
interacting ultracold particles, with variable interaction strength. The manifested advantages and 
deep physical insights that can be gained through the employment of the results of our study for a 
comprehensive understanding of the nature of the quantum states of interacting many-particle systems, via 
analysis of the all-order (that is 1st, 2nd and 3rd) momentum correlation functions for three bosonic atoms
in a three well confinement, are illustrated and discussed in the context of time-of-flight inteferometric 
interrogations of the interaction-strength-induced emergent quantum phase transition from the Mott insulating 
phase to the superfluid one. Furthermore, we discuss that our inteferometric interrogations establish strong 
analogies with the quantum-optics interference of three photons, including the aspects of genuine 
three-photon interference, which are focal to explorations targeting the development and implementation of 
quantum information applications and quantum computing.
\end{abstract}

\maketitle


\section{Introduction}
\label{intr}

Theoretical and experimental access to many-body correlations is essential in elucidating the properties and
underlying physics of strongly interacting systems \cite{cira12,garc14}. In the framework of ultracold atoms,
the quantum correlations in momentum space associated with bosonic or fermionic neutral atoms trapped in 
optical tweezers (with a finite number $N$ of particles \cite{prei19,berg19,bech20}) or in extended optical 
lattices (with control of the 1D, 2D, or 3D dimensionality \cite{grei02,gerb05,gerb05.2,clem18,clem19}) 
are currently attracting significant experimental attention, empowered 
\cite{bech20,prei19,berg19,clem18,clem19,hodg17} by advances in single-atom-resolved detection methods 
\cite{ott16}.

In this paper, we derive explicit analytic expressions for the 3rd-, 2nd-, and 1st-order momentum 
correlations of 3 ultracold bosonic atoms trapped in an optical trap of 3 wells in a linear arrangement 
(denoted as 3b-3w). Compared to the case of 2 particles in 2 wells (2p-2w) 
\cite{bran17,bran18,yann19.1,yann19.2}, 
a complete Hubbard-model treatment of momentum correlations (as a function of the 
interparticle interaction) for the 3b-3w case increases the complexity and effort involved, by an order of 
magnitude, because of the larger Hilbert space and the larger number of states, i.e., a total of 10 states 
instead of 4, including the excited states which are long-lived \cite{joch15} for trapped ultracold atoms. 
Therefore, demonstrating that this complexity of the theoretical treatment can be handled in an 
efficient manner through the use of algebraic computer languages constitutes an important step toward the 
implementation of the bottom-up approach for simulating many-body physics with ultracold atoms. In this 
respect, the statement above parallels earlier observations that three-particle entanglement extends 
two-particle entanglement in a nontrivial way \cite{zeil99,cira00,yann19.3}.

\textcolor{black}{
Compared to the standard numerical treatments \cite{galle15,rave17,shib72,call87,dago94}
of the Hubbard model,} 
the advantage of our algebraic treatment
is the ability to produce in closed analytic form cosinusoidal/sinusoidal expressions of the many-body wave 
function and the associated momentum correlations of all orders; see for example Eqs.\ 
(\ref{wfbexpr}), (\ref{2ndbexpr}), and (\ref{frstbexpr}), which codify the main results of our paper. 
Due to recent experimental advances in tunability and control of a system of a few ultracold atoms trapped in 
finite optical lattices (referred to also as optical tweezers), such momentum correlations can be measured directly
in time-of-flight experiments \cite{berg19,prei19,bech20} and their experimental cosinusoidal diffraction patterns 
are revealing direct analogies with the quantum optics of massless photons \cite{bran18,yann19.1,prei19}. 

\textcolor{black}{
In this context, this paper aims at researchers actively engaged in experimental and theoretical investigations 
of the properties of (finite) quantum few-body systems, as well as those aiming to understand many-body quantum 
systems through bottom-up hierarchical modeling of trapped finite ultracold-atom assemblies with deterministically  
controllable increased size and complexity; see, e.g., 
Refs.\ \cite{kauf14,kauf18,berg19,prei19,bech20,joch15,sowi16,zinn14}. 
Indeed, we target researchers in these fields by providing finger-print characteristics to aid the design, 
diagnostics, and interpretation of experiments,} 
\textcolor{black}{
as well as by giving benchmark results \cite{note9} for comparisons with future theoretical treatments.} 
We foresee these as important merits that will contribute to future impact of our work.

\textcolor{black}{
In addition, the availability of the complete analytic set of momentum correlations enabled us to reveal and 
explore two major physical aspects of the 3b-3w ultracold-atom system, namely: (i) Signatures of an emergent 
quantum phase transition \cite{note7}, from a Superfuid phase to a Mott-insulator phase -- here the designation 
'emergent' is used to indicate the gradual emergence of a phase transition in a finite system as the system size 
is increased to infinity \cite{note7}, alternatively  termed as 'inter-phase crossover' --  and 
(ii) Analogies between the interference properties of three trapped ultracold atom systems with quantum-optics 
three-photon interference. These aspects are elaborated in some detail immediately below.
}

{\it (i) Signatures of emergent Superfluid to Mott transition:\/}
The sharp superfluid-to-Mott transition has been observed in extended optical lattices with trapped 
ultracold bosonic 
alkali atoms ($^{87}$Rb) \cite{grei02}, as well as with excited $^4$He$^*$ bosonic atoms \cite{clem18}. In these 
experiments, after a time-of-flight (TOF) expansion, the single-particle momentum (spm) density (1st-order 
momentum correlation) was recorded. An oscillating spm-density provides a hallmark of a superfluid phase, 
associated with a maximum uncertainty regarding a particle's site occupation; this happens for the 
non-interacting case when the particles are fully delocalized. On the other hand, a featureless spm-density is
the hallmark of being deeply in the Mott-insulator phase when all particles are fully localized on the
lattice sites exhibiting no fluctuations in the site occupancies.

\textcolor{black}{
Here, we show that the 1st-order momentum correlations for the 3b-3w system vary smoothly, alternating as a 
function of the Hubbard $\cu$ between a featureless profile and that resulting from the sum of two cosine terms;
such profile alternations may provide signatures of an emerging superfluid to Mott-insulator phase crossing.
The periods of the cosine terms depend on the inverse of the lattice constant $d$ and its double $2d$ ($d$ being 
the nearest-neighbor interwell distance). We note that for extended lattices only the $\cos(dk)$ term has been
theoretically specified \cite{gerb05.2,seng05,triv09} 
with perturbative $1/\cu$ approaches, and that our non-perturbative
results suggest that all cosine terms with all possible interwell distances in the argument should in general 
contribute.}
 
Furthermore, we show that the correspondence between the featureless profiles and the interaction strength is
not a one-to-one correspondence. Indeed, we show that a 
featureless spm-density can correspond to different strengths of the interaction, depending on the sign of the 
interaction (repulsive versus attractive) and the precise Hubbard state under consideration (ground state or 
one of the excited states). {\it For a unique characterization of a phase regime, both the 2nd-order and the 
3rd-order momentum correlations beyond the spm-density are required.\/}

{\it (ii) Analogies with quantum-optics three-photon interference:\/}
Recent experimental \cite{prei19,berg19,lege04,gerr15.1,gerr15.2,tamm18.1,tamm19} and theoretical 
\cite{bran17,bran18,bonn18,tamm18.2,yann19.1,yann19.2,yann19.3} advances have ushered a new research direction 
regarding investigations of higher-order quantum interference resolved at the level of the intrinsic microscopic 
variables that constitute the single-particle wave packet of the interfering particles. These intrinsic 
variables are pairwise conjugated; they are the single-particle momenta ($k$'s) and mutual distances ($d$'s) for
massive localized particles \cite{prei19,berg19,bran17,bran18,bonn18,yann19.1,yann19.2,yann19.3} and the 
frequencies ($\omega$'s) and relative time delays ($\tau$'s) for massless photons 
\cite{lege04,gerr15.1,gerr15.2,tamm18.1,tamm18.2,tamm19}. 

For the case of two fermionic or bosonic ultracold atoms, we investigated in Ref.\ \cite{yann19.1} this 
correspondence in detail and we proceeded to establish a complete analogy between the cosinusoidal patterns 
(with arguments $\propto kd$ or $\propto \omega \tau$) of the second-order $(k_1,k_2)$ correlation
maps for the two trapped atoms (determined experimentally through TOF measurements \cite{prei19,berg19}) with the
landscapes of the two-photon ($\omega_1,\omega_2)$ interferograms \cite{gerr15.1,gerr15.2,tamm19}. In addition, 
we demonstrated that the Hong-Ou-Mandel (HOM) \cite{hom87} single-occupancy coincidence probability at
the detectors, $P_{11}$ (which relates to the celebrated HOM dip for total destructive interference, i.e., 
when $P_{11}=0$), corresponds to a double integral over the momentum variables $(k_1,k_2)$ 
of a specific term contributing to the 
full correlation map, in full analogy with the treatment of the optical ($\omega_1,\omega_2)$ interferograms in 
Ref.\ \cite{gerr15.1}. Due to this summation over the intrinsic momentum (or frequency for photons) variables, 
the information contained in the HOM dip is limited compared to the full correlation map. Precise analogs of the 
original optical HOM dip (with $P_{11}$ varying as a function of relative time delay or separation between
particles) have also been experimentally realized using the interference of massive particles, 
i.e., two colliding electrons \cite{taru98,jonc12,bocq13} or two colliding $^4$He atoms \cite{lope15}. For the
case of two ultracold atoms trapped in two optical tweezers, analogs of the $P_{11}$ coincidence probability
can be determined via {\it in situ\/} measurements, as a function of the time evolution of the system 
\cite{kauf14,yann19.1} or the interparticle interaction \cite{bran18,yann19.1}.
 
In this paper, we establish for the 3b-3w case the full range of analogies between the TOF spectroscopy 
\cite{note3},
as well as the {\it in-situ\/} measurements, of localized massive particles and the multi-photon interference 
in linear optical networks \cite{agar15,tamm18.1,tamm18.2,tamm19}, 
paying attention in particular to the mutual interparticle interactions which are 
absent for photons. These analogies encompass extensions of the 2p-2w analogies mentioned above, i.e., 
correlation maps dependent on three momentum variables $(k_1,k_2,k_3)$ for massive particles versus 
interferograms with three frequency variables $(\omega_1,\omega_2,\omega_3)$ for massless photons, and the HOM 
$P_{111}$ coincidence probability for three particles versus that for three photons. Most importantly, however,
these analogies include highly nontrivial aspects beyond the reach of two-photon (or two-particle) 
and one-photon (or one-particle) interferences,
such as genuine three-photon interference \cite{agne17,mens17} which cannot be determined from the 
knowledge solely of the lower two-photon and one-photon interferences.

\begin{figure}[t]
\includegraphics[width=8cm]{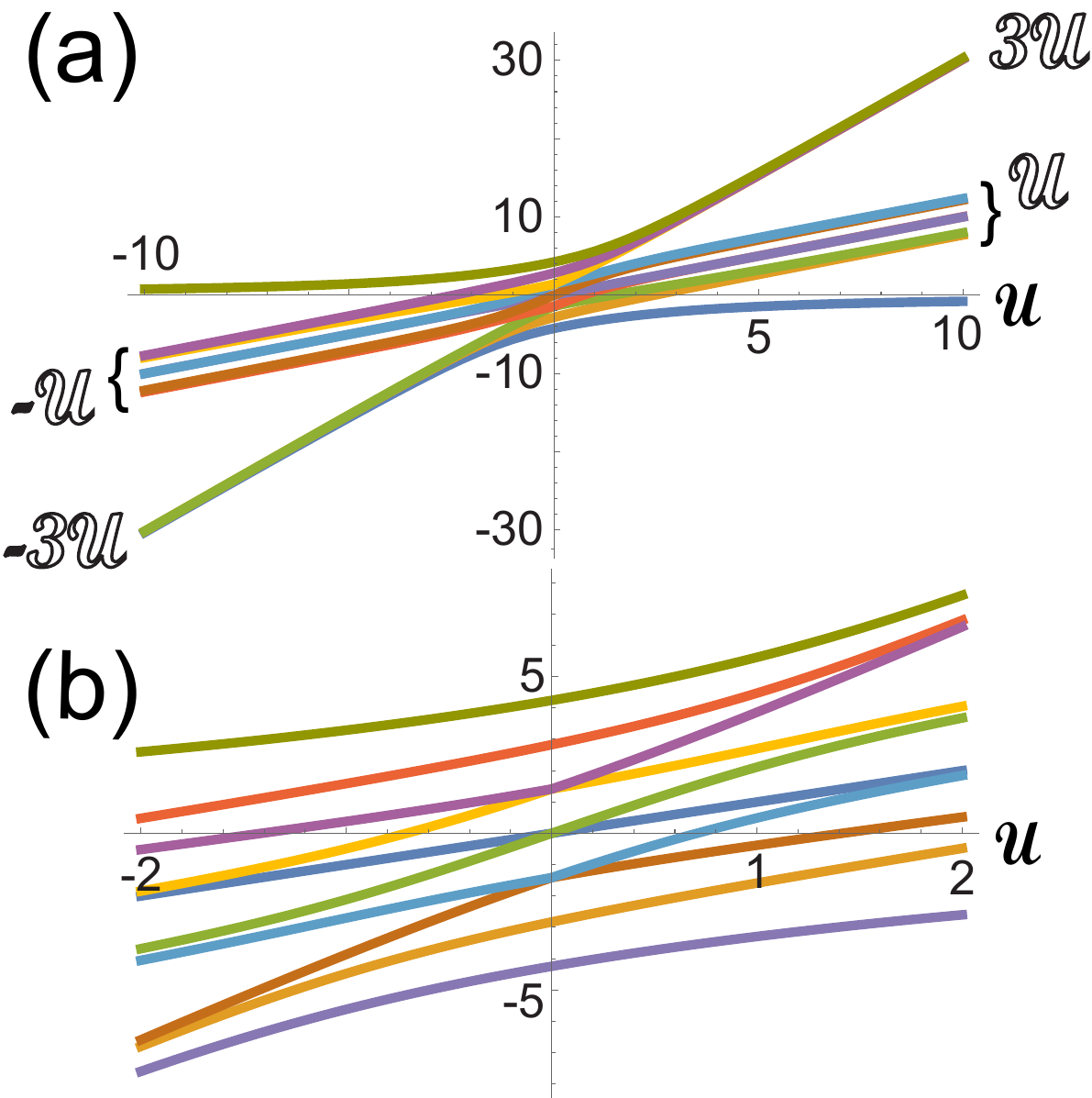}
\caption{Spectrum of the ten bosonic eigenvalues in Eq.\ (\ref{eigvalb}) as a function of $\cu$ (horizontal axis).
(a) This frame (with the extended $-10 \leq \cu \leq 10$ scale) illustrates the convergence to the three values 
of zero [ground state ($\cu >0$) or highest excited state ($\cu < 0$)], $\pm |\cu|$ (six excited states),
and $\pm 3 |\cu|$ (ground state and two excited states for $\cu < 0$).
(b) A more detailed view in the range $-2 \leq \cu \leq 2$.
Taking into consideration the three energy crossings at $\cu=0$, the corresponding eigenstates 
are labeled in ascending energy order as 
$i=1$, $2$, $3r(4l)$, $4r(3l)$, $5r(6l)$, $6r(5l)$, $7r(8l)$, $8r(7l)$, $9$, $10$,
where ``$r$'' means ``right'' for the region of positive $\cu$ and ``$l$'' means ``left'' for the region of
negative $\cu$. 
} 
\label{feigvalb}
\end{figure}

\subsection {Plan of paper}
\label{plan}

Following the introductory section where we defined the aims of this work, we introduce in Sec.\ \ref{3b-hb} the 
linear three-site Hubbard model and its analytic solution for three spinless ultracold bosonic atoms. We display 
the spectrum of the ten bosonic eigenvalues of the Hubbard model for both attractive and repulsive interatomic 
interactions (Fig.\ \ref{feigvalb}), and discuss in detail: (1) the infinite repulsive or attractive interaction 
limit, and (2)  the non-interacting limit. In Sec.\ \ref{hordcorr} we outline the general definition and relations 
pertaining to higher-order correlations in momentum space. 

In the following several sections we give explicit 
analytic results and graphical illustrations pertaining to momentum correlation functions of the various orders, 
starting from the third-order, since the lower-order are obtained from the third-order one by integration over the 
unresolved momentum variables [see, e.g., Eq.\ (\ref{2nd}) for the second-order momentum correlation]. The 
third-order momentum correlations for 3 bosons in 3 wells, with explicit discussion of the infinite-interaction 
(repulsive or attractive) limit is given in Sec.\ \ref{3rdcorrUpmI} (see Fig.\ \ref{f3rdcorrb}),
followed by explicit results for the non-interacting limit in Sec.\ \ref{3rdcorrU0}. Sec.\ \ref{s3rdanyu} is 
devoted to a presentation and discussion of results for the third-order momentum correlations for 3 bosons in 3 
wells as a function of the strength of the inter-atom interaction over the whole range, from highly attractive to 
highly repulsive (see momentum correlation maps in Fig.\ \ref{f3rdcorrbst2}).
Next we discuss in Sec.\ \ref{s2ndanyu} the second-order momentum correlation as a 
function of the interparticle interaction; see momentum correlation maps for the whole interaction range in  
Fig.\ \ref{f2ndcorrbst2}.

The first-order momentum correlation, obtained via integration of the second-order one over the momentum of one of 
the atoms, is discussed as a function of inter-atom interaction strength in Sec.\ \ref{s1stanyu}, 
with a graphic  illustration in Fig.\ \ref{f1stcorrbst2} for the first-excited state of 3 bosons in 3 
wells, illustrating transition as a function of interaction strength from localized to superfluid behavior.  
Sec.\ \ref{sign} is devoted to a detailed study of the quantum phase transition from localized to 
superfluid behavior, as deduced from inspection of the first-order correlation function for the ground state of 3 
bosons in 3 wells (Fig.\ \ref{corrbst1}, top row), and further elucidated and elaborated 
with the use of second-order (Fig.\ \ref{corrbst1}, middle row), and third-order 
(Fig.\ \ref{corrbst1}, bottom row) momentum correlation maps. Further discussion of the 
quantum phase transition through analysis of site occupancies and their fluctuations for the ground and 
first-excited states as a function of the interparticle interactions, illuminating the connection between the 
quantum phase-transition from superfluid (phase coherent) to localized (incoherent) states, and the phase-number 
(site occupancy) uncertainty principle, is illustrated in Fig.\ \ref{focc}. 

Sec.\ \ref{anal} expounds on analogies with three-photon interference in quantum optics, including {\it genuine\/}
three-photon interference. We summarize the contents of the paper in Sec.\ \ref{summ}, 
closing with a comment concerning the expected relevance of the all-order momentum-space correlations for the 
3 bosons in 3 wells as an alternative route to exploration with massive particles of aspects pertaining to the 
boson sampling problem \cite{aaar13} and its extensions, which are serving as a major topic (see, e.g., Refs. 
\cite{tamm15,tamm15.1,tich14,lain14,wals19}) in quantum-optics investigations as an intermediate step towards 
the implementation of a quantum computer. 

\textcolor{black}{
Appendix \ref{a11} and Appendix \ref{a12} complement Sec.\  
\ref{eigvecuinf} and Sec.\ \ref{eigvecu0}, respectively, by listing the Hubbard eigenvectors of the 
remaining eight excited states not discussed in the main text (where, as above-mentioned, we focus on the ground 
and first-excited states). In addition, regarding again the remaining eight excited states not discussed in the 
main text, Appendix \ref{a1} and Appendix \ref{a2} complement Sec.\ \ref{3rdcorrUpmI} and Sec.\ \ref{3rdcorrU0}, 
respectively, by listing the corresponding three-body wave functions. Specifically, Appendices \ref{a11} and 
\ref{a1} focus on the limit of infinite repulsive or attractive interaction, whereas Appendices \ref{a12} and 
\ref{a2} focus on the noninteracting case. 
}
The last three appendices give details of the all-order correlation 
functions as a function of the interaction strength for the remaining eight states not discussed in the main text. 

\section{The linear three-site Hubbard model and its analytic solution for three spinless ultracold bosonic atoms}
\label{3b-hb}

Numerical solutions for small Hubbard clusters are readily available in the literature. Here we present a
compact analytic exposition for all the 10 eigenvalues and eigenstates of the linear three-bosons/three-site 
Hubbard Hamiltonian. Such analytic solutions, involving both the ground and excited states, are needed to 
further obtain the characteristic cosinusoidal or sinusoidal expressions for the associated third-, second-, 
and first-order momentum correlations.

The following {\it ten\/} primitive kets form a basis that spans the many-body Hilbert space of three spinless 
bosonic atoms distributed over three trapping wells:
\begin{align}
\begin{split}
& 1 \rightarrow \booo, \\ 
& 2 \rightarrow \btoz, \; 3 \rightarrow \btzo, \; 4 \rightarrow \botz, \\ 
& 5 \rightarrow \bzto, \; 6 \rightarrow \bozt, \; 7 \rightarrow \bzot, \\
& 8 \rightarrow \bhzz, \; 9 \rightarrow \bzhz, \; 10 \rightarrow \bzzh.
\end{split}
\label{3b-kets}
\end{align}
The kets used above are of a general notation $|n_1,n_2,n_3\rangle$, where $n_i$ (with $i=1,2,3$) denotes the 
particle occupancy at the $i$th well.   
We note that there is only one primitive ket (No. 1) with all three wells being singly-occupied. The case of
doubly-occupied wells is represented by 6 primitives kets (Nos. 2$-$7). Finally, there are 3 primitive kets
(Nos. 8$-$10) that represent triply-occupied wells.

The Bose-Hubbard Hamiltonian for 3 spinless bosons trapped in 3 wells in a linear arrangement is given by
\begin{align}
H_B=-J(\hat{b}^\dagger_1 \hat{b}_2 + \hat{b}^\dagger_2 \hat{b}_3 + h.c.)+ \frac{U}{2}\sum_{i=1}^3 n_i(n_i-1),
\label{3b-hub}
\end{align}
where $n_i=\hat{b}^\dagger_i \hat{b}_i$ is the occupation operator per site. $J$ is the hopping (tunneling) 
parameter and the Hubbard $U$ can be positive (repulsive interaction), vanishing (noninteracting), or negative 
(attractive interaction). 
 
Using the capabilities of the SNEG \cite{sneg} program in conjunction with the MATHEMATICA \cite{math18} 
algebraic language, one can write the following matrix Hamiltonian for the spinless three-boson Hubbard problem:
\begin{widetext}
\begin{align}
\begin{split}
H_b=\left(
\begin{array}{cccccccccc}
 0 & 0 & -\sqrt{2} J & -\sqrt{2} J & -\sqrt{2} J & -\sqrt{2}
   J & 0 & 0 & 0 & 0 \\
 0 & U & -J & -2 J & 0 & 0 & 0 & -\sqrt{3} J & 0 & 0 \\
 -\sqrt{2} J & -J & U & 0 & 0 & 0 & 0 & 0 & 0 & 0 \\
 -\sqrt{2} J & -2 J & 0 & U & 0 & 0 & 0 & 0 & -\sqrt{3} J &
   0 \\
 -\sqrt{2} J & 0 & 0 & 0 & U & 0 & -2 J & 0 & -\sqrt{3} J &
   0 \\
 -\sqrt{2} J & 0 & 0 & 0 & 0 & U & -J & 0 & 0 & 0 \\
 0 & 0 & 0 & 0 & -2 J & -J & U & 0 & 0 & -\sqrt{3} J \\
 0 & -\sqrt{3} J & 0 & 0 & 0 & 0 & 0 & 3 U & 0 & 0 \\
 0 & 0 & 0 & -\sqrt{3} J & -\sqrt{3} J & 0 & 0 & 0 & 3 U & 0
   \\
 0 & 0 & 0 & 0 & 0 & 0 & -\sqrt{3} J & 0 & 0 & 3 U \\
\end{array}
\right)
\end{split}
\label{3b-mat}
\end{align}  
\end{widetext}

The eigenvalues (in units of $J$) of the bosonic matrix Hamiltonian in Eq.\ (\ref{3b-mat}) are:
\begin{align}
\begin{array}{ll}
E_1 = \; { ^6{\cal R} }^b_1         & \;\;\;\;\;\; E_6 = \; { ^3{\cal R} }^b_2\;(\cu) \\
E_2 = \; { ^3{\cal R} }^b_1         & \;\;\;\;\;\; E_7 = \; { ^6{\cal R} }^b_4 \\
E_3 = \; { ^6{\cal R} }^b_2         & \;\;\;\;\;\; E_8 = \; { ^6{\cal R} }^b_5\\
E_4 = \; { ^6{\cal R} }^b_3         & \;\;\;\;\;\; E_9 = \; { ^3{\cal R} }^b_3 \\
E_5 = \; \cu \;({ ^3{\cal R} }^b_2) & \;\;\;\;\;\; E_{10} = \; { ^6{\cal R} }^b_6,
\end{array}
\label{eigvalb}
\end{align}
\textcolor{black}{
where $\cu=U/J$.
} 
For $E_5$ and $E_6$, the quantities without parentheses apply for $\cu >0$ and those within 
parentheses for $\cu<0$. The expressions for the remaining eigenvalues apply for any $\cu$, negative or 
positive. ${ ^6{\cal R} }^b_i$, $i=1,\ldots,6$ denote in ascending order (for any $\cu$, negative or positive) 
the six real roots of the sixth-order polynomial
\begin{widetext}
\begin{align}
\begin{split}
 P^b_6(x) = & x^6 - 9\cu x^5 + (30\cu^2-22) x^4 \\ 
& + (144\cu-46\cu^3) x^3 + (76 - 314 \cu^2 + 33 \cu^4) x^2 -
(252 \cu - 264 \cu^3 + 9 \cu^5) x - (72 - 180 \cu^2 + 72 \cu^4),
\end{split}
\label{6pb}
\end{align}
\end{widetext}
and ${^3{\cal R} }^b_i$, $i=1,2,3$ denote in ascending order (for any $\cu$, negative or positive) the three real 
roots of the third-order polynomial
\begin{align}
P^b_3(x) = x^3 - 5 \cu x^2 + (7 \cu^2 - 8) x + 18 \cu - 3 \cu^3. 
\label{3pb}
\end{align}

\begin{table}[b]
\caption{\label{tcorr}}
\textcolor{black}{
Correspondence of the energy eigenvalues of the Hubbard matrix Hamiltonian [Eq.\ (\ref{3b-mat})] at the double
degeneracies at $\cu=0$; see Fig.\ \ref{feigvalb}.} 
\begin{ruledtabular}
\begin{tabular}{ccc|ccc}
$E_3(\cu>0)$ & $\Longleftrightarrow$  & $E_4(\cu<0)$ & $E_4(\cu>0)$ & $\Longleftrightarrow$  & $E_3(\cu<0)$ \\
$E_5(\cu>0)$ & $\Longleftrightarrow$  & $E_6(\cu<0)$ & $E_6(\cu>0)$ & $\Longleftrightarrow$  & $E_5(\cu<0)$ \\ 
$E_7(\cu>0)$ & $\Longleftrightarrow$  & $E_8(\cu<0)$ & $E_8(\cu>0)$ & $\Longleftrightarrow$  & $E_7(\cu<0)$ \\
\end{tabular}
\end{ruledtabular}
\end{table}

\textcolor{black}{
At $\cu=0$, a smooth crossing of eigenvalues implies the correspondence displayed in TABLE \ref{tcorr},
associated with the double degeneracies $E_3(\cu=0) = E_4(\cu=0)$, $E_5(\cu=0)= E_6(\cu=0)$, 
and $E_7(\cu=0) = E_8(\cu=0)$. 
}
These remarks are reflected in the choice of online colors (or shading in the 
print grayscale version) for the $\cu>0$ and $\cu<0$ segments of the curves in Fig.\ 
\ref{feigvalb}, where the bosonic eigenvalues listed in Eq.\ (\ref{eigvalb}) are plotted as a function of $\cu$.
Note further that the ordering between $E_4$ and $E_5$ is interchanged for $|\cu| \geq 3\sqrt{2}=4.24264$ [not
visible in Fig.\ \ref{feigvalb}(a) due to the scale of the figure]. 
\textcolor{black}{
In the following, the corresponding 
Hubbard eigenstates are labeled in ascending energy order as
$i=1$, $2$, $3r(4l)$, $4r(3l)$, $5r(6l)$, $6r(5l)$, $7r(8l)$, $8r(7l)$, $9$, $10$,
where ``$r$'' means ``right'' for the region of positive $\cu$ and ``$l$'' means ``left'' for the region of
negative $\cu$.
}

The 10 normalized eigenvectors $\phi^b_i(\cu)$, with $i=1,\ldots,10$, of the bosonic matrix Hamiltonian in Eq.\ 
(\ref{3b-mat}) have the general form
\begin{align}
\begin{split} 
\phi&^b_i(\cu)= \\
\{ & \bc_{111}(\cu),\bc_{210}(\cu),\bc_{201}(\cu),\bc_{120}(\cu),\bc_{021}(\cu),\\
& \bc_{102}(\cu),\bc_{012}(\cu),\bc_{300}(\cu),\bc_{030}(\cu),\bc_{003}(\cu) \}.
\end{split}
\label{phiU} 
\end{align}

Because the algebraic expressions for the $\bc_{ijk}$'s for an arbitrary $\cu$ are very long and complicated,
we explicitly list in this paper the Hubbard eigenvectors only for the characteristic limits of infinite 
repulsive and attractive interaction ($\cu \rightarrow \pm \infty$) and for the non-interacting case ($\cu=0$).
\textcolor{black}{
Specifically, for the reader's convenience, we list in the main text only the Hubbard eigenvectors for the 
ground- and first-excited states; see Sec.\ \ref{eigvecuinf} and Sec.\ \ref{eigvecu0}. The eigenvectors for the
remaining 8 excited states are given in Appendix \ref{a11} (for $\cu \rightarrow \pm \infty$) and 
Appendix \ref{a12} (for $\cu=0$). 
}
     
\subsection{The infinite repulsive or attractive interaction ($\cu \rightarrow \pm \infty$) limit}
\label{eigvecuinf}

For large values of $|\cu|$ ($\cu \rightarrow \pm \infty$), the ten bosonic eigenvalues in Eq.\ (\ref{eigvalb}) 
(in units of $J$) are well approximated by the simpler expressions:
\begin{align}
\begin{split}
E_1^{+\infty} (E_{10}^{-\infty})= &\; -8/\cu + 20/\cu^3 \\
E_2^{+\infty} (E_9^{-\infty})   = &\; \cu\mp\sqrt{5}-3/(4\cu) \\
E_3^{+\infty} (E_8^{-\infty})   = &\; \cu\mp\sqrt{5}+33/(20\cu) \\
E_4^{+\infty} (E_7^{-\infty})   = &\; \cu+1/(5\cu) \\
E_5^{+\infty} (E_6^{-\infty})   = &\; \cu \\
E_6^{+\infty} (E_5^{-\infty})   = &\;  \cu\pm\sqrt{5}-3/(4\cu) \\
E_7^{+\infty} (E_4^{-\infty})   = &\;  \cu\pm\sqrt{5}+33/(20\cu) \\
E_8^{+\infty} (E_3^{-\infty})   = &\; 3\cu+3/(2\cu)-9/(4\cu^3) \\
E_9^{+\infty} (E_2^{-\infty})   = &\; 3\cu+3/(2\cu)+3/(4\cu^3) \\
E_{10}^{+\infty} (E_1^{-\infty})= &\; 3\cu+3/\cu+7/(2\cu^3),
\end{split}
\label{eigvalb2}
\end{align}
where symbols $E_i^{+\infty}$ without a parenthesis and the upper signs in $\mp$ and $\pm$ refer to the positive 
limit $\cu \rightarrow +\infty$, and those ($E_i^{-\infty}$) within a parenthesis and the lower signs in $\mp$ 
and $\pm$ refer to the negative limit $\cu \rightarrow -\infty$. 

From the above, one sees that for large $\pm|\cu|$ the bosonic eigenvalues are organized in three groups:
a high-energy (low-energy) group of three eigenvalues around $\pm 3 |\cu|$ (triply occupied sites, see below), a 
middle-energy group of six eigenvalues around $\pm|\cu|$ (doubly occupied sites, see below), and a single 
negative and lowest (positive and highest) eigenvalue approaching zero (singly occupied sites, see below). Fig.\ 
\ref{feigvalb} illustrates this behavior.

The corresponding eigenvectors at $\cu \rightarrow +\infty$ and $\cu \rightarrow -\infty$ for the ground and
first-excited states are given by
\begin{align}
\begin{split}
\phi^{b,+\infty}_1  &= \{1,0,0,0,0,0,0,0,0,0\}\\
\phi^{b,-\infty}_1 &= \{0,0,0,0,0,0,0,0,1,0\}
\end{split}
\label{phi1}
\end{align}
\begin{align}
\begin{split}
\phi^{b,+\infty}_2 &=
\left\{0,-\frac{1}{2},-\frac{1}{2\sqrt{5}},-\frac{1}{\sqrt{5}},\frac{1}{\sqrt{5}},\frac{
   1}{2 \sqrt{5}},\frac{1}{2},0,0,0\right\}\\ 
\phi^{b,-\infty}_2 &= \{0,0,0,0,0,0,0,-\frac{1}{\sqrt{2}},0,\frac{1}{\sqrt{2}} \}
\end{split}
\label{phi2}
\end{align}

\textcolor{black}{
The eigevectors for the remaining 8 excited states are listed in Appendix \ref{a11}. 
}
Note that the eigenvectors in Eqs. (\ref{phi1}) and (\ref{phi2}) and in Appendix A are grouped
in pairs ($+\infty$, $-\infty$), which are displayed using a common equation number

The eigenvectors at $\cu \rightarrow +\infty$ and $\cu \rightarrow -\infty$ are pairwise related 
as follows:
\begin{align}
\begin{array}{ll}
\phi^{b,+\infty}_1 = - \phi^{b,-\infty}_{10} & \;\;\;\;\;\; \phi^{b,+\infty}_6 = - \phi^{b,-\infty}_9 \\
\phi^{b,+\infty}_2 = - \phi^{b,-\infty}_5    & \;\;\;\;\;\; \phi^{b,+\infty}_7 = - \phi^{b,-\infty}_8 \\
\phi^{b,+\infty}_3 = - \phi^{b,-\infty}_4    & \;\;\;\;\;\; \phi^{b,+\infty}_8 = \phi^{b,-\infty}_3 \\
\phi^{b,+\infty}_4 =  \phi^{b,-\infty}_7     & \;\;\;\;\;\; \phi^{b,+\infty}_9 = \phi^{b,-\infty}_2 \\
\phi^{b,+\infty}_5 =  \phi^{b,-\infty}_6     & \;\;\;\;\;\; \phi^{b,+\infty}_{10} = \phi^{b,-\infty}_1
\end{array}.
\label{phirelUpmI}
\end{align}
The pairs in Eq.\ (\ref{phirelUpmI}) correspond to states with the same absolute eigenvalues $|E_i^{+\infty}|$
and $|E_j^{-\infty}|$ (with $i,j=1,\ldots,10$) given in Eq.\ (\ref{eigvalb2}).  

\subsection{The noninteracting ($\cu =0$) limit}
\label{eigvecu0}

When $\cu=0$, the polynomial-root eigenvalues listed in Eq.\ (\ref{eigvalb}) simplify to
\begin{align}
\begin{array}{ll}
E_1 = \; - 3 \sqrt{2}  & \;\;\;\;\;\; E_6 = \; 0  \\
E_2 = \; - 2 \sqrt{2}  & \;\;\;\;\;\; E_7 = \; \sqrt{2} \\
E_3 = \; - \sqrt{2}    & \;\;\;\;\;\; E_8 = \; \sqrt{2} \\
E_4 = \; - \sqrt{2}    & \;\;\;\;\;\; E_9 = \; 2 \sqrt{2} \\
E_5 = \; 0                & \;\;\;\;\;\; E_{10} = \; 3 \sqrt{2}
\end{array},
\label{eigvalb3}
\end{align}
  
The $\cu=0$ Hubbard ground-state eigenvector is given by 
\begin{align}
\begin{split}
\phi_1^{b,\cu=0} =\left\{\frac{\sqrt{3}}{4},
   \frac{\sqrt{\frac{3}{2}}}{4},
   \frac{\sqrt{3}}{8},
   \frac{\sqrt{3}}{4},
   \frac{\sqrt{3}}{4},
   \frac{\sqrt{3}}{8},
   \frac{\sqrt{\frac{3}{2}}}{4},
   \frac{1}{8},\frac{1}{2\sqrt{2}},
   \frac{1}{8}\right\}
\end{split}
\label{eigvecU0st1}
\end{align} 
whereas the first-excited state is represented by the eigenvector
\begin{align}
\begin{split}
& \phi_2^{b,\cu=0}= \\
& \left\{0,-\frac{1}{2},-\frac{1}{4
   \sqrt{2}},-\frac{1}{2 \sqrt{2}},\frac{1}{2
   \sqrt{2}},\frac{1}{4
   \sqrt{2}},\frac{1}{2},-\frac{\sqrt{\frac{3}{2}}}{4
   },0,\frac{\sqrt{\frac{3}{2}}}{4}\right\}.
\end{split}
\label{eigvecU0st2}
\end{align} 

\textcolor{black}{
The eigenvectors for the remaining 8 excited states are listed in Appendix \ref{a12}.
}

\begin{figure*}[t]
\includegraphics[width=17.5cm]{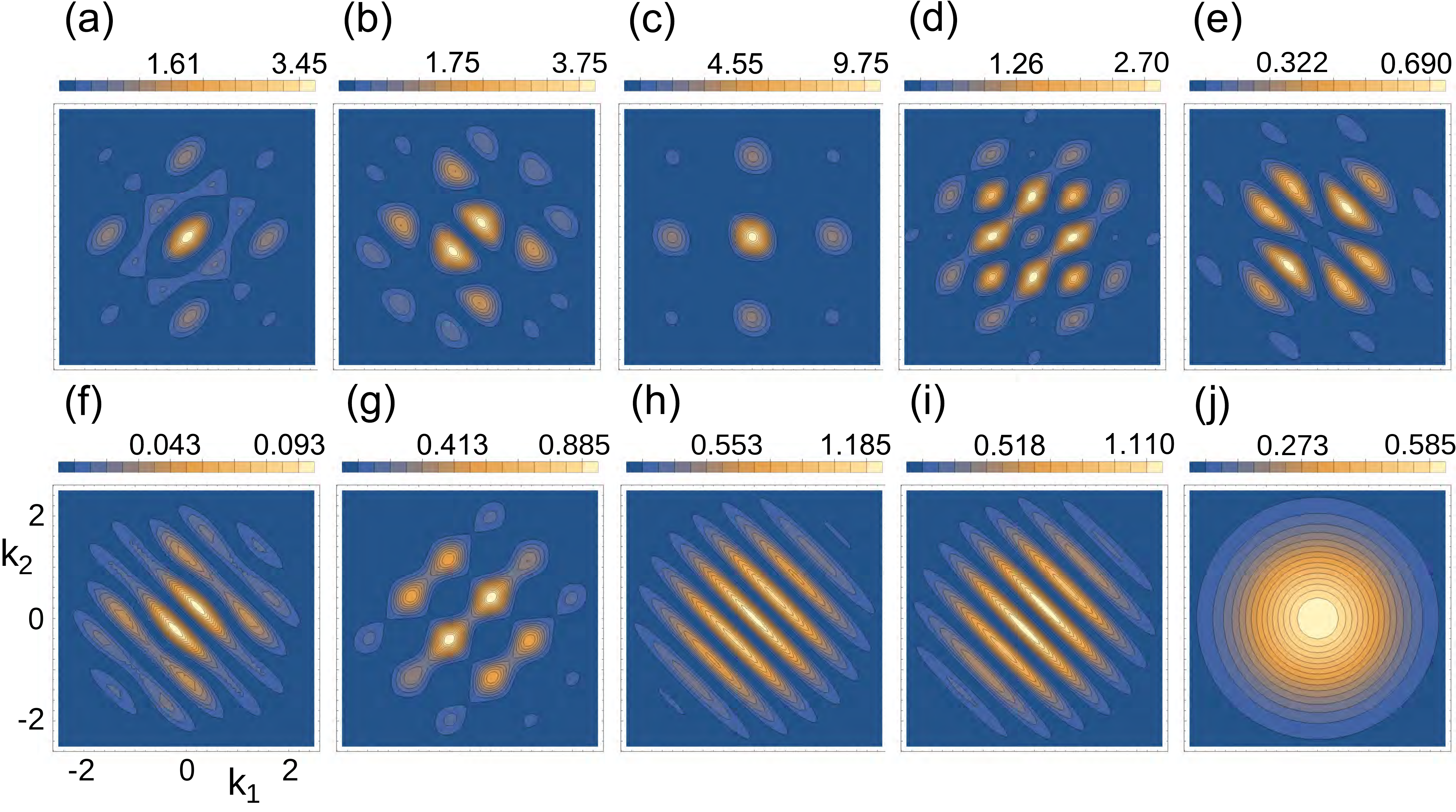}
\caption{Cuts ($k_3=0$) of 3rd-order momentum correlation maps, $^3{\cal G}_i^{b,+\infty}=
\Phi_i^{b,+\infty}\Phi_i^{b,+\infty,*}$, corresponding to the momentum-space wave functions for three bosons 
in three wells [see Eqs.\ (\ref{phibUpmI_1})-(\ref{phibUpmI_2}) 
and Eqs.\ (\ref{phibUpmI_3})-(\ref{phibUpmI_10}), top lines].
(a) Ground state ($i=1$). (b) First-excited sate ($i=2$). (c) Second-excited sate ($i=3$).
(d) Third-excited sate ($i=4$). (e) Fourth-excited sate ($i=5$). (f) Fifth-excited sate ($i=6$).
(g) Sixth-excited sate ($i=7$). (h) Seventh-excited sate ($i=8$). (i) Eighth-excited sate ($i=9$).
(j) Ninth-excited sate ($i=10$).
The choice of parameters is: interwell distance $d=3.8$ $\mu$m and 
spectral width of single-particle distribution in momentum space [see Eq.\ (\ref{psikd})] 
being the inverse of $s=0.5$ $\mu$m. 
\textcolor{black}{
The correlation functions $^3{\cal G}_i^{b,+\infty}(k_1,k_2,k_3=0)$ (map landscapes) are given in units of 
$\mu$m$^3$ according to the color bars on top of each panel, and the momenta $k_1$ and $k_2$ are in units of 
1/$\mu$m. 
}
The value of the plotted correlation functions was multiplied by a factor of 10 to achieve better contrast 
for the map features. 3rd-order momentum correlation maps for the infinite attractive limit are not 
explicitly plotted due to the equalities between pairs of the Hubbard eigenvectors at 
$\cu \rightarrow -\infty$ and $\cu \rightarrow +\infty$; see Eq.\ (\ref{phirelUpmI}) for the detailed
association of states.}
\label{f3rdcorrb}
\end{figure*}

\section{Higher-order correlations in momentum space: Outline of general definitions}
\label{hordcorr}

To motivate our discussion about momentum-space correlation functions, it is convenient to recall that, 
usually, a configuration-interaction (CI) calculation (or other exact diagonalization schemes used for 
solution of the microscopic many-body Hamiltonian) yields a many-body wave function expressed in position 
coordinates. 
Then the $N$th-order {\it real space\/} density, $\rho(x_1,x_1^\prime,x_2,x_2^\prime,...,x_N,x_N^\prime)$,
for an $N$-particle system is defined as the product of the many-body wave function 
$\Psi(x_1, x_2, \ldots,x_N)$ and its complex conjugate $\Psi^*(x_1^\prime, x_2^\prime, \ldots,x_N^\prime)$ 
\cite{lowd55}. The $i$th-order density function (with $i \leq N$) is defined as an integral over $\rho$ taken 
over the coordinates $x_{i+1},\ldots,x_N$ of $N-i$ particles, i.e.,
\begin{align}
\begin{split}
&\rho_i(x_1,x_1^\prime,x_2,x_2^\prime, \ldots ,x_i,x_i^\prime)=\\
& \int dx_{i+1} \dots dx_N \rho(x_1,x_1^\prime,..,x_i,x_i^\prime,x_{i+1},x_{i+1},...x_N,x_N).
\end{split}
\label{rhoi}
\end{align} 

To obtain the $i$th-order real space correlation, one simply sets the prime coordinates in Eq.\ (\ref{rhoi}) to
be equal to the corresponding unprimed ones, 
\begin{align}
^i{\cal G}(x_1,x_2,...,x_i)=\rho_i(x_1,x_1,x_2,x_2,...,x_i,x_i).
\label{g1sp}
\end{align}

Knowing the real-space density, one can obtain the corresponding higher-order momentum correlations through a 
Fourier transform \cite{bran17,bran18,yann19.1,alvi12}
\begin{align}
\begin{split}
&^i{\cal G} (k_1,k_2,\ldots,k_i) = \\
& \frac{1}{4\pi^2} \int e^{ i k_1 ( x_1-x_1') } e^{i k_2 ( x_2-x_2')}\ldots
e^{i k_i ( x_i-x_i')} \\
& \times  \rho_i (x_1, x_1', x_2, x_2',\ldots, x_i, x_i')  
dx_1 dx_1' dx_2 dx_2'\ldots dx_i dx_i',
\end{split}
\label{tbmc}
\end{align}

In this paper, we obtain directly an expression for the momentum-space $N$-body wave function 
corresponding to the Hubbard model Hamiltonian. This circumvents the need for the above Fourier-transform. 
Instead, consistent with the Fourier-transform relation [Eq.\ (\ref{tbmc}) above], 
the highest-order $N$th-order momentum correlation function is given by the modulus square
\begin{align}
^N{\cal G}(k_1,k_2,...,k_N)=|\Phi(k_1,k_2,...,k_N)|^2,
\label{gnmom}
\end{align}
and, successively, any lower $(N-i)$th-order (with $i=1,\ldots,N-1$) momentum correlation is obtained through 
an integration of the higher $(N-i+1)$th-order correlation over the $k_{N-i+1}$ momentum.  

\section{Third-order momentum correlations for 3 bosons in 3 wells: 
The infinite-interaction limit ($\cu \rightarrow \pm \infty$)}
\label{3rdcorrUpmI}

\textcolor{black}{
To derive the all-order momentum correlations, we augment the finite-site Hubbard model as follows:
Each boson in any of the three wells is represented by a single-particle localized orbital having the form
of a displaced Gaussian function \cite{bran17,bran18,yann19.1,yann19.3}, which in the real configuration space 
has the form
\begin{equation}
\psi_j(x) = \frac{1}{ (2 \pi)^{1/4} \sqrt{s} } \exp \left[ - \frac{(x-d_j)^2}{4s^2} \right].
\label{psixd}
\end{equation}
In Eq.\ (\ref{psixd}), $d_j$ ($j=1,2,3$) denotes the position of each of the three wells and $2s$ is the 
width of the Gaussian function in real configuration space.
}
\textcolor{black}{
In this way, the structure (interwell distances) and the spatial profile of the orbitals of the
trapped particles enter in the augmented Hubbard model.
In momentum space, the corresponding orbital $\psi_j(k)$ is given by the Fourier transform of $\psi_j(x)$, 
namely, $\psi_j(k)=(1/\sqrt{2\pi})\int_{-\infty}^\infty \psi_j(x)\exp(ikx)dx$. Performing this Fourier 
transform, one finds
\begin{equation}
\psi_j(k) = \frac{2^{1/4}\sqrt{s}}{\pi^{1/4}} e^{-k^2 s^2} e^{i d_j k}.
\label{psikd}
\end{equation}
Naturally the spectral witdth of the orbital's profile in the momentum space is $1/s$.
}

\textcolor{black}{
In using orbitals localized on each well, our treatment of the augmented Hubbard trimer is similar to 
Coulson's treatment of the Hydrogen molecule \cite{coul41}. In broader terms, our use of localized orbitals 
(atomic orbitals) belongs to the general methodology in chemistry known as LCAO-MO (linear combination of atomic
orbitals $-$  molecular orbitals \cite{szabobook,wiki2}). 
}

\textcolor{black}{
We stress that the cosinusoidal/sinusoidal dependencies of the momentum correlations derived here [and their 
coefficients $\cc$'s, $\cb$'s, and $\ca$'s; see Eqs.\ (\ref{wfbexpr}), (\ref{2ndbexpr}), and (\ref{frstbexpr}) 
below] do not depend on the precise profile of the atomic orbital, as noted already in Ref.\ \cite{coul41}, 
where the general symbol $\mathfrak{A}(k)$ was used for the Fourier transform of $\psi_0(x)$ at $d_0=0$. 
For the Hydrogen molecule an obvious choice is a Slater-type orbital (see Eqs.\ (35) and (36) in Ref.\ 
\cite{coul41}). The reason behind this behavior is the so-called shift property \cite{shifttt} of the Fourier 
transform, which applies to a displaced profile (centered at $d_j \neq 0$); it states that
\begin{align}
\mathfrak{F}[\psi_j(x)] = \mathfrak{F}[\psi_0(x)] \exp(ikd_j) = \mathfrak{A}(k)\exp(ikd_j),  
\label{shift}
\end{align}
where $\mathfrak{F}$ denotes the Fourier-transform operation \cite{shifttt}. The Fourier-transformed profile 
$\mathfrak{A}(k)$ at the initial site factors out in all expressions of the momentum correlations.
The Gaussian profile (also used in aforementioned experimental publications \cite{prei19,berg19,bech20,bonn18})
in our paper was used for convenience; it is an obvious approximation for the lowest single-particle 
level in a deep potential \cite{note5} approaching a harmonic trap in the framework of experiments on neutral 
ultracold atoms \cite{note6}.  
}

\textcolor{black}{
For a discussion of the comparison, for the entire range of interatomic interactions, $\cu$, between exact 
microscopic diagonalization of the Hamiltonian (configuration interaction, CI) calculations, results of the
augmented Hubbard-model, and measurements from trapped ultracold-atoms experiments, see Ref.\ \cite{note8}.
}

With the help of the single-boson orbitals in Eq.\ (\ref{psikd}), each basis ket in Eq.\ (\ref{3b-kets}) can
be mapped onto a wave function of the three single-particle momenta $k_1$, $k_2$, and $k_3$. For each ket, this 
wave function naturally is a permanent built from the three bosonic orbitals. For a general eigenvector 
solution of the Hubbard Hamiltonian, the corresponding wave function $\Phi^b_i(k_1, k_2, k_3)$ 
(with $i=1,\ldots,10$) in momentum space is a sum over such permanents, and the associated third-order 
correlation function is simply the modulus square, i.e., 
\begin{align}
^3{\cal G}^b_i (k_1,k_2,k_3)=|\Phi^b_i(k_1,k_2,k_3)|^2.
\label{3gdef}
\end{align} 
Because the expressions for the third-order correlations can become very long and cumbersome, 
for bookkeeping purposes, we found advantageous to display and characterize instead the three-body wave functions 
$\Phi^b_i(k_1,k_2,k_3)$ themselves. Then the associated third-order correlations can be calculated using 
Eq.\ (\ref{3gdef}).

Below, in Eqs.\ (\ref{phibUpmI_1})-(\ref{phibUpmI_2}), we list without commentary the momentum-space wave 
functions, $\Phi^{b,\pm\infty}_1(k_1,k_2,k_3)$ and $\Phi^{b,\pm\infty}_2(k_1,k_2,k_3)$, associated with the 
Hubbard eigenvectors, $\phi^{b,\pm\infty}_1$ and $\phi^{b,\pm\infty}_2$, respectively
[see Eqs.\ (\ref{phi1})-(\ref{phi2})], at the limits of infinite repulsive or attractive strength (i.e., for 
$\cu \rightarrow \pm \infty$). The commentary integrating these wave functions into the broader scheme of their 
evolution as a function of any interaction strength $-\infty < \cu < +\infty$ is left for Sec.\ \ref{s3rdanyu} 
below. The three-body wave functions for the remaining 8 excited states are listed in Appendix \ref{a1}.
Note that the wave functions in Eqs. (\ref{phibUpmI_1}) and (\ref{phibUpmI_2}) below and in Appendix C are grouped
in pairs ($+\infty$, $-\infty$), which are displayed using a common equation number

Assuming that the wells are linearly placed at $d_1 = -d$, $d_2 = 0$, and $d_3 = d$,
these momentum-space wave functions at $\cu \rightarrow \pm \infty$ are as follows:
\begin{widetext}
\begin{align}
\begin{split}
\Phi^{b,+\infty}_1(k_1, k_2, k_3) & = 
\frac { 2 \times 2^{1/4} } { \sqrt{3} \pi^{3/4} } s^{3/2} e^{ -(k_1^2+k_2^2+k_3^2)s^2 } 
[ \cos( d(k_1-k_2) ) + \cos( d(k_1-k_3) ) + \cos( d(k_2-k_3) ) ], \\
\Phi^{b,-\infty}_1(k_1, k_2, k_3) & = 
\left( \frac{2}{\pi} \right)^{3/4} s^{3/2} e^{ -(k_1^2+k_2^2+k_3^2)s^2 }.
\end{split}
\label{phibUpmI_1}
\end{align}
\begin{align}
\begin{split}
\Phi^{b,+\infty}_2(& k_1, k_2, k_3) =  
   \frac{ i 2^{3/4} } { 5 \sqrt{3} \pi ^{3/4} } s^{3/2} e^{ -(k_1^2+k_2^2+k_3^2)s^2 } \\ 
   & \times \left[ \sqrt{5} \sin (d(-k_1+k_2+k_3))+\sqrt{5} \sin (d(k_1+k_2-k_3)) \right.
   +\sqrt{5} \sin (d(k_1-k_2+k_3)) \\
   &\;\; +5 \sin (d(k_1+k_2)) +5 \sin (d(k_1+k_3)) + 5 \sin (d(k_2+k_3)) +2 \sqrt{5} \sin (d k_1)
   \left. +2 \sqrt{5} \sin (d k_2)+2 \sqrt{5} \sin (dk_3)\right], \\
\Phi^{b,-\infty}_2(& k_1, k_2, k_3) =  
    \frac{ 2 i 2^{1/4} } {\pi ^{3/4} } s^{3/2} e^{ -(k_1^2+k_2^2+k_3^2)s^2 }
   \sin (d(k_1+k_2+k_3)).
\end{split}
\label{phibUpmI_2}
\end{align}
\end{widetext}

Plots for the corresponding 3rd-order momentum correlations $^3{\cal G}^{b,+\infty}_i (k_1,k_2,k_3)$,
with $i=1,\ldots,10$ [see Eq.\ (\ref{3gdef})], are presented in Fig.\ \ref{f3rdcorrb}.
We note that we do not explicitly plot the 3rd-order momentum correlations for the limit of infinite
attraction ($\cu \rightarrow -\infty$) because $^3{\cal G}^{b,-\infty}_i (k_1,k_2,k_3)=
^3{\cal G}^{b,+\infty}_j (k_1,k_2,k_3)$ for the pairs $(i=1,j=10)$, $(i=2,j=9)$, $(i=3,j=8)$, $(i=4,j=3)$,
$(i=5,j=2)$, $(i=6,j=5)$, $(i=7,j=4)$, $(i=8,j=7)$, $(i=9,j=6)$, and $(i=10,j=1)$ due to the equalities 
between eigenvectors listed in Eq.\ (\ref{phirelUpmI}).  

{\it Explicit expression for the third-order correlation $^3{\cal G}^{b,+\infty}_1 (k_1,k_2,k_3)$.\/}
Because of the special role played by the ground state $\phi^{b,+\infty}_1=\booo$ at infinite repulsion, we
explicitly list below the corresponding third-order correlation function, i.e.,
\begin{align}
\begin{split}
&^3{\cal G}^{b,+\infty}_1 (k_1,k_2,k_3)=|\Phi^{b,+\infty}_1(k_1,k_2,k_3)|^2=\\
& \frac{2 \sqrt{2}}{3 \pi^{3/2}} s^3 e^{-2 s^2 (k_1^2+k_2^2+k_3^2)}
 \big\{3 + 2 \cos (d (k_1+k_2-2 k_3)) \\
&   +2 \cos (d (k_2+ k_3 -2 k_1))
   +2 \cos (d (k_1+k_3 - 2 k_2))\\
&  +\cos (2 d (k_1- k_2))
   +2 \cos (d (k_1- k_2))\\
&   +\cos (2 d (k_1- k_3))
   +2 \cos (d (k_1- k_3))\\
&   +\cos (2 d (k_2- k_3))
   +2 \cos (d (k_2-k_3)) \big\}.
\end{split}
\label{3gUIst1}
\end{align}

It is worth noting that the expression (\ref{3gUIst1}) above for 3 bosons is similar to the third-order
correlation for the triplet states (with total spin $S=3/2$ and spin projections $S_z=3/2$ or $S_z=1/2$) for
3-fermions trapped in 3 wells, except that in the fermionic case the sign in front of the cosine terms with
only 2 momenta in the cosine argument is negative; see Refs.\ \cite{yann19.3,prei19}

\section{Third-order momentum correlations for 3 bosons in 3 wells: 
The non-interacting limit $\cu=0$}
\label{3rdcorrU0}

Assuming that the wells are linearly placed at $d_1=-d$, $d_2=0$, and $d_3=d$, the noninteracting ground-state 
three-boson wave function in momentum space is given by
\begin{widetext}
\begin{align}
\begin{split}
& \frac{(2 \pi )^{3/4}}{s^{3/2}} e^{(k_1^2+k_2^2+k_3^2)s^2} \Phi_1^{b,\cu=0}(k_1,k_2,k_3) = 
   1 + 2 \sqrt{2} \cos (d {k_1}) \cos (d {k_2}) \cos (d {k_3}) \\
&  +2 \cos (d {k_1}) \cos(d {k_2})
   +2 \cos (d {k_1}) \cos (d {k_3})+\sqrt{2} \cos (d {k_1})
   +2 \cos (d {k_2}) \cos (d {k_3})+\sqrt{2} \cos (d {k_2})
   +\sqrt{2} \cos (d {k_3}) 
\end{split}
\label{phibU0_1}
\end{align}
The above takes also the form of the general expression (\ref{wfbexpr}) below, i.e.,
\begin{align}
\begin{split}
& \frac{(2 \pi )^{3/4}}{s^{3/2}} e^{(k_1^2+k_2^2+k_3^2)s^2} \Phi_1^{b,\cu=0}(k_1,k_2,k_3) = 
   1 + \sqrt{2} \big( \cos (d k_1) + \cos (d k_2) + \cos (d k_3) \big) \\
&  + \cos [d (k_1-k_2)] + \cos [d (k_1-k_3)] + \cos [d (k_2-k_3)] 
   + \cos [d (k_1+k_2)] + \cos [d (k_1+k_3)] + \cos [d (k_2+k_3)] \\
&  + \frac{1}{\sqrt{2}} \big(\cos [d (k_1+k_2-k_3)] + \cos [d (k_1-k_2+k_3)] + \cos [d (-k_1+k_2+k_3)] 
   + \cos [d (k_1+k_2+k_3)]\big)
\end{split}
\label{phibU0_1_2}
\end{align}

For the first-excited state, the three-boson noninteracting wave function in momentum space at $\cu=0$ was 
found to be
\begin{align}
\begin{split}
& \frac{-i (2 \pi )^{3/4} \sqrt{3}}{s^{3/2}} e^{(k_1^2+k_2^2+k_3^2)s^2} \Phi_2^{b,\cu=0}(k_1,k_2,k_3) = 
   2 \big(\sin (d k_1) + \sin (d k_2) + \sin (d k_3) \big) \\
&  + 2 \sqrt{2} \big(\sin [d (k_1+k_2)] + \sin [d (k_1+k_3)] + \sin [d (k_2+k_3)] \big) \\
&  + \sin [d (k_1-k_2+k_3)] + \sin [d (-k_1+k_2+k_3)] + \sin [d (k_1+k_2-k_3)] 
   + 3 \sin [d (k_1+k_2+k_3)].
\end{split}
\label{phibU0_2}
\end{align}
\end{widetext}

The noninteracting three-body wave functions for the remaining 8 excited states are listed in Appendix 
\ref{a2}.

\section{Third-order momentum correlations for 3 bosons in 3 wells 
as a function of the strength of the interaction $\cu$}
\label{s3rdanyu}

The general cosinusoidal (or sinusoidal) expression of third-order correlations is 
too cumbersome and lengthy to be displayed in
print in a paper. Instead, as mentioned earlier, we give here the general expression for the three-boson wave 
function $\Phi_i^b(k_1,k_2,k_3)$ (with $i=1,\ldots,10$) calculated in the momentum space. Then the third-order 
momentum correlations are obtained simply as the modulus square of this wave function [see Eq.\ (\ref{3gdef})]. 

Using MATHEMATICA, we found that the general cosinusoidal (or sinusoidal) expression of the three-body wave 
function has the form:
\begin{align}
\begin{split}
& \Phi_j^b(k_1,k_2,k_3) = p^j s^{3/2} e^{-(k_1^2+k_2^2+k_3^2)s^2} \\
& \times \{  {\cal C}_0^j + {\cal C}_1^j ({\cal F}(dk_1)+{\cal F}(dk_2)+{\cal F}(dk_3)) \\
& + {\cal C}_{1-1}^j ( {\cal F}[d(k_1-k_2)] + {\cal F}[d(k_1-k_3)] + {\cal F}[d(k_2-k_3)] )\\
& + {\cal C}_{1+1}^j ( {\cal F} [d(k_1+k_2)] + {\cal F} [d(k_1+k_3)] + {\cal F} [d(k_2+k_3)] )\\
& +  {\cal C}_{1+1-1}^j ( {\cal F}[d(k_1+k_2-k_3)]+{\cal F}[d(k_1-k_2+k_3)] \\
&  +  {\cal F}[d(-k_1+k_2+k_3)] ) +  {\cal C}_{1+1+1}^j {\cal F}[d(k_1+k_2+k_3)]  \},
\end{split}
\label{wfbexpr}
\end{align}
where $p^j=1$ and ${\cal F}$ stands for ``$\cos$'' for the states 
$j=1,3r(4l),4r(3l),7r(8l),8r(7l),10$; $p^j=i$ (here $i^2=-1$; it is not an index) and 
${\cal F}$ stands for ``$\sin$'' for the remaining states $j=2,5r(6l),6r(5l),9$. 
\textcolor{black}{The ${\cal C}_0$ coefficient 
denotes an ${\cal F}$-independent term. The subscripts $1$, $1\pm1$, and $1+1\pm1$ in the other $\cc$ 
coefficients reflect the number of terms in the argument of the ${\cal F}$ functions and
the sign in front of each of them (without consideration of any ordering of the $k_1$, $k_2$, and $k_3$ momentum 
variables).} 

\begin{figure}[b]
\includegraphics[width=7.2cm]{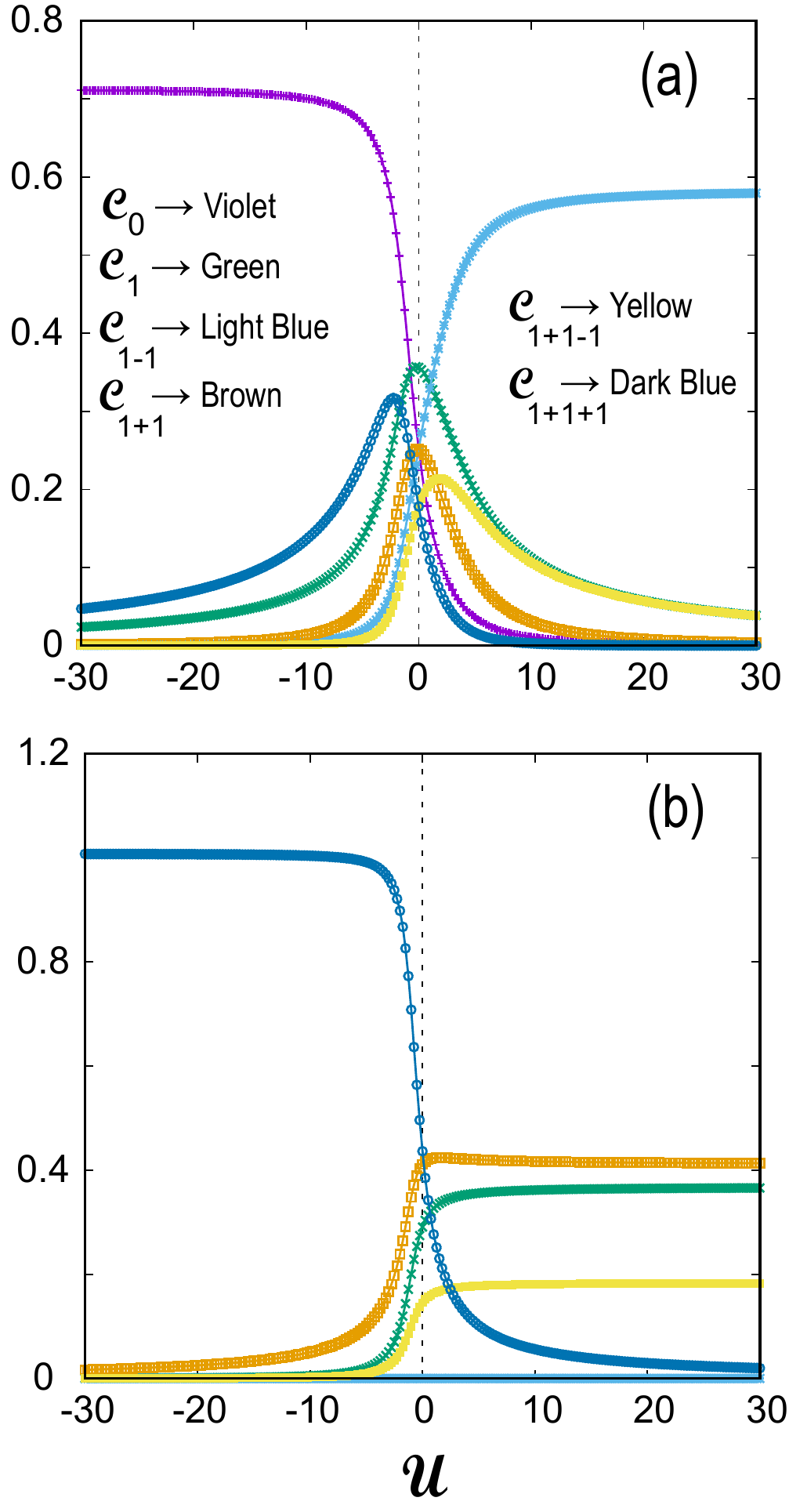}\\
\caption{
\textcolor{black}{
The six different ${\cal C}$-coefficients (dimensionless) [see Eq.\ (\ref{wfbexpr})] for the 2 
lowest-in-energy eigenstates of 3 bosons trapped in 3 linearly arranged wells as a function of $\cu$
(dimensionless).
}
(a) Ground state ($i=1$). (b) First-excited state ($i=2$). See text for a detailed description. 
For a description of the remaining eight excited states, see Appendix \ref{a3rd}.
The choice of online colors is as follows: 
$\cc_0 \rightarrow$ Violet, $\cc_1 \rightarrow$ Green, $\cc_{1-1} \rightarrow$ Light Blue,
$\cc_{1+1} \rightarrow$ Brown, $\cc_{1+1-1} \rightarrow$ Yellow, $\cc_{1+1+1} \rightarrow$ Dark Blue.
\textcolor{black}{
For the print grayscale version, the positioning (referred to as \#$n$, with $n=1,2,3,\dots$)
of the curves from top to bottom at the point $\cu = -6$ 
is as follows: (a) $\cc_0 \rightarrow$ \#1, $\cc_1 \rightarrow$ \#3, $\cc_{1-1} \rightarrow$ \#5,
$\cc_{1+1} \rightarrow$ \#4, $\cc_{1+1-1} \rightarrow$ \#6, $\cc_{1+1+1} \rightarrow$ \#2 and
(b) $\cc_0=0$, $\cc_1 \rightarrow$ \#3, $\cc_{1-1}=0$,
$\cc_{1+1} \rightarrow$ \#2, $\cc_{1+1-1} \rightarrow$ \#4, $\cc_{1+1+1} \rightarrow$ \#1.}
}
\label{ccoef}
\end{figure}

In general, there are 14 cosinusoidal (or sinusoidal) terms and 6 distinct 
$\cu$-dependent coefficients ${\cal C}$'s 
for a given state in expression (\ref{wfbexpr}). 
We note that ${\cal C}_0 \equiv 0$ and ${\cal C}_{1-1} \equiv 0$ for any $\cu$ for all
the states of the second group above for which ${\cal F} \equiv \sin$. 
The ${\cal C}$-coefficients for the 2 lowest-in-energy eigenstates are plotted in Fig.\
\ref{ccoef} as a function of $\cu$. The corresponding explicit numerical values can be found in a data file
included in the supplemental material \cite{supp}.
 
\begin{figure*}[t]
\includegraphics[width=17.5cm]{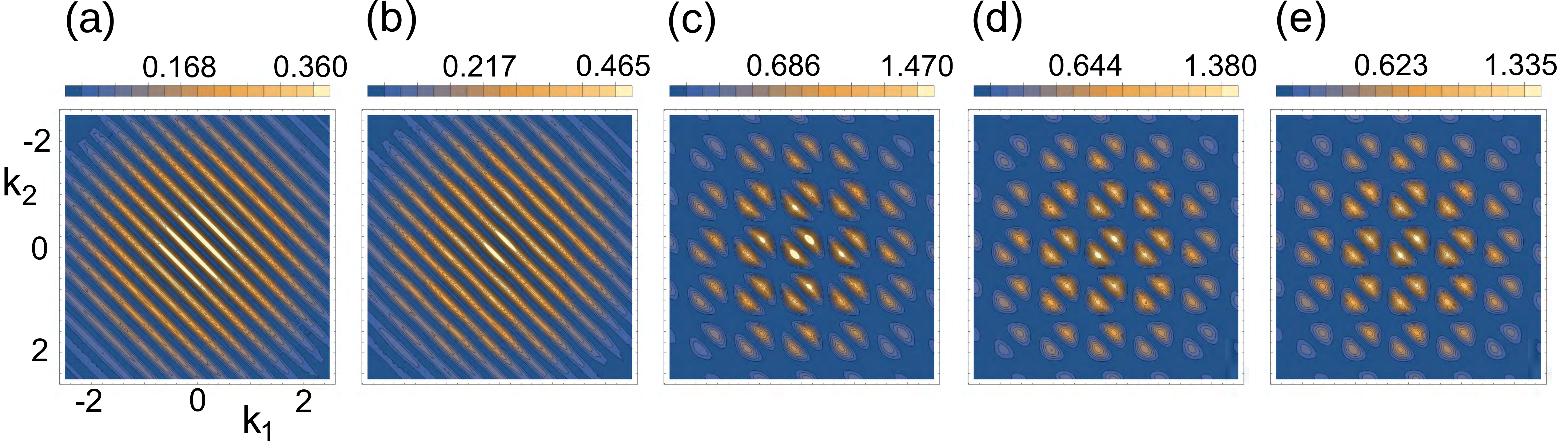}
\caption{Cuts ($k_3=0$) of 3rd-order momentum correlation maps for the first-excited state of 3 bosons in 3 wells
[see Eqs.\ (\ref{3gdef}) and (\ref{wfbexpr}) with $i=2$].
(a) $\cu=-200$. (b) $\cu=-10$. (c) $\cu=0$. (d) $\cu=10$. (e) $\cu=200$.
The choice of parameters is: interwell distance $d=7$ $\mu$m and spectral width of single-particle distribution 
in momentum space [see Eq.\ (\ref{psikd})] being the inverse of $s=0.35$ $\mu$m. The correlation functions 
$^3{\cal G}_i^b(k_1,k_2,k_3=0)$ (map landscapes) are given in units of $\mu$m$^3$ 
according to the color bars on top of each panel, and the momenta $k_1$ and $k_2$ are in units of 1/$\mu$m.
The value of the plotted correlation functions was multiplied by a factor of 10 to achieve better contrast 
for the map features.}
\label{f3rdcorrbst2}
\end{figure*}

\begin{table*}[t]
\caption{\label{bgsu0}
\textcolor{black}{
The 9 distinct coefficients at $\cu=0$ present in Eq.\ (\ref{2ndbexpr}) in the case of the ground state.}} 
\begin{ruledtabular}
\begin{tabular}{ccccccccc}
$\cb^{1,\cu=0}_0$ & $\cb^{1,\cu=0}_1$ & $\cb^{1,\cu=0}_2$ & $\cb^{1,\cu=0}_{1-1}$ & $\cb^{1,\cu=0}_{2-2}$ &
$\cb^{1,\cu=0}_{2-1}$ & $\cb^{1,\cu=0}_{1+1}$ & $\cb^{1,\cu=0}_{2+2}$ & $\cb^{1,\cu=0}_{2+1}$ \\ \hline
$2/\pi=$ & $2\sqrt{2}/\pi=$ & $1/\pi=$ & $2/\pi=$ & $1/(4\pi)=$ &
$1/(\sqrt{2}\pi)=$ & $2/\pi=$ & $1/(4\pi)=$ & $1/(\sqrt{2}\pi)=$ \\
$ 0.63662$ & $0.90032$ & $0.31831$ & $0.63662$ & $0.07958$ &
$0.22508$ & $0.63662$ & $0.07958$ & $0.22508$
\end{tabular}
\end{ruledtabular}
\end{table*}

{\it The ground state (state denoted as $i=1$ for $-\infty < \cu < +\infty$):\/}
For $\cu \rightarrow -\infty$, it is seen from the panel (a) in Fig.\ \ref{ccoef} that only the constant 
coefficient $\cc^{1,-\infty}_0=(2/\pi)^{3/4}=0.7127$ survives in expression (\ref{wfbexpr}); the ground-state 
in momentum space is given by the second expression in Eq.\ (\ref{phibUpmI_1}). 
It is a simple Gaussian distribution associated with a Bose-Einstein condensate, reflecting
the fact that all three bosons are localized in the middle well and occupy the same orbital; the corresponding
Hubbard eigenvector is given by $\phi^{b,-\infty}_1$ [second line in Eq.\ (\ref{phi1})] which contains only a
single component from the primitive kets listed in Eq.\ (\ref{3b-kets}), i.e., the basis ket No. 9 
$\rightarrow |030\rangle$.  

For $\cu =0$, all 6 coefficients, $\cc^{1,\cu=0}$'s, are present, and their numerical values from the frame (a) in 
Fig.\ \ref{ccoef} agree with the corresponding algebraic expressions for $\Phi_1^{b,\cu=0}(k_1,k_2,k_3)$ in
Eq.\ (\ref{phibU0_1_2}).

For $\cu \rightarrow +\infty$, only the coefficient $\cc^{1,+\infty}_{1-1}=
2\times2^{1/4}/(\sqrt{3}\pi^{3/4}) = 0.5819$ survives in expression (\ref{wfbexpr});
see again panel (a) in Fig.\ \ref{ccoef}. The ground-state in momentum space comprises three cosinusoidal
terms and is given by the first expression in Eq.\ (\ref{phibUpmI_1}). This form corresponds to the Hubbard 
eigenvector $\phi^{b,+\infty}_1$ [first line in Eq.\ (\ref{phi1})] which contains only a single component from 
the primitive kets listed in Eq.\ (\ref{3b-kets}), i.e., the basis ket No. 1 $\rightarrow |111\rangle$. 

As mentioned earlier, the primitive ket $|111\rangle$ represents a case where all three wells are singly occupied.
Thus it enables a direct mapping to quantum-optics investigations of the frequency-resolved interference of 
three temporally distinguishable photons prepared in three separate fibers (tritter) 
\cite{tamm19} [recall the analogies
\cite{yann19.1}: particle momentum ($k$) $\leftrightarrow$ photon frequency ($\omega/c$) and interwell distance 
($d$) $\leftrightarrow$ time-delay between single photons ($\tau c$)].

{\it The first excited state (state denoted as $i=2$ for $-\infty < \cu < +\infty$):\/}
For $\cu \rightarrow -\infty$ only the coefficient 
${\cc}^{2,-\infty}_{1+1+1}= 2\times 2^{1/4}/\pi^{3/4} = 1.0079$ survives 
in expression (\ref{wfbexpr}) [see frame (b) in Fig.\ \ref{ccoef}]; 
the corresponding state, $\phi^{b,-\infty}_2$ [second line in Eq.\ (\ref{phi2})], is a NOON state of the form 
$(-|300\rangle + |003\rangle)/\sqrt{2}$, and the corresponding wave function in momentum
space is given by the second expression in Eq.\ (\ref{phibUpmI_2}), {\it which includes a single sin term only\/}. 

For $\cu=0$, four coefficients are present, namely ${\cc}^{2,\cu=0}_1$, ${\cc}^{2,\cu=0}_{1+1}$,
${\cc}^{2,\cu=0}_{1+1-1}$, and ${\cc}^{2,\cu=0}_{1+1+1}$. Their numerical values from frame (b) in 
Fig.\ \ref{ccoef} agree with the corresponding algebraic expressions for $\Phi_2^{b,\cu=0}(k_1,k_2,k_3)$ in
Eq.\ (\ref{phibU0_2}).

For $\cu \rightarrow +\infty$ only three coefficients, 
${\cc}^{2,+\infty}_1=2\times2^{3/4}/(\sqrt{15}\pi^{3/4})=0.3680$, 
${\cc}^{2,+\infty}_{1+1}=2^{3/4}/(\sqrt{3}\pi^{3/4})=0.4115$, and
${\cc}^{2,+\infty}_{1+1-1}=2^{3/4}/(\sqrt{15}\pi^{3/4})=0.1840$, 
survive in expression (\ref{wfbexpr}) [see frame (b)
in Fig.\ \ref{ccoef}]; the corresponding state, $\phi^{b,+\infty}_2$ [first line in Eq.\ (\ref{phi2})] consists
of all 6 primitive kets [see Eq.\ (\ref{3b-kets})] representing exclusively doubly-occupied wells, and the 
corresponding wave function in momentum space has 9 {\it sinusoidal\/} terms and is given by the first 
expression in Eq.\ (\ref{phibUpmI_2}).  

In the main text of this paper, we restrict the  $\cu$-evolution of the $\cc(\cu)$'s coefficients in Eq.\
(\ref{wfbexpr}) to the two lowest-in-energy states. Indeed the ground state and the first excited state
are the natural candidates for initial experiments. For example, for the case of two and three ultracold fermions
($^6$Li atoms), see Ref.\ \cite{berg19} and Ref.\ \cite{prei19}, respectively; for recent experiments focused on 
the ground state of large bosonic Hubbard systems, see Refs.\ \cite{grei02} and \cite{gerb05.2} ($^{87}$Rb atoms)
and Ref.\ \cite{clem18,clem19} ($^4$He$^*$ atoms). In the case of trapped ultracold atoms other excited states
are in principle accessible. Thus in anticipation of future experimental activity, we complete in Appendix 
\ref{a3rd} the description of the details of the $\cu$-evolution of the $\cc(\cu)$'s for the 
remaining eight excited states. 

Fig.\ \ref{f3rdcorrbst2} illustrates visually for the first-excited state ($i=2$) the $\cu$-evolution of the 
third-order correlation maps described by expressions (\ref{3gdef}) and (\ref{wfbexpr}) when $i=1$. The maps for
5 characteristic values of $\cu$ are plotted, namely, $\cu=-200$, $-10$, $0$, $10$, and $200$. 
Corresponding illustrations for the ground state are left for Sec.\ \ref{sign}. 

\begin{figure}[t]
\includegraphics[width=7.2cm]{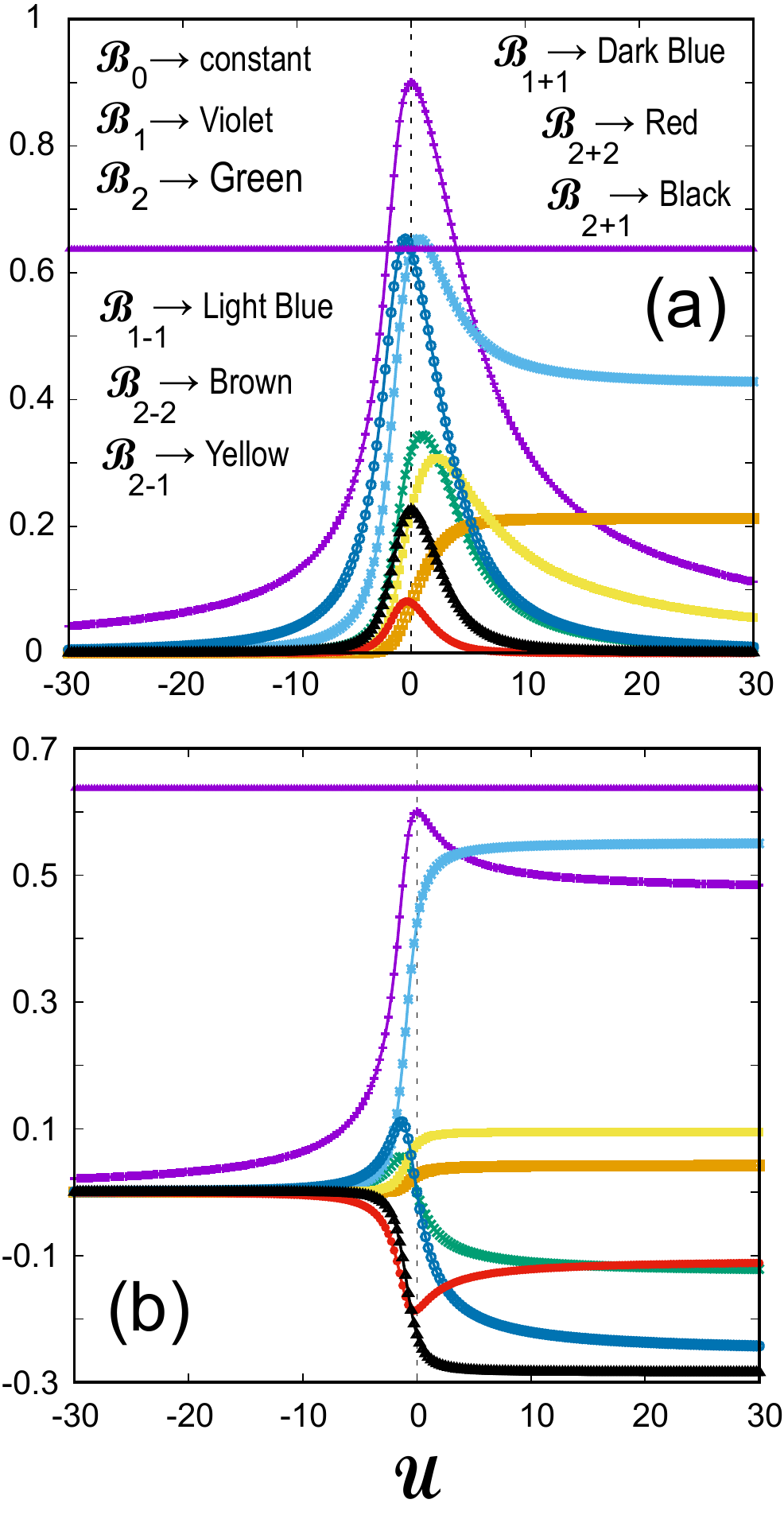}\\
\caption{The ${\cal B}$-coefficients (dimensionless) [see Eq.\ (\ref{2ndbexpr})] for the 2 lowest-in-energy 
eigenstates of 3 bosons trapped in 3 linearly arranged wells as a function of the interaction strength $\cu$
(dimensionless). (a) ground 
state ($i=1$). (b) First-excited state ($i=2$). See text for a detailed description. 
The choice of online colors is as follows: 
$\cb_0 \rightarrow$ Constant (Violet), $\cb_1 \rightarrow$ Second Violet,
$\cb_2 \rightarrow$ Green, $\cb_{1-1} \rightarrow$ Light Blue, $\cb_{2-2} \rightarrow$ Brown,
$\cb_{2-1} \rightarrow$ Yellow, $\cb_{1+1} \rightarrow$ Dark Blue,
$\cb_{2+2} \rightarrow$ Red, $\cb_{2+1} \rightarrow$ Black.
For the print grayscale version, the positioning (referred to as \#$n$, with $n=1,2,3,\dots$)
of the curves from top to bottom at the point $\cu = +2$
is as follows: (a) $\cb_0 {\rm (constant)} \rightarrow$ \#3, $\cb_1 \rightarrow$ \#1, $\cb_2 \rightarrow$ \#5,
$\cb_{1-1} \rightarrow$ \#2, $\cb_{2-2} \rightarrow$ \#8, $\cb_{2-1} \rightarrow$ \#6, 
$\cb_{1+1} \rightarrow$ \#4, $\cb_{2+2} \rightarrow$ \#9, $\cb_{2+1} \rightarrow$ \#7 and
(b) $\cb_0 {\rm (constant)} \rightarrow$ \#1, $\cb_1 \rightarrow$ \#2, $\cb_2 \rightarrow$ \#6,
$\cb_{1-1} \rightarrow$ \#3, $\cb_{2-2} \rightarrow$ \#5, $\cb_{2-1} \rightarrow$ \#4,
$\cb_{1+1} \rightarrow$ \#7, $\cb_{2+2} \rightarrow$ \#8, $\cb_{2+1} \rightarrow$ \#9. 
For a description of the remaining eight excited states, see Appendix \ref{a2nd}.
}
\label{bcoef}
\end{figure}

\begin{figure*}[t]
\includegraphics[width=17.5cm]{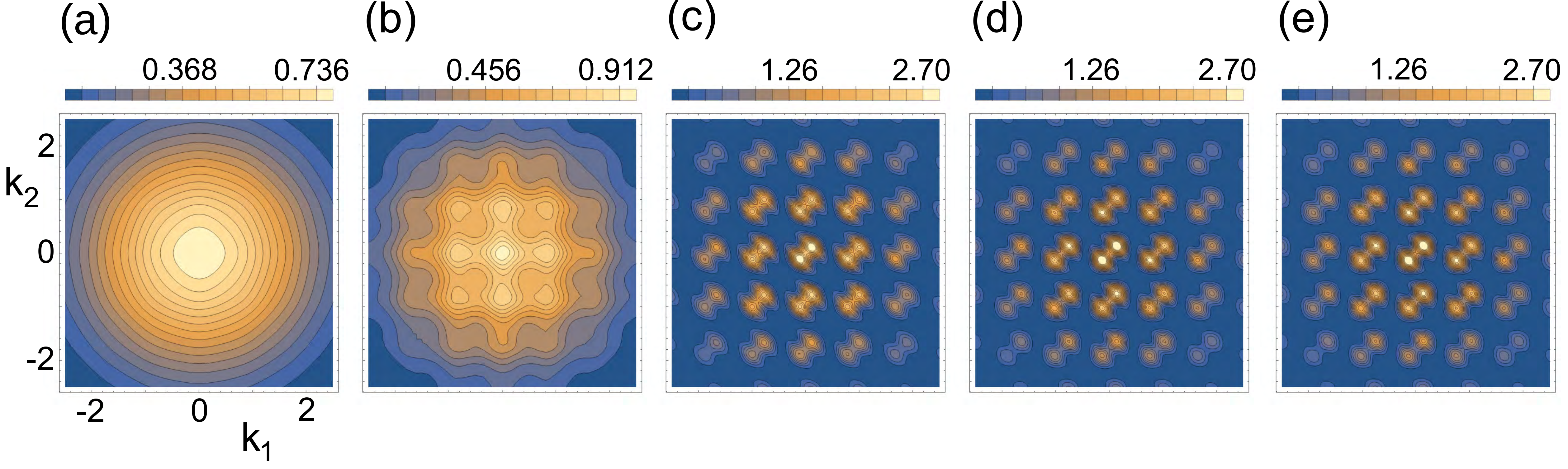}
\caption{2nd-order momentum correlation maps for the first-excited state of 3 bosons in 3 wells
[see Eq.\ (\ref{2ndbexpr}) with $i=2$]. 
(a) $\cu=-200$. (b) $\cu=-10$. (c) $\cu=0$. (d) $\cu=10$. (e) $\cu=200$.
The choice of parameters is: interwell distance $d=7$ $\mu$m and spectral width of single-particle 
distribution in momentum space [see Eq.\ (\ref{psikd})] being the inverse of $s=0.35$ $\mu$m. 
The correlation functions $^2{\cal G}_i^{b}(k_1,k_2)$ (map landscapes) are given in units of $\mu$m$^2$ 
according to the color bars on top of each panel, and the momenta $k_1$ and $k_2$ are in units of 1/$\mu$m.
The value of the plotted correlation functions was
multiplied by a factor of 10 to achieve better contrast for the map features.}
\label{f2ndcorrbst2}
\end{figure*}

\begin{table*}[t]
\caption{\label{bst2u0}
\textcolor{black}{
The 7 distinct coefficients at $\cu=0$ present in Eq.\ (\ref{2ndbexpr}) in the case of the 1st-excited state.}}
\begin{ruledtabular}
\begin{tabular}{ccccccccc}
$\cb^{2,\cu=0}_0$ & $\cb^{2,\cu=0}_1$ & $\cb^{2,\cu=0}_2$ & $\cb^{2,\cu=0}_{1-1}$ & $\cb^{2,\cu=0}_{2-2}$ &
$\cb^{2,\cu=0}_{2-1}$ & $\cb^{2,\cu=0}_{1+1}$ & $\cb^{2,\cu=0}_{2+2}$ & $\cb^{2,\cu=0}_{2+1}$ \\ \hline
$2/\pi=$ & $4\sqrt{2}/\pi=$ & 0  & $4/(3\pi)=$ & $1/(12\pi)=$ &
$1/(3\sqrt{2}\pi)=$ &  0  & $-7/(12\pi)=$ & $-1/(\sqrt{2}\pi)=$ \\
$ 0.63662$ & $0.60021$ & ~~  & $0.42441$ & $0.026526$ &
$0.075026$ &  ~~  & $-0.185681$ & $-0.22508$
\end{tabular}
\end{ruledtabular}
\end{table*}

\begin{table*}[t]
\caption{\label{bst2uinf}
\textcolor{black}{
The 9 distinct coefficients at $\cu \rightarrow +\infty$ present in Eq.\ (\ref{2ndbexpr}) in the case of the 
1st-excited state.}}
\begin{ruledtabular}
\begin{tabular}{ccccccccc}
$\cb^{2,+\infty}_0$ & $\cb^{2,+\infty}_1$ & $\cb^{2,+\infty}_2$ & $\cb^{2,+\infty}_{1-1}$ & 
$\cb^{2,+\infty}_{2-2}$ & $\cb^{2,+\infty}_{2-1}$ & $\cb^{2,+\infty}_{1+1}$ & 
$\cb^{2,+\infty}_{2+2}$ & $\cb^{2,+\infty}_{2+1}$ \\ \hline
$2/\pi$ & $2\sqrt{5}/(3\pi)$ & $-2/(5\pi)$ & $26/(15\pi)$ & $2/(15\pi)$ &
$2/(3\sqrt{5})$ & $-4/(5\pi)$ & $-1/(3\pi)$ & $-2/(\sqrt{5}\pi)$ 
\end{tabular}
\end{ruledtabular}
\end{table*}

\section{Second-order momentum correlations for 3 bosons in 3 wells 
as a function of the strength of the interaction $\cu$}
\label{s2ndanyu}

The second-order correlations are obtained through an integration of the third-order ones over the third 
momentum variable $k_3$, i.e.,
\begin{align}
^2{\cal G}_i^{b}(k_1,k_2)=\int^{\infty}_{-\infty}\; ^3{\cal G}_i^{b}(k_1,k_2,k_3)dk_3,
\label{2nd}
\end{align}
with $i=1,\ldots,10$.

Using MATHEMATICA and neglecting the terms that vanish as $e^{-\gamma d^2/s^2}$ (for arbitrary $\gamma >0$ and 
$d^2/s^2 >> 1$), we found that the second-order correlations are given by the following general expression
\begin{align}
\begin{split}
& ^2{\cal G}_i^{b}(k_1,k_2)= s^2 e^{-2(k_1^2+k_2^2)s^2} \\
& \times \{  {\cal B}_0^i + {\cal B}_{1}^i (\cos(dk_1)+\cos(dk_2)) \\
& + {\cal B}_{2}^i (\cos(2dk_1)+\cos(2dk_2)) \\
& + {\cal B}_{1-1}^i \cos[d(k_1-k_2)] + {\cal B}_{2-2}^i \cos[2d(k_1-k_2)] \\
& + {\cal B}_{2-1}^i (\cos[d(k_1-2k_2)]+\cos[d(2k_1-k_2)])\\
& + {\cal B}_{1+1}^i \cos[d(k_1+k_2)] + {\cal B}_{2+2}^i \cos[2d(k_1+k_2)] \\
& +  {\cal B}_{2+1}^i (\cos[d(k_1+2k_2)]+\cos[d(2k_1+k_2)])  \}.
\end{split}
\label{2ndbexpr}
\end{align}

\textcolor{black}{The $\cb_0$ coefficient denotes a $\cos$-independent term. The subscripts $1$, $2$, $1\pm1$, 
$2\pm1$, and $2\pm2$ in the other $\cb$ coefficients reflect the number of terms in the argument of the 
$\cos$ functions (one or two) and the factor of $\pm1$ or $\pm2$ in front of $k_1$ or $k_2$ (without 
consideration of any ordering of $k_1$ and $k_2$).}   

Including the constant term, there are 13 sinusoidal terms, but only 9 distinct coefficients in Eq.\ 
(\ref{2ndbexpr}). The first coefficient above is a constant, i.e., ${\cal B}^i_0=2/\pi \approx 0.63662$ for all 
ten eigenstates. The remaining 8 ${\cal B}$-coefficients in Eq.\ (\ref{2ndbexpr}) are $\cu$-dependent.
These $\cu$-dependent ${\cal B}$-coefficients for the 2 lowest-in-energy eigenstates are plotted in Fig.\ 
\ref{bcoef} as a function of $\cu$. The corresponding explicit numerical values can be found in a data file 
included in the supplemental material \cite{supp}. 
Note that expression (\ref{2ndbexpr}) has a total of 13 different cosine terms. 

{\it The ground state (state denoted as $i=1$ for $-\infty < \cu < +\infty$):\/} 
For $\cu \rightarrow -\infty$ only the constant term, ${\cal B}^1_0$ survives; see the frame (a) in Fig.\
\ref{bcoef}. The ground state is the triply occupied middle well [see the Hubbard eigenvector in the second line
of Eq.\ (\ref{phi1})]. In this case, the second-order correlation function is
\begin{align}
^2{\cal G}_1^{b,-\infty}(k_1,k_2)= \frac{2}{\pi}s^2 e^{-2(k_1^2+k_2^2)s^2}.
\label{3b2ndggsUmI}
\end{align}

\textcolor{black}{
In the noninteracting case ($\cu=0$), for which the Hubbard eigenvector is given by Eq.\ (\ref{eigvecU0st1}), 
all 13 cosinusoidal terms and 9 distinct coefficients (listed in TABLE \ref{bgsu0}) are present in Eq.\ 
(\ref{2ndbexpr}), in agreement with frame (a) of Fig.\ \ref{bcoef}. 
}

For $\cu \rightarrow +\infty$, three terms survive, including the constant one; see frame (a) in Fig.\ 
\ref{bcoef}. In this case, the ground state is that of all three wells being singly occupied. In this case, the 
second-order correlation function acquires a simple expression
\begin{align}
\begin{split}
^2{\cal G}_1^{b,+\infty}& (k_1,k_2) = \frac{2}{3\pi} s^2 e^{-2(k_1^2+k_2^2)s^2} \{ 3 \\
& + 2 \cos[d(k_1-k_2)] + \cos[2d(k_1-k_2)] \}.
\end{split}
\label{3b2ndggsUI}
\end{align}

It is interesting to note that the second-order correlation function for three fermions with parallel spins
trapped in three wells in the limit $\cu \rightarrow +\infty$ is given by the same expression as that in
Eq.\ (\ref{3b2ndggsUI}), but with the 2 and 1 coefficients in front of the $\cos[d(k_1-k_2)]$ and 
$\cos[2d(k_1-k_2)]$ terms being replaced by their negatives, $-2$ and $-1$, respectively (see Eq.\ (9) and
TABLE I (row for $i=3$) in Ref.\ \cite{yann19.3}). This naturally is a reflection of the different quantum 
statistics between bosons and fermions.  

Fig.\ \ref{f2ndcorrbst2} illustrates for the first-excited state the $\cu$-evolution of the second-order 
correlation maps described by expression (\ref{2ndbexpr}) when $i=2$. The maps for 5 specific values of $\cu$
are plotted, namely, $\cu=-200$, $-10$, $0$, $10$, and $200$. 

{\it The first excited state (state denoted as $i=2$ for $-\infty < \cu < +\infty$):\/}
For $\cu \rightarrow -\infty$ only the constant term, ${\cal B}^2_0=2/\pi$, survives; the corresponding state 
is a NOON state of the form $(-|300\rangle + |003\rangle)/\sqrt{2}$. In this case, the second-order 
correlation function is again
\begin{align}
^2{\cal G}_2^{b,-\infty}(k_1,k_2)= \frac{2}{\pi}s^2 e^{-2(k_1^2+k_2^2)s^2}.
\label{3b2ndgst2UmI}
\end{align}

\textcolor{black}{
In the noninteracting case ($\cu=0$), for which the Hubbard eigenvector is given by Eq.\ (\ref{eigvecU0st2}), 10
cosinusoidal terms and 7 distinct coefficients (listed in TABLE \ref{bst2u0}) are present in Eq.\ (\ref{2ndbexpr}), 
in agreement with frame (b) of Fig.\ \ref{bcoef}. 
}

\textcolor{black}{
For $\cu \rightarrow +\infty$, all 13 sinusoidal terms survive in expression (\ref{2ndbexpr}); 
the corresponding state is given by the first expression in Eq.\ (\ref{phi2}). 
For this case, we give the 9 distinct coefficients in TABLE \ref{bst2uinf}.
}

These results are in agreement with the $\cu$-dependence portrayed in frame (b) of Fig.\ \ref{bcoef}.

For a description of the remaining eight excited states, see Appendix \ref{a2nd}.

\begin{figure}[t]
\includegraphics[width=7.2cm]{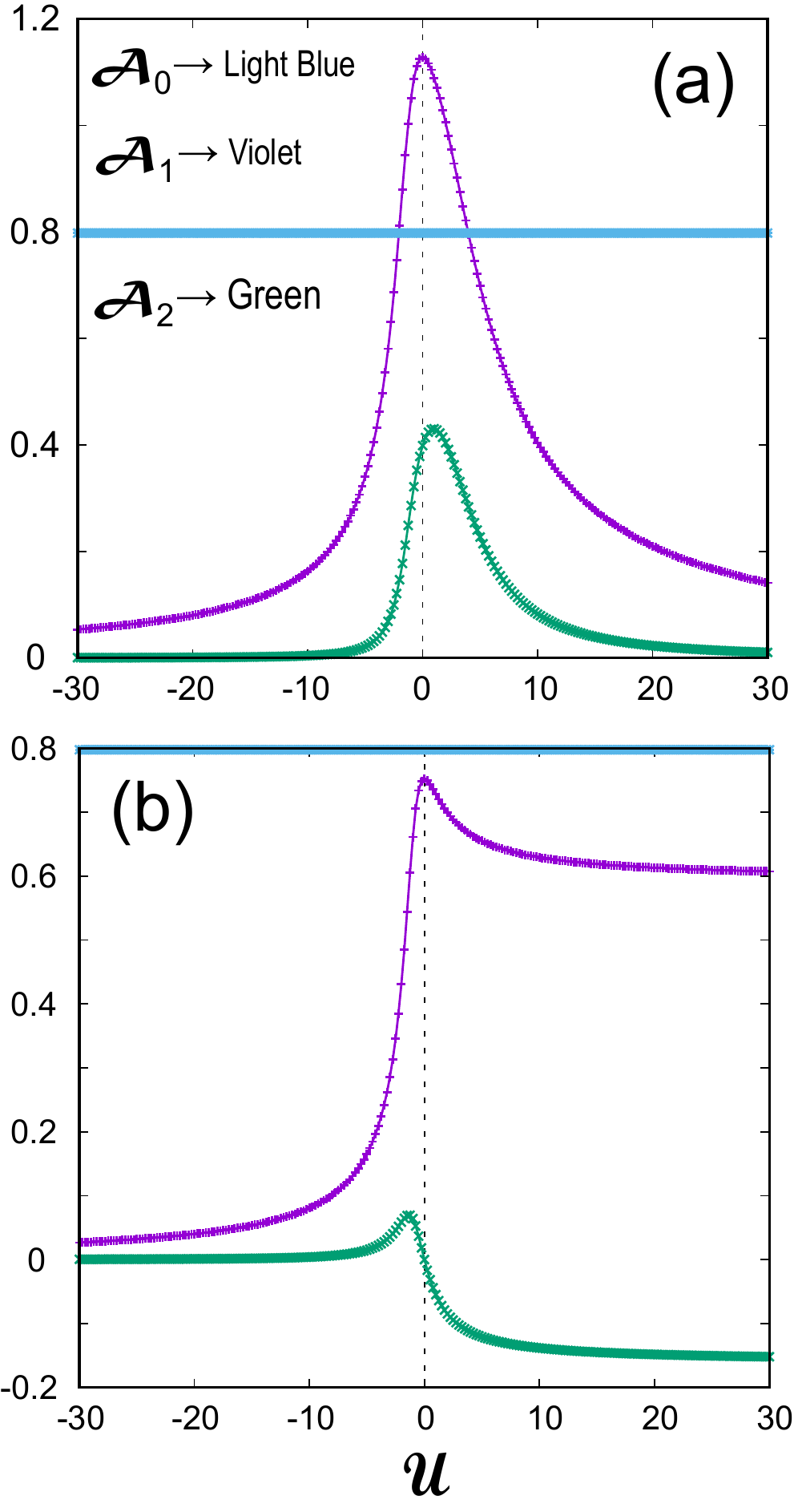}
\caption{The ${\cal A}$-coefficients (dimensionless) [see Eq.\ (\ref{frstbexpr})] for the 2 lowest-in-energy 
eigenstates of 3 bosons trapped in 3 linear wells as a function of the interaction strength $\cu$ (dimensionless). 
(a) ground state ($i=1$). (b) First-excited state ($i=2$). See text for a detailed description. 
The choice of online colors is as follows:
$\ca_0 \rightarrow$ Constant (Light Blue), $\ca_1 \rightarrow$ Violet, $\ca_2 \rightarrow$ Green.
For the print grayscale version, excluding the top constant $\ca_0$ horizontal line, the positioning of the two
remaining curves in both frames is: $\ca_1 \rightarrow$ upper curve, $\ca_2 \rightarrow$ lower curve. 
For a description of the remaining eight excited states, see Appendix \ref{a1st}.
}
\label{acoef}
\end{figure}

\begin{figure*}[t]
\includegraphics[width=17.5cm]{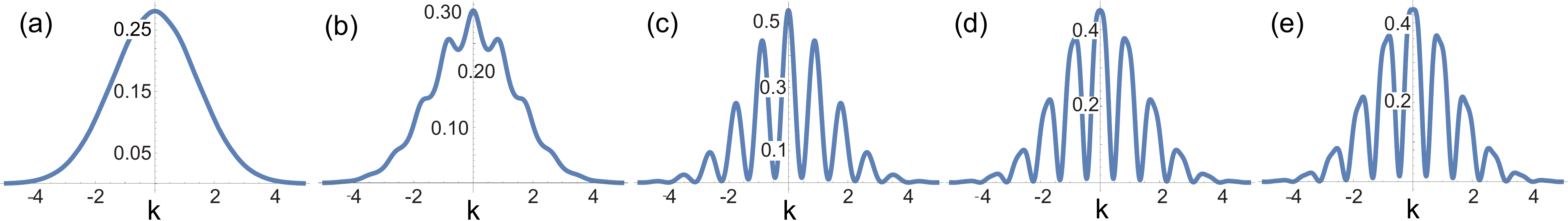}
\caption{1st-order momentum correlation plots for the first-excited state of 3 bosons in 3 wells.
From left to right: (a) $\cu=-200$, (b) $\cu=-10$, (c) $\cu=0$, (d) $\cu=10$, and (e) $\cu=200$ 
[see Eq. (\ref{frstbexpr}) with $i=2$]. 
The correlation functions $^1{\cal G}_i^b(k)$ (vertical axes) are given in units of $\mu$m, and 
the momenta $k$ are in units of 1/$\mu$m.
The choice of parameters is: interwell distance $d=7$ $\mu$m and width of single-particle distribution in
momentum space [see Eq.\ (\ref{psikd})] 
\textcolor{black}{
being the inverse of $s=0.35$ $\mu$m.
}
}
\label{f1stcorrbst2}
\end{figure*}

\section{First-order momentum correlations for 3 bosons in 3 wells 
as a function of the strength of the interaction $\cu$}
\label{s1stanyu}

The first-order correlations are obtained through an integration of the second-order ones [see Eq.\ 
(\ref{2ndbexpr})] over the second momentum variable $k_2$, i.e.,
\begin{align}
^1{\cal G}_i^{b}(k)=\int^{\infty}_{-\infty}\; ^2{\cal G}_i^{b}(k,k_2)dk_2,
\label{frst}
\end{align}
with $i=1,\ldots,10$.
 
Exploiting the computational abilities of MATHEMATICA and neglecting  terms that vanish as $e^{-\gamma d^2/s^2}$ 
(for arbitrary $\gamma >0$ and $d^2/s^2 >> 1$), one can find that the first-order correlations are given by the 
following general expression
\begin{align}
^1{\cal G}_i^{b}(k)= s e^{-2 k^2 s^2} \{ {\cal A}^i_0 + {\cal A}^i_1 \cos(dk) + {\cal A}^i_2 \cos(2dk) \}.
\label{frstbexpr}
\end{align}

${\cal A}^i_0=\sqrt{2/\pi} \approx 0.797885$ above is $\cu$-independent for all ten eigenstates. The remaining two 
coefficients in Eq.\ (\ref{frstbexpr}), $\ca^i_1$ and $\ca^i_2$ are $\cu$-dependent for 9 out of the ten 
eigenstates. These $\cu$-dependent ${\cal A}$-coefficients for the 2 lowest-in-energy eigenstates are 
plotted as a function of $\cu$ in Fig.\ \ref{acoef}. 
The corresponding explicit numerical values can be found in a data file included in the supplemental material 
\cite{supp}. 

{\it The ground state (state denoted as $i=1$ for $-\infty < \cu < +\infty$):\/}
For $\cu \rightarrow -\infty$, it is seen from frame (a) in Fig.\ \ref{acoef} that only the constant
coefficient $\ca^1_0$ survives in expression (\ref{frstbexpr}), i.e., the first-order correlation
(single-particle density) in momentum space is devoid of any oscillatory structure, being given simply by a
Gaussian distribution function,
\begin{align}
^1{\cal G}_1^{b,-\infty}(k)= \sqrt{ \frac{2}{\pi}} s e^{-2 k^2 s^2}.
\label{frstggsUmI}
\end{align}
This structureless distribution corresponds to a photonic triple-slit experiment where Young's \cite{youn04}
``which way'' question, related to the source of the particle detected 
with a time-of-flight measurement, can be answered with a 100\%
certainty as being one single well (zero quantum fluctuations in the single-particle occupation number per site).
Indeed, the corresponding ground-state Hubbard eigenvector is given by $\phi^{b,-\infty}_1$ [second line in Eq.\ 
(\ref{phi1})] which contains only one triply-occupied component from the primitive kets listed in Eq.\ 
(\ref{3b-kets}), i.e., the basis ket No. 9 $\rightarrow |030\rangle$. 

For the non-interacting case ($\cu=0$), all 3 coefficients survive [see frame (a) in Fig.\ \ref{acoef}]; 
specifically one has: 
\begin{align}
^1{\cal G}_1^{b,\cu=0}(k)=\sqrt{ \frac{2}{\pi} } s e^{-2 k^2 s^2} \{1 + \sqrt{2} \cos(dk) + 
\frac{1}{2} \cos(2dk) \}.
\label{frstggsU0}
\end{align}

Expression (\ref{frstggsU0}) exhibits a highly oscillatory interference pattern. It corresponds to the  ground 
state given by the Hubbard eigenvector in Eq.\ (\ref{eigvecU0st1}), which is often described as a bosonic 
superfluid. Indeed the quantum fluctuations in the single-particle occupation number per site are strongest and 
the single-particle bosonic orbitals are maximally delocalized over all three sites.

For $\cu \rightarrow +\infty$, it is seen from frame (a) in Fig.\ \ref{acoef} that again only the 
$\cu$-independent coefficient $\ca^1_0$ survives in expression (\ref{frstbexpr}), i.e., the first-order correlation
(single-particle density) in momentum space is devoid of any oscillatory structure, being given simply by a
Gaussian distribution function like in Eq.\ (\ref{frstggsUmI}), i.e.,
\begin{align}
^1{\cal G}_1^{b,+\infty}(k) = {^1{\cal G}}_1^{b,-\infty}(k).
\label{frstggsUI}
\end{align}
Again, this structureless distribution corresponds to a photonic triple-slit experiment where Young's 
\cite{youn04} ``which way'' question, related to the source of the particle detected 
with a time-of-flight measurement, can be answered with a 100\%
certainty as being one single well (zero quantum fluctuations in the single-particle occupation number per site).
Indeed, the corresponding ground-state Hubbard eigenvector is given by $\phi^{b,+\infty}_1$ [first line in Eq.\
(\ref{phi1})] which contains only the singly-occupied component from the primitive kets listed in Eq.\
(\ref{3b-kets}), i.e., the basis ket No. 1 $\rightarrow |111\rangle$. The implications of the above results 
encoded in Eqs.\ (\ref{frstggsUmI}), (\ref{frstggsU0}), and (\ref{frstggsUI}) regarding phase transitions will be 
discussed below in Sec.\ \ref{sign}.

{\it The first excited state (state denoted as $i=2$ for $-\infty < \cu < +\infty$):\/}
For $\cu \rightarrow -\infty$, it is seen from frame (b) in Fig.\ \ref{acoef} that only the constant
coefficient $\ca^1_0$ survives in expression (\ref{frstbexpr}), i.e., the first-order correlation
(single-particle density) in momentum space is devoid of any oscillatory structure, being given simply by a
Gaussian distribution function,
\begin{align}
^1{\cal G}_2^{b,-\infty}(k)= \sqrt{ \frac{2}{\pi}} s e^{-2 k^2 s^2}.
\label{frstgUmIst2}
\end{align}

In this case, this structureless distribution does not correspond to zero quantum fluctuations in the 
single-particle occupation number per site (see detailed discussion in Sec.\ \ref{sign} below).
Indeed, the corresponding Hubbard eigenvector is given by $\phi^{b,-\infty}_2$ [second line in Eq.\
(\ref{phi2})] which is a NOON state spread over two sites. i.e., it is a superposition of the two basis
kets No. 8 $\rightarrow |300\rangle$ and No. 10 $\rightarrow |003\rangle$. 

For the non-interacting case ($\cu=0$), 2 coefficients survive [see frame (b) in Fig.\ \ref{acoef}];
specifically one has:
\begin{align}
^1{\cal G}_2^{b,\cu=0}(k)=\sqrt{ \frac{2}{\pi} } s e^{-2 k^2 s^2} \{1 + \frac{2\sqrt{2}}{3} \cos(dk)\}.
\label{frstgU0st2}
\end{align}
Expression (\ref{frstgU0st2}) exhibits a highly oscillatory interference pattern. It corresponds to the
state given by the Hubbard eigenvector in Eq.\ (\ref{eigvecU0st2}).

For $\cu \rightarrow +\infty$, all 3 coefficients survive [see frame (b) in Fig.\ \ref{acoef}], one of them
being negative; specifically one has:
\begin{align}
^1{\cal G}_2^{b,+\infty}(k)=\sqrt{ \frac{2}{\pi} } s e^{-2 k^2 s^2} \{1 + \frac{\sqrt{5}}{3} \cos(dk)
-\frac{1}{5} \cos(2dk) \}.
\label{frstgUIst2}
\end{align}

Expression (\ref{frstgUIst2}) exhibits a highly oscillatory interference pattern. It corresponds to the
state given by the Hubbard eigenvector in the first line of Eq.\ (\ref{phi2}), which consists exclusively
of double-single occupancy components [basis kets No. 2 to No. 7; see Eq.\ (\ref{3b-kets})]  

Fig.\ \ref{f1stcorrbst2} illustrates for the first-excited state the $\cu$-evolution of the first-order 
correlations described by expression (\ref{frstbexpr}) when $i=2$. The cases for 5 characteristic values
of $\cu$ are plotted, namely, $\cu=-200$, $-10$, $0$, $10$, and $200$.

For a description of the remaining eight excited states, see Appendix \ref{a1st}.

\begin{figure*}[t]
\includegraphics[width=17.5cm]{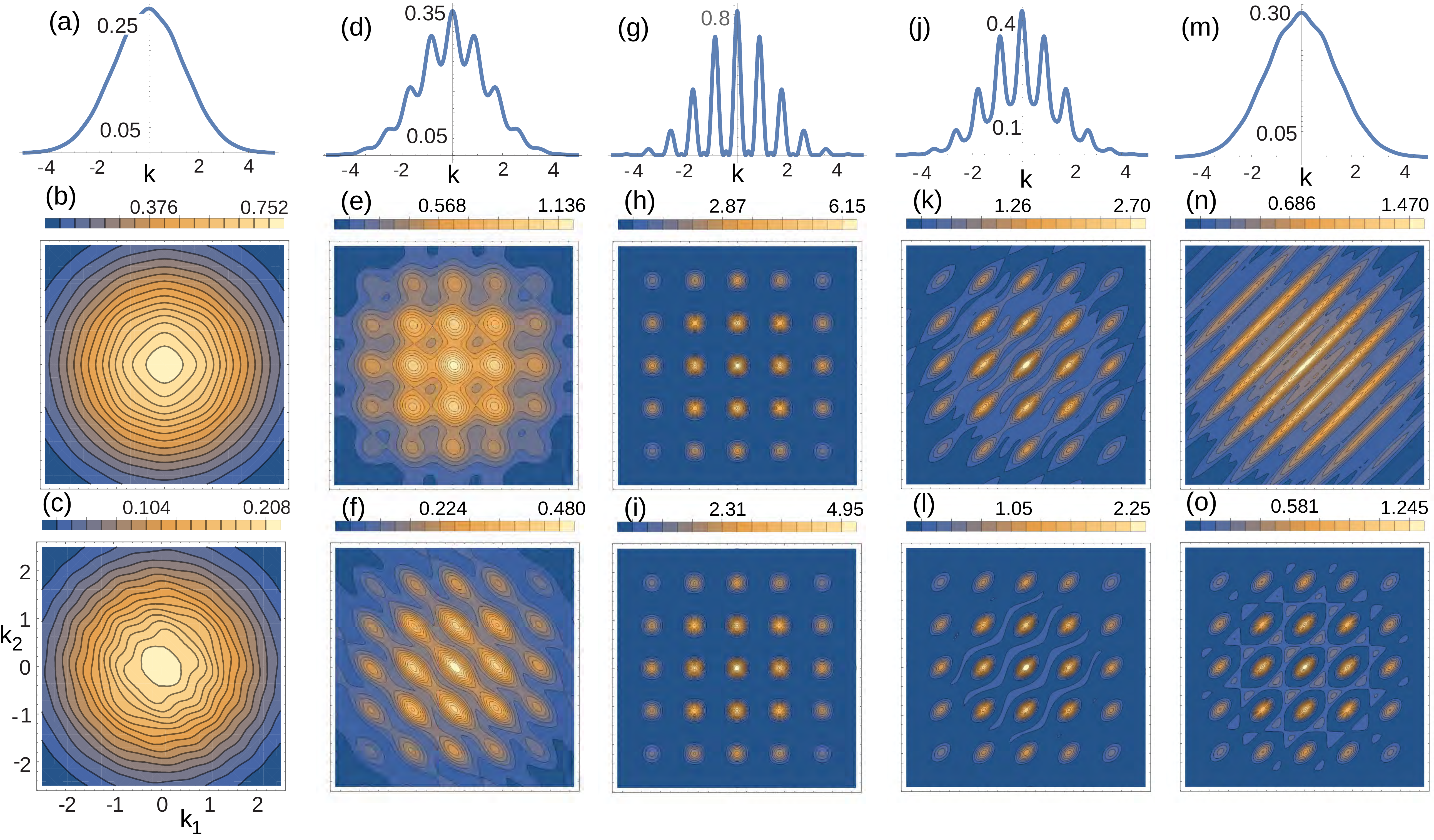}
\caption{Momentum correlation plots and maps for the ground state of 3 bosons in 3 wells.
Top row (a,d,g,j,m): 1st-order correlations $^1{\cal G}_i^b(k)$ (vertical axes) in units of $\mu$m.
Middle row (b,e,h,k,n): 2nd-order correlations $^2{\cal G}_i^b(k_1,k_2)$ in units of $\mu$m$^2$
according to the color bars on top of each panel.
Bottom row (c,f,i,l,o): 3rd-order (cuts at $k_3=0$) correlations $^3{\cal G}_i^b(k_1,k_2,k_3=0)$ 
in units of $\mu$m$^3$ according to the color bars on top of each panel.
The momenta $k$, $k_1$, and $k_2$ are in units of 1/$\mu$m. From left to right column: 
$\cu=-200$, $-10$, $0$, $10$, and $300$ [see Eqs. (\ref{frstbexpr}) and (\ref{2ndbexpr}) 
with $i=1$, as well as Eqs.\ (\ref{3gdef}) and (\ref{wfbexpr}) with $i=1$]. 
The choice of parameters is: interwell distance $d=7$ $\mu$m and 
spectral width of single-particle distribution in momentum space [see Eq.\ (\ref{psikd})] 
being the inverse of $s=0.35$ $\mu$m. 
The value of the plotted correlation functions in the bottom two 
rows was multiplied by a factor of 10 to achieve better contrast for the map features.}
\label{corrbst1}
\end{figure*}

\textcolor{black}{
\section{Signatures of emergent quantum phase transitions}
\label{sign}
}

The system of 3 bosons in 3 wells is a building block of bulk-size systems containing a large number of bosons
(e.g., $^{87}$Rb or $^4$He$^*$ atoms) in 3D, 2D, and 1D optical lattices. Such bulk-like systems have been 
available already for some time and several physical aspects of them have been explored experimentally 
\cite{grei02,gerb05,gerb05.2,clem18,clem19,bloc05}, accompanied by theoretical studies \cite{seng05,triv09}. 
In particular, of direct interest to this paper are the observations, obtained through time-of-flight 
measurements, of the superfluid to Mott insulator phase transition \cite{grei02,gerb05,gerb05.2,clem18,clem19}
(in 3D lattices), and of the second-order particle interference \cite{bloc05} 
(in 1D lattices) in analogy with a quantal extension of Hanburry Brown-Twiss-type optical interference.

The detailed algebraic analysis of all-order correlations presented earlier for the system of 3 bosons in 3 wells
provides the tools for exploring these major physical aspects (quantum phase transitions and quantum-optics 
analogies) in the context of a finite-size system. In this respect, it is a first step towards the deciphering of
the evolution of these aspects as the system size increases from a few particles to the thermodynamic limit. In 
this section, we analyze the signatures for quantum phase transitions that appear already in the case of a finite
system as small as 3 bosons. 

We begin by collecting in a single figure (Fig.\ \ref{corrbst1}) and for the ground state of the 3 bosons-3 wells
systems all three levels of correlations as a function of the interaction strength $\cu$ (with $\cu=-200$, $-10$, 
0, 10, and 300). For large $\cu$ ($\cu=300$, describing very strong repulsive interparticle interaction), 
the system's ground-state Hubbard eigenvector is very close to
the single ket No. 1 $\rightarrow |111\rangle$ [see $\phi_1^{b,+\infty}$ in Eq.\ (\ref{phi1})] which describes
exclusively singly-occupied sites. For 3 bosons in 3 wells, the state $|111\rangle$ is the analog of the Mott 
insulator phase, familiar from bulk systems.  
The associated three-body wave function is well approximated by the permanent 
$\Phi^{b,+\infty}_1(k_1,k_2,k_3)$ [see Eq.\ (\ref{phibUpmI_1})] formed from the three localized orbitals 
$\psi_j(k)$ in Eq.\ (\ref{psikd}). 

A crucial observation is that the corresponding single-particle momentum density 
(first-order correlation) portrayed
in frame (m) of Fig.\ \ref{corrbst1} (in top row) is structureless and devoid of any oscillatory pattern, in
contrast to fully developed oscillations present in the single-particle density of the non-interacting ground
state [see frame (g) in top row of Fig.\ \ref{corrbst1}]. As was the case with the bulk systems, this 
structureless pattern in the first-order correlation can thus be used as a signature of the Mott insulator even in
the case of a small system.     

In analogy with the interpretation for bulk systems, the appearance of oscillations in the non-interacting case
can be associated with the spreading of the single-particle orbitals over all the three sites (three wells).
Namely, for $\cu=0$, the lowest energy single-particle wave function of the tight-binding Hamiltonian (in matrix
representation)
\begin{align}
H_{b,{\rm TB}}^{\rm sp} =
-J \left(
\begin{array}{ccc}
 0 & 1 & 0 \\
 1 & 0 & 1 \\
 0 & 1 & 0 \\
\end{array}
\right)
\label{hbsp}
\end{align} 
is a molecular orbital which is expressed as a coherent linear superposition of all three localized atomic 
orbitals $\psi_j(k)$ [with $j=1,2,3$, see Eq.\ (\ref{psikd})], namely 
\begin{align}
\begin{split}
\psi_{\rm MO}(k)& =\frac{\psi_1(k)}{2} + \frac{\psi_2(k)}{\sqrt{2}} + \frac{\psi_3(k)}{2} \\
& = \frac{2^{1/4}\sqrt{s}}{\pi^{1/4}} e^{-k^2 s^2} 
\left( \frac{e^{-idk}}{2} + \frac{1}{\sqrt{2}} + \frac{e^{idk}}{2} \right).
\end{split}
\label{mo}
\end{align}
Then the three-body wave function is constructed by triply occupying this molecular orbital, i.e., it is
given by the Bose-Einstein-condensate product 
\begin{align}
\Phi^{b,\cu=0}_1(k_1,k_2,k_3) = \psi_{\rm MO}(k_1)\psi_{\rm MO}(k_2)\psi_{\rm MO}(k_3).
\label{phibU0_1_3} 
\end{align}
Eq.\ (\ref{phibU0_1_3}) above
equals expression (\ref{phibU0_1}) derived by us earlier (see Sec.\ \ref{3rdcorrU0}) as the $\cu=0$ 
limit of the solution of the Bose-Hubbard Hamiltonian [Eq.\ (\ref{3b-hub})], obtained through the matrix 
representation [Eq.\ (\ref{3b-mat})] in the 10-ket basis [Eq.\ (\ref{3b-kets})] for the problem of three bosons 
trapped in three-wells.

Because of the molecular orbital in Eq.\ (\ref{mo}), which expresses the delocalization of the single-particle
wave functions over the whole system, the three-body wave function $\Phi^{b,\cu=0}_1(k_1,k_2,k_3)$ can be
characterized as describing a superfluid phase in analogy with the bulk case \cite{grei02,fish89}. 
The natural difference of course is that in the bulk case the superfluid to Mott-insulator transition happens 
abruptly at $\cu= z \times 5.8$ \cite{fish89}, with $z$ being the number of next neighbors of a lattice site, 
whereas for the small finite system this transition is not sharp but proceeds continuously as a function of $\cu$. 
Some steps of this smooth evolution are illustrated in frame (g) ($\cu=0$), frame (j) ($\cu=10$), and 
frame (m) ($\cu=300$) of Fig.\ \ref{corrbst1} (top row).

Furthermore, another aspect from the bulk studies that is relevant to our 3-boson results is the determination,
made deeply in the Mott-insulator region, of a small oscillatory contribution to the single-particle density 
superimposed on the structureless background \cite{gerb05,gerb05.2,seng05,triv09}. This contribution \cite{note1}
was found to vary as $\propto -2\sum_{\nu=x,y,z} \cos(k_\nu d)/\cu$, as obtained via perturbative (or related) 
approaches around $\cu \rightarrow +\infty$. 
Our exact algebraic expression for $^1{\cal G}_i^{b}(k)$ [Eq.\ (\ref{frstbexpr})], which is valid for
any $\cu$, contains a second term $\cos(2dk)$ in addition to the $\cos(dk)$ term. Deeply in the Mott-insulator
regime, however, there is agreement at the qualitative level between our result and the bulk one, because the
coefficient $\ca^1_2$ vanishes much faster than the coefficient $\ca^1_1$ as $\cu \rightarrow +\infty$ as 
is revealed by an inspection of the curves in frame (a) of Fig.\ \ref{acoef}. 

At the non-interacting limit ($\cu=0$), however, this second term cannot be neglected [see frame (a) in 
Fig.\ \ref{acoef} and Eq.\ (\ref{frstggsU0})]. In this limit, its effect is to narrow the width of the 
cosinusoidal peaks at $k=2\pi j/d$, with $j=0,1,2,\ldots$. From this, one can conjecture \cite{yannun} that for 
larger systems with $N$ bosons, all cosine terms of the form $\cos(ndk)$ with $n=1,2,3,\ldots,N-1$ (corresponding 
to all possible interwell distances) will contribute. The summation of many of such terms will enhance further
the shrinking of the width of the main peaks, while it will give a practically vanishing result in the
in-between regions. Thus the main peaks will acquire the shape of sharp spikes as was indeed observed 
\cite{grei02} in the bulk systems.  

\begin{figure*}[t]
\includegraphics[width=13.0cm]{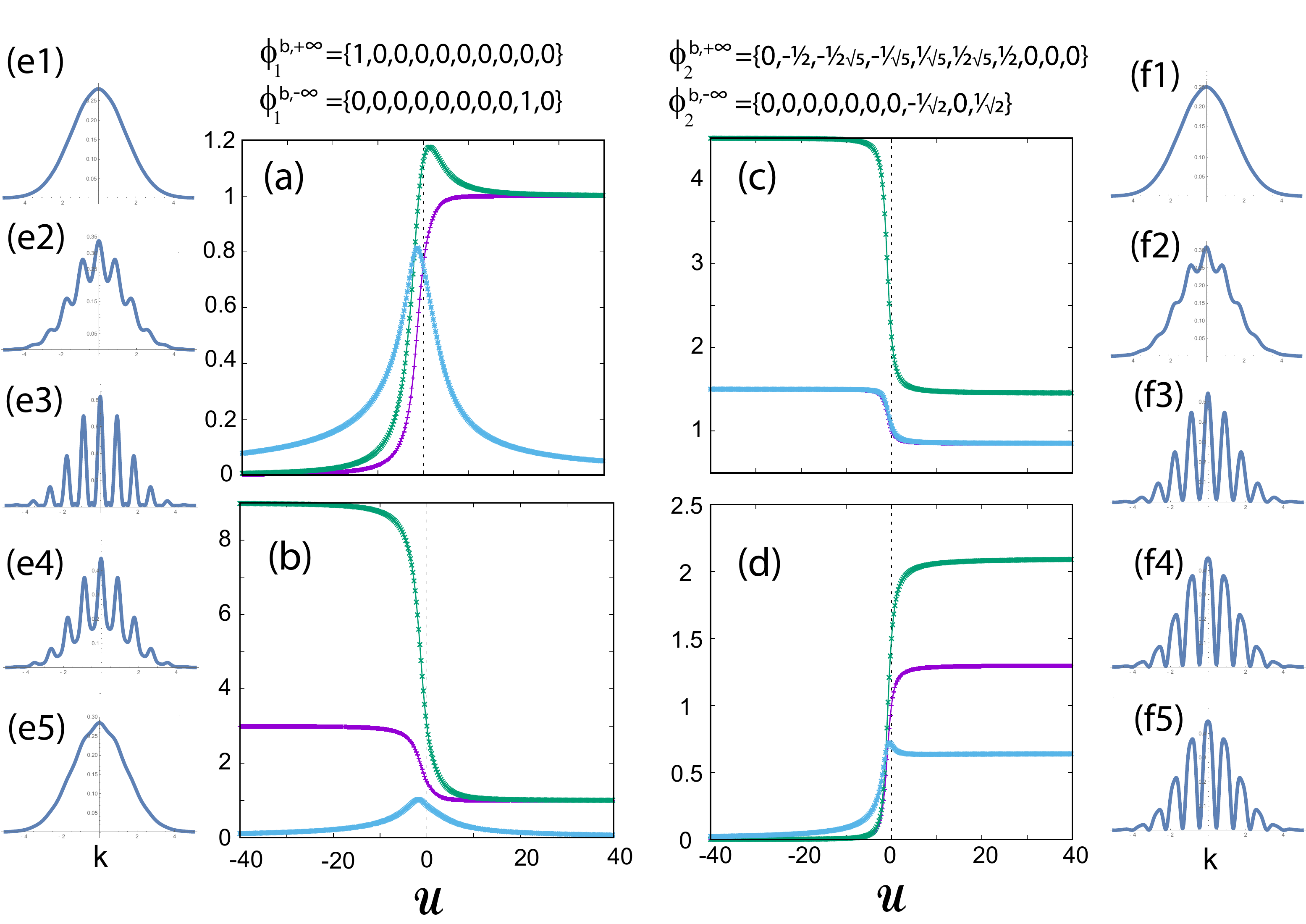}
\caption{(a,b,c,d) Site occupations (vertical axes, dimensionless) and 
their fluctuations (vertical axes, dimensionless) for the ground, $\phi_1^b(\cu)$ (a,b), and 
first-excited, $\phi_2^b(\cu)$ (c,d), states as a function of the strength $\cu$ of the interaction.
Panels (a) and (c) refer to the left site (well), whereas panels (b) and (d) refer to the middle site (well).   
Violet color (midle curve at $\cu=+40$): site occupations. Green color (upper curve at $\cu=+40$): 
expectation value of the square of the site number operator. 
Light blue color (lower curve at $\cu=+40$): standard deviation. Note that the middle and lower
curves in frame (c) coincide for all practical purposes.
(ex,fx) The first-order correlations for the ground and first-excited states, respectively, for
five characteristic values : $\cu=-200$ (x=1) (close to $\rightarrow -\infty$), -10 (x=2), 0 (x=3), 10 (x=4), 
and 300 (x=5) (close to $\rightarrow +\infty$). The first-order correlations $^1{\cal G}^b(k)$ (vertical axes) 
are in units of $\mu$m and the momenta $k$ are in units of 1/$\mu$m. The choice of parameters for the 
correlations is: interwell distance $d=7$ $\mu$m and spectral width of single-particle distribution in momentum 
space [see Eq.\ (\ref{psikd})] being the inverse of $s=0.35$ $\mu$m.
}   
\label{focc}
\end{figure*}

In the present paper, we cover the full range of interaction strengths, from infinite attraction
($\cu \rightarrow -\infty$) to infinite repulsion ($\cu \rightarrow +\infty$). Following the sequence of 
frames from the third to the first frame in Fig.\ \ref{corrbst1} (top row), it is seen that a structureless 
single-particle momentum density emerges also in the limit $\cu \rightarrow -\infty$; for intermediate negative 
values of $\cu$, the weight of the oscillatory pattern decreases gradually as the absolute value $|\cu|$ 
increases. However,
based on our full solution of the 3 bosons-3 wells Hubbard system, it is apparent that this succession (i.e., 
from the third to the first frame of Fig.\ \ref{corrbst1}) does not reflect a transition from a superfluid to a 
Mott-insulator phase. Indeed, the Hubbard ground-state eigenvector for $\cu \rightarrow -\infty$ is given by 
$\phi_1^{b,-\infty}$ in the second line of Eq.\ (\ref{phi1}), which can properly be characterized as a 
Bose-Einstein condensate; namely, this ground state consists only of a single basis ket (No. 9 $\rightarrow 
|030\rangle$) that represents a triply occupied atomic orbital $\psi_2(k)$ [see Eq.\ (\ref{psikd})] located in 
the middle well. 
 
{\it The caveat from the discussion above is that the first-order correlation does not uniquely characterize the
associated many-body state.\/} This is not an uncommon occurrence, as can be also seen from an inspection of
Fig.\ \ref{f1stcorrbst2}, which illustrates a succession of $^1{\cal G}_2^{b}(k)$'s for the first excited state. 
Indeed, the single-particle momentum density in frame (a) in Fig.\ \ref{f1stcorrbst2} (case of $\cu=-200$) 
is structureless; however, the corresponding Hubbard eigenvector is very well approximated by $\phi_2^{b,-\infty}$ 
in the second line of Eq.\ (\ref{phi2}). Naturally, this eigenvector represents a many-body state that is neither
a Mott insulator nor a Bose-Einstein condensate. Rather it represents a $(-|300\rangle+|003\rangle)\sqrt{2}$
NOON state; the family of NOON states are a focal point in quantum-optics investigations \cite{oubook,shihbook}.

For a complete characterization of the many-body state under consideration, additional information, beyond the 
first-order correlations, is needed. A natural candidate to this effect are
the maps for the second-order (Sec.\ \ref{s2ndanyu}) and third-order (Sec.\ \ref{s3rdanyu}) correlations 
investigated earlier. For example, in the case of the structureless single-particle momentum density cases
discussed above [i.e., frame (m) in Fig.\ \ref{corrbst1} (top row),  frame (a) in Fig.\ \ref{corrbst1} 
(top row), and frame (a) in Fig.\ \ref{f1stcorrbst2}], all three corresponding third-order correlation maps are 
drastically different [compare frame (c) in Fig.\ \ref{corrbst1} (bottom row), 
frame (o) in Fig.\ \ref{corrbst1} (bottom row), and frame (a) in Fig.\ \ref{f3rdcorrbst2}]. 

Note that the information provided by second-order correlation maps only is still not sufficient for the 
full characterization of the underlying many-body
state. Indeed, the second-order correlation maps in frame (b) of Fig.\ \ref{corrbst1} (second row)
(case of the ground state at $\cu=-200$) is very similar to that in frame (a) of Fig.\ \ref{f2ndcorrbst2}
(case of the first-excited state at $\cu=-200$).

We stress again at this point that Figs.\ \ref{f3rdcorrbst2}, \ref{f2ndcorrbst2}, \ref{f1stcorrbst2}, 
and Fig.\ \ref{corrbst1} illustrate graphically the ability of our methodology to determine all three levels of
momentum correlations and their evolution as a function of the interaction strength $\cu$, from the attractive to
the repulsive regime, and thus to provide the tools for a complete characterization of the underlying many-body
states.
 
Before leaving this section, we found it worthwhile to explicitly investigate the conjecture that vanishing 
fluctuations in the site occupations are always associated with a structureless first-order momentum correlation.
To this effect, we plot in Fig.\ \ref{focc} the site occupation, 
$\langle \phi^b_j(\cu) |n_i| \phi^b_j(\cu) \rangle$ [the site number operator $n_i=\hat{b}^\dagger_i \hat{b}_i$; 
see below Eq.\ (\ref{3b-hub})], the expectation value of the square of the site number operator, 
$\langle \phi^b_j(\cu) |n_i^2| \phi^b_j(\cu) \rangle$, and the standard deviation,
$\sqrt{\langle \phi^b_j(\cu) |n_i^2| \phi^b_j(\cu) \rangle-\langle \phi^b_j(\cu) |n_i| \phi^b_j(\cu) \rangle^2}$ 
for the ground ($j=1$) and first-excited ($j=2$) states and for the left $(i=1)$ and middle ($i=2$) sites (wells).
As already noted in the introductory section of this paper, the connection between the fluctuations in 
site-occupation and the appearance of structural patterns (or the lack thereof) in the first-order momentum 
correlations is a manifestation of the connection between the quantum phase-transition from superfluid (coherent)
to localized (incoherent) states, and the quantum uncertainty relation connecting the fluctuations in phase  
and number (site-occupancy). 

From an inspection of the four panels (a,b,c,d) in Fig.\ \ref{focc}, one concludes that indeed in all four panels 
an oscillatory pattern in the single-particle momentum density [see subpanels (e2,e3,e4) and (f2,f3,f4,f5)] is 
accompanied by a nonvanishing fluctuation in the site occupations. However, a structureless single-particle 
momentum density is not always associated with a vanishing fluctuation; see the case of the NOON state 
$\phi^{b,-\infty}_2$ [Fig.\ \ref{focc}(c)] for which the standard deviation of the left well is 3/2, whereas the 
corresponding single-particle momentum density [Fig.\ \ref{focc}(f1)] is structureless.    

Finally, we mention that temperature effects on the quantum phase transitions in bosonic gases trapped in
optical lattices have recently attracted some attention (see, e.g., Refs.\ \cite{lu06,jin19}). 
Our beyond-mean-field theoretical approach can be generalized \cite{yannun} to account for such effects, 
but this falls outside the scope of the present paper. 

\section{Analogies with three-photon interference in quantum optics}
\label{anal}

In this section, we elaborate on the analogies between our results for the system of 3 massive bosons trapped
in 3 wells with the three-photon interference in quantum optics, which is an area of frontline research
activities \cite{spag13,tich14,agar15,agne17,mens17,tamm18.1,tamm18.2,tamm19}. Such three-photon interference 
investigations fall into two major categories: (1) Those that employ a tritter \cite{note2} to produce a
scattering event between three photons impinging on the input ports of a tritter and which measure coincidence 
probabilities for the photons exiting the three output ports \cite{spag13,tich14,agar15,agne17,mens17}.
At the abstract theoretical level, the scattering event is described by a unitary scattering matrix. The
coincidence probabilities are denoted as $P_{111}$ (one photon in each one of the output ports), $P_{210}$
(two photons in the first port and a single photon in the second port), $P_{300}$ (three photons in the first
port), etc..., and they are apparently a direct generalization of the $P_{11}$ and $P_{20}$ coincidence
probabilities familiar from the celebrated HOM \cite{hom87} two-photon interference experiment.
Variations in the $P_{ijk}$, with $i,j,k=0,1,2,3$ and $i+j+k=3$, probabilities are achieved through control of 
the time delays between photons and other parameters of the tritter. (2) Those that resolve the intrinsic 
conjugate variables underlying the wave packets of the impinging photons on the tritter (i.e., frequency, 
$\omega$, and time delay, $\tau$) \cite{tamm18.1,tamm18.2,tamm19}; for earlier two-photon interference 
investigations in this category, see Refs.\ \cite{lege04,gerr15.1,gerr15.2}. This category of experiments 
produces spectral correlation landscapes as a function of the three frequencies 
$\omega_1$, $\omega_2$, and $\omega_3$.    

\begin{figure}[t]
\includegraphics[width=7.5cm]{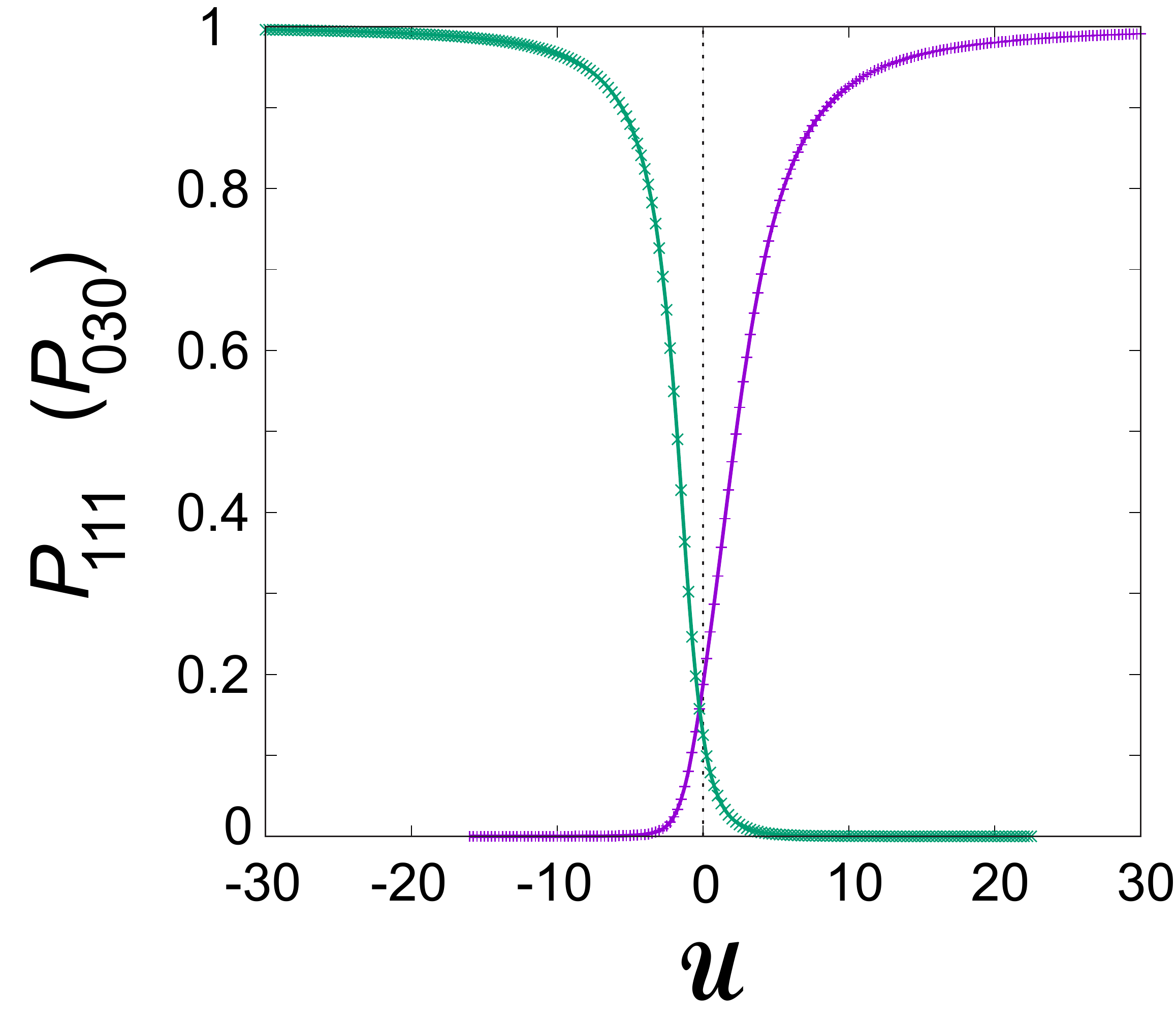}
\caption{The (dimensionless) Hong-Ou-Mandel-type probabilities $P_{111}$ (violet, right curve) and $P_{030}$ 
(green, left curve) associated with the Hubbard ground-state vector $\phi^b_1(\cu)$ as a function of the 
interaction strength $\cu$ (dimensionless).}
\label{fphom}
\end{figure}

In the case of the 3 bosons in 3 wells, the quantum-optics category (1) above finds an analog to 
{\it in situ\/} experiments and their theoretical treatments. Indeed, the analogs of the three-photon wave 
function in the output ports are the vector solutions [stationary or time-dependent (not considered in this 
paper)] of the Hubbard Hamiltonian matrix in Eq.\ (\ref{3b-mat}); compare the general form of the Hubbard
vector solutions [Eq.\ (\ref{phiU}) in Sec.\ \ref{3b-hb}] to Eq.\ (5) for the three-photon output state from a 
tritter in Ref.\ \cite{agar15}. Control of these Hubbard vector solutions is achieved through 
variation of the interaction parameter $\cu$ and the choice of a ground or excited state. For example,
choosing the ground-state vector, the probability for finding only one boson in each well is given
by the modulus square of the $\cu$-dependent coefficient in the Hubbard eigenvector [Eq.\ (\ref{phiU})] in 
front of the basis ket No. 1 $\rightarrow |111 \rangle$, i.e., $P_{111}(\cu)=|\bc_{111}(\cu)|^2$; naturally
$P_{030}(\cu)=|\bc_{030}(\cu)|^2$.

In Fig.\ \ref{fphom}, we plot the $P_{111}(\cu)$ and $P_{030}(\cu)$ probabilities associated with the Hubbard 
ground-state eigenvector $\phi^b_1(\cu)$. This figure is reminiscent of Fig.\ 2 in Ref.\ \cite{spag13} 
(see also Fig.\ 3 in Ref.\ \cite{agar15}). It is interesting to note that the three-photon state $\bhzz$ 
(experimentally realized in Ref.\ \cite{spag13}) is described in quantum optics as a ``three-photon bosonic
coalescence'', whereas for atomic and molecular physics a description as a micro Bose-Einstein condensate 
appears to come naturally in mind.

Note, further, that the $P_{ijk}$'s in 
Ref.\ \cite{spag13} depend on two parameters, instead of a single one. For the case of 3 massive bosons in
3 wells, a second parameter becomes relevant by considering the time evolution of the Hubbard vector solutions
\cite{yannun}; see Refs.\ \cite{yann19.1,kauf14} for the consideration of the time-evolution in the case of
2 massive bosons in 2 wells. Note further that, in quantum optics, two fully overlapping photons are
described as perfectly {\it indistinguishable\/}, whereas two non-overlapping photons are described as perfectly
{\it distinguishable\/} \cite{spag13,tich14}. In the context of the present study for 3 massive trapped bosons
(which uses the assumption $d^2/s^2 >> 1$), an example of the former is the ket No. 9 $\rightarrow |030 \rangle$,
whereas an example of the latter is the ket No. 1 $\rightarrow |111 \rangle$. A double-single occupancy
ket, like ket No. 2 $\rightarrow |210 \rangle$, can be referred to as a mode with two indistinguishable and one
distinguishable bosons \cite{tich14}. 

The analogy between the two-photon optical HOM formalism and the vector solutions of the Hubbard 
theoretical modeling for 2 bosons (or 2 fermions) in 2 wells was reported earlier in Refs.\ 
\cite{bran18,yann19.1}

Furthermore, in the case of the 3 bosons in 3 wells, the quantum-optics category (2) above finds an analog to
time-of-flight experiments and their theoretical treatments.  This analogy derives from the following
correspondence (revealed in Ref.\ \cite{yann19.1})
\begin{align}
\begin{split}
k & \longleftrightarrow \omega/c \\
d & \longleftrightarrow \tau c \\
kd & \longleftrightarrow \omega\tau.
\end{split}
\label{kdwt}
\end{align}

As was done \cite{yann19.1} for the case of 2 massive trapped particles versus two interfering photons, this 
correspondence can be used to establish a complete analogy between the cosinusoidal patterns of all three orders 
of momentum correlation functions presented in this paper for 3 massive and trapped bosons (and which can be 
determined experimentally through time-of-flight measurements \cite{prei19}) with the landscapes 
\cite{tamm18.1,tamm19} of the frequency-resolved three-photon interferograms (which are a function of the three 
photon frequencies, $\omega_1$, $\omega_2$, and $\omega_3$). For example the interferograms in Fig.\ 3 of
Ref.\ \cite{tamm19} are analogous to the map in Fig.\ \ref{corrbst1} [frame (i), bottom row] of the 
$k_3=0$ cut of the third-order momentum correlation associated with 3 non-interacting trapped massive bosons.
A difference to keep in mind is that in this paper the interwell distances were taken to be equal, whereas the 
time delays in Ref.\ \cite{tamm19} are unequal. 

Furthermore, Eq.\ (S1) in the Supplemental Material of Ref.\ \cite{tamm19} which describes the three-photon 
output wave function at the detectors, $\psi(\omega_1, \omega_2, \omega_3)$, is a permanent of the three 
single-photon wave functions $\chi_j(\omega_i)=E_j(\omega_i)\exp(-i \omega_i t_j)$, with $i,j=1,2,3$, where $t_j$
denotes time instances [corresponding to the position of each well in our single-particle orbitals displayed in
Eq.\ (\ref{psikd})]. As a result, for $E_1(\omega_i)=E_2(\omega_i)=E_2(\omega_i)=E(\omega_i)$ and $t_1=-\tau$, 
$t_2=0$, and $t_3=\tau$, it reduces exactly to the form of the three-body wave function 
$\Phi^{b,+\infty}_1(k_1,k_2,k_3)$ [see top line in Eq.\ (\ref{phibUpmI_1})] in this paper which is associated 
with the case of the three singly-occupied wells, i.e., the Hubbard solution at infinite repulsion, $|111\rangle$
(perfectly distinguishable bosons).

A central focus in the recent quantum-optics literature has been the demonstration of genuine three-photon
interference \cite{agne17,mens17}, that is interference effects that cannot be inferred by a knowledge of the 
one- and two-photon interference patterns. In the language of many-body literature for massive particles, this
is equivalent to isolating the {\it connected\/} terms, ${\cal G}_{\rm con}$, in the total third-order 
correlations by subtracting the {\it disconnected\/} ones, ${\cal G}_{\rm dis}$. Reflecting its name, the 
disconnected contribution to the total third-order correlation consists of products of the first- and 
second-order correlations. 

For the case of 3 perfectly distinguishable bosons in 3 wells 
(described by the ket $|111\rangle$), one can observe that
the first-order momentum correlation given in Eqs.\ (\ref{frstggsUI}) and (\ref{frstggsUmI}) does not contain any
cosine (or sine) terms, whereas the second-order momentum correlation given in Eq.\ (\ref{3b2ndggsUI}) contains
consine terms with two momenta in the cosine arguments. As a result, the connected part of the 
third-order momentum correlations [see Eq.\ (\ref{3gUIst1})] is necessarily reflected in the 
cosine terms having an argument that depends on all three single-particle momenta $k_1$, $k_2$, and $k_3$.
Another way to view the above remarks is that the genuine three-body interference involves a total phase  
$\varphi$ which is the sum of three partial phases $\varphi_1$, $\varphi_2$, and $\varphi_3$, associated with the
individual bosons, i.e., $\varphi=\varphi_1+\varphi_2+\varphi_3$. Such a triple phase (referred to also as a 
triad phase) has been prominent in the quantum-optics literature \cite{agne17,mens17} regarding genuine 
three-photon interference.

Specifically, the disconnected part of the third-order correlation for 3 bosons is given by the expression
\begin{align}
\begin{split}
& ^3{\cal G}_{\rm dis}^b(k_1,k_2,k_3)=-2 {^1{\cal G}^b(k_1)} {^1{\cal G}^b(k_2)} {^1{\cal G}^b(k_3)} + \\
& {^1{\cal G}^b(k_1)} {^2{\cal G}^b(k_2,k_3)} + {^1{\cal G}^b(k_2)} {^2{\cal G}^b(k_1,k_3)} + \\
& {^1{\cal G}^b(k_3)} {^2{\cal G}^b(k_1,k_2)}.
\end{split}
\label{gdis}
\end{align} 

We can apply the above expression immediately to the case of the Hubbard ground-state eigenvector $\booo$ (limit 
of infinite repulsion, 3 perfectly distinguishable bosons), because we have derived explicit algebraic 
expressions for the corresponding third-order [Eq.\ (\ref{3gUIst1})], second-order [Eq.\ (\ref{3b2ndggsUI})], 
and first-order momentum correlations [Eqs.\ (\ref{frstggsUI}) and (\ref{frstggsUmI})]. 
Indeed one finds for the connected correlation part
\begin{align}
\begin{split}
& ^3{\cal G}_{1,\rm con}^{b,+\infty}(k_1,k_2,k_3)=\\
& ^3{\cal G}_1^{b,+\infty}(k_1,k_2,k_3) - ^3{\cal G}_{1,\rm dis}^{b,+\infty}(k_1,k_2,k_3) =\\
& \frac{4 \sqrt{2}}{3 \pi^{3/2}} s^3 e^{-2 s^2 (k_1^2+k_2^2+k_3^2)}
 \big\{\cos (d (k_1+k_2-2 k_3)) \\
&  + \cos (d (k_2+ k_3 -2 k_1))
   + \cos (d (k_1+k_3 - 2 k_2)) \big\}.
\end{split}
\label{gcon}
\end{align}
 
It is worth noting that the result in Eq.\ (\ref{gcon}) above for the connected correlation part for 3 perfectly
distinguishable bosons in 3 wells coincides with the corresponding result \cite{prei19,yann19.3} for 3 perfectly 
distinguishable {\it fully spin polarized fermions\/} in 3 wells.

\section{Summary}
\label{summ}
In this paper, we develop and expand a formalism and a theoretical framework, which, with the use of an 
algebraic-language computations tool (MATHEMATICA \cite{math18}), allows us to derive explicit analytic 
expressions for all three orders (third, 
second, and first) of momentum-space correlations for 3 interacting ultracold bosonic atoms confined in 3 optical 
wells in a linear geometry. This 3b-3w system was modeled as a three-site Bose-Hubbard Hamiltonian whose 10
eigenvectors were mapped onto first-quantization three-body wave functions in momentum space by: (1) associating 
the bosons with the Fourier transforms of displaced Gaussian functions centered on each well, and (2) constructing 
the permanents associated with the basis kets of the Hubbard Hilbert space by using the Fourier transforms of the 
displaced Gaussians describing the trapped bosons.
The 3rd-order momentum-space correlations are the modulus square of such three-body wave functions, and the
second- and first-order correlations are derived through successive integrations over the unresolved momentum
variables. This methodology applies to all bosonic states with strong \cite{note4} or without entanglement, and 
\textcolor{black}{
does not rely on the standard Wick's factorization scheme, employed in earlier studies (see, e.g., Refs,\ 
\cite{gome06,hodg13,schm17}) of higher-order momentum correlations for expanding or colliding Bose-Einstein 
condensates of ultracold atoms.   
}

The availability of such explicit analytic correlation functions will greatly assist in the analysis of
anticipated future TOF measurements with few ($N > 2$) ultracold atoms trapped in optical lattices, following 
the demonstrated feasibility of determining higher-than-first-order momentum correlation functions 
via single-particle detection in the case of $N=2$ fermionic $^6$Li atoms \cite{berg19}, $N=3$ fully 
spin-polarized fermionic $^6$Li atoms \cite{prei19}, and a large number of bosonic $^4$He$^*$ atoms \cite{clem19}.

The availability of the complete set of all-order momentum correlations enabled us to reveal and explore in detail
two major physical aspects of the 3b-3w ultracold-atom system:
(I) That a small system of only 3 bosons exhibits indeed an embryonic behavior akin to an emergent superfluid to 
Mott transition and (II) That both the {\it in situ\/} and TOF spectroscopies of the 3b-3w system exhibit 
analogies with the quantum-optics three-photon interference, including the aspects of genuine three-photon
interference which cannot be understood from the knowledge of the lower second- and first-order correlations
alone \cite{agne17,mens17}.

The superfluid to Mott-insulator transition in extended optical lattices \cite{grei02,gerb05,gerb05.2} was 
explored based on the variations in the shape of the first-order momentum correlations. For the 3b-3w system, we 
reported clear variations of the first-order momentum correlations, from being oscillatory with a period that 
depends on the inter-well distance, characteristic of a coherent state of a superfluid phase with multiple site 
occupancies by each of the trapped ultracold bosonic atoms (high site-occupancy uncertainty), to a structureless
shape characteristic of localized states, (see below) with low site-occupancy uncertainty and consequent high 
phase-uncertainty (incoherent phase). Furthermore, we also concluded
that the first-order momentum correlations are not sufficient to characterize uniquely
the underlying nature of a state of the 3b-3w system. To this effect, knowledge of all three orders of 
correlations is needed. Indeed, a structureless first-order correlation relates to three different 3b-3w
states, i.e., the $|030\rangle$ ground state at $\cu \rightarrow -\infty$ (Bose-Einstein condensate), the 
$|111\rangle$ ground state at $\cu \rightarrow +\infty$ (Mott insulator), and the 
$(-|300\rangle + |003\rangle)/\sqrt{2}$ first-excited state at $\cu \rightarrow -\infty$ (NOON state).

Concerning the quantum optics analogies, we established that {\it in situ\/} measurements of the site 
occupation probabilities as a function of $\cu$, $P_{ijk}(\cu)$ (with $i,j=1,\ldots,3$), provide analogs of the 
celebrated HOM coincidence probabilities for three photons at the output ports of a tritter as discussed in 
Refs.\ \cite{spag13,agar15}. We further established that the momentum-space all-order correlations for the
3b-3w system parallel the frequency-resolved interferograms of distinguishable photons as explored in
Refs.\ \cite{tamm18.1,tamm18.2,tamm19}. The analogies with the genuine three-photon interference were 
established in the framework of the many-body theoretical concepts of disconnected versus connected
correlation terms.   

To achieve simplicity in this paper, we assumed throughout that the interwell separation is much larger than 
the width of the single-particle Gaussian function in the real configuration space, 
i.e., $d^2/s^2 >> 1$ (see Sec. \ref{3rdcorrUpmI}). This is equivalent to considering localized bosons with vanishing
overlaps (distinguishable bosons in different wells) or unity overlaps (indistinguishable bosons in the same 
well); indeed the overlap of two single-particle wave functions according to Eq.\ (\ref{psikd}) is given by 
$S=e^{-d^2/(8s^2)}$. Considering cases with small, but finite $S$, which represent partial indistinguishability 
\cite{tich14}, complicates substantially the analytic results \cite{yannun}.

Finally, we note here that our all-order momentum-space correlations for the 3b-3w system can contribute an 
alternative way to study and explore with massive particles aspects of the boson sampling problem \cite{aaar13},
and in particular its extension to the multiboson correlation sampling \cite{tamm15,tamm15.1}. We note that
boson sampling problems have become a major focus [see, e.g., Refs. \cite{tamm15,tamm15.1,tich14,lain14,wals19}] 
in quantum-optics investigations because they are considered to be an intermediate step on the road towards the 
implementation of the quantum computer.

\section{Acknowledgments} 
This work has been supported by a grant from the Air Force Office of Scientific Research
(AFOSR, USA) under Award No. FA9550-15-1-0519. Calculations were carried out at the GATECH Center for
Computational Materials Science.

\appendix

\section{
\textcolor{black}{
Hubbard eigenvectors: The infinite repulsive or attractive interaction ($\cu \rightarrow \pm \infty$) 
limit for the remaining eight excited states}}
\label{a11}

This Appendix complements Sec.\ \ref{eigvecuinf} by listing without commentary the Hubbard eigenvectors 
of the remaining eight excited states not discussed in the main text.

\begin{align}
\begin{split}
\phi^{b,+\infty}_3 &=
\left\{0,\frac{1}{2},\frac{1}{2
   \sqrt{5}},\frac{1}{\sqrt{5}},\frac{1}{\sqrt{5}},\frac{1
   }{2 \sqrt{5}},\frac{1}{2},0,0,0\right\} \\
\phi^{b,-\infty}_3 &= \{0,0,0,0,0,0,0,\frac{1}{\sqrt{2}},0,\frac{1}{\sqrt{2}} \}
\end{split}
\label{phi3}
\end{align}
\begin{align}
\begin{split}
\phi^{b,+\infty}_4 &=
\left\{0,0,\sqrt{\frac{2}{5}},-\frac{1}{\sqrt{10}},-\frac{
   1}{\sqrt{10}},\sqrt{\frac{2}{5}},0,0,0,0\right\} \\
\phi^{b,-\infty}_4 &= 
\left\{0,-\frac{1}{2},-\frac{1}{2
   \sqrt{5}},-\frac{1}{\sqrt{5}},-\frac{1}{\sqrt{5}},-\frac
   {1}{2 \sqrt{5}},-\frac{1}{2},0,0,0\right\} 
\end{split}
\label{phi4}
\end{align}
\begin{align}
\begin{split}
\phi^{b,+\infty}_5 &=
\left\{0,0,-\sqrt{\frac{2}{5}},\frac{1}{\sqrt{10}},-\frac{
   1}{\sqrt{10}},\sqrt{\frac{2}{5}},0,0,0,0\right\} \\
\phi^{b,-\infty}_5 &=
\left\{0,\frac{1}{2},\frac{1}{2\sqrt{5}},\frac{1}{\sqrt{5}},
         -\frac{1}{\sqrt{5}},-\frac{1}{2\sqrt{5}},-\frac{1}{2},0,0,0\right\}
\end{split}
\label{phi5}
\end{align}
\begin{align}
\begin{split}
\phi^{b,+\infty}_6 &= 
\left\{0,-\frac{1}{2},\frac{1}{2
   \sqrt{5}},\frac{1}{\sqrt{5}},-\frac{1}{\sqrt{5}},-\frac
   {1}{2 \sqrt{5}},\frac{1}{2},0,0,0\right\} \\
\phi^{b,-\infty}_6 &=
\left\{0,0,-\sqrt{\frac{2}{5}},\frac{1}{\sqrt{10}},-\frac{
   1}{\sqrt{10}},\sqrt{\frac{2}{5}},0,0,0,0\right\}
\end{split}
\label{phi6}
\end{align}
\begin{align}
\begin{split}
\phi^{b,+\infty}_7 &= 
\left\{0,\frac{1}{2},-\frac{1}{2
   \sqrt{5}},-\frac{1}{\sqrt{5}},-\frac{1}{\sqrt{5}},-\frac
   {1}{2 \sqrt{5}},\frac{1}{2},0,0,0\right\} \\
\phi^{b,-\infty}_7 &=
\left\{0,0,\sqrt{\frac{2}{5}},-\frac{1}{\sqrt{10}},-\frac{
   1}{\sqrt{10}},\sqrt{\frac{2}{5}},0,0,0,0\right\} 
\end{split}
\label{phi7}
\end{align}
\begin{align}
\begin{split}
\phi^{b,+\infty}_8 &= 
\left\{0,0,0,0,0,0,0,\frac{1}{\sqrt{2}},0,\frac{1}{\sqrt{2
   }}\right\} \\
\phi^{b,-\infty}_8 &=
\left\{0,-\frac{1}{2},\frac{1}{2
   \sqrt{5}},\frac{1}{\sqrt{5}},\frac{1}{\sqrt{5}},\frac{1
   }{2 \sqrt{5}},-\frac{1}{2},0,0,0\right\} 
\end{split}
\label{phi8}
\end{align}
\begin{align}
\begin{split}
\phi^{b,+\infty}_9 &= 
\left\{0,0,0,0,0,0,0,-\frac{1}{\sqrt{2}},0,\frac{1}{\sqrt{
   2}}\right\} \\
\phi^{b,-\infty}_9 &=
\left\{0,\frac{1}{2},-\frac{1}{2\sqrt{5}},-\frac{1}{\sqrt{5}},\frac{1}{\sqrt{5}},\frac{
   1}{2 \sqrt{5}},-\frac{1}{2},0,0,0\right\} \\
\end{split}
\label{phi9}
\end{align}
\begin{align}
\begin{split}
\phi^{b,+\infty}_{10} &= \{0,0,0,0,0,0,0,0,1,0\}\\
\phi^{b,-\infty}_{10} &= \{-1,0,0,0,0,0,0,0,0,0\}
\end{split}
\label{phi10}
\end{align}

\section{
\textcolor{black}{
Hubbard eigenvectors: The noninteracting ($\cu =0$) limit for the remaining eight excited states}}
\label{a12}

Because of the three pairwise degeneracies [see Eq.\ (\ref{eigvalb3})], care must be used when 
determining the six eigenvectors  3, 4, 5, 6, 7, and 8 at $\cu=0$. The proper Hubbard eigenvectors listed below 
were determined by taking the limit $\cu \rightarrow 0+$. For the eigenvectors No. 3, 4, 7, and 8, the
associated algebraic formulas are lengthy, and as a result we give below the numerical expressions of
these eigenvectors. Eigenvector No. 5 is $\cu$-independent.

\begin{align}
\begin{split}
& \phi_{3r(4l)}^{b,\cu=0}= \\
& \{-0.553362,0.189903,-0.419079,0.142399,0.142399,\\
& -0.419079,0.189903,0.232583,0.348804,0.232583\},
\end{split}
\label{eigvecU0st3}
\end{align} 
\begin{align}
\begin{split}
& \phi_{4r(3l)}^{b,\cu=0}= \\
& \{-0.079316,0.346679,0.165823,-0.205481,-0.205481,\\
& 0.165823,0.346679,0.424594,-0.503325,0.424594\},
\end{split}
\label{eigvecU0st4}
\end{align}
\begin{align}
\begin{split}
& \phi_{5r(6l)}^{b,\cu=0} = \phi_5^{b,+\infty} =\phi_6^{b,-\infty}, 
\end{split}
\label{eigvecU0st5}
\end{align}  
\begin{align}
\begin{split}
& \phi_{6r(5l)}^{b,\cu=0}= \\
& \left\{0,0,\frac{\sqrt{\frac{3}{5}}}{4},
\frac{\sqrt{\frac{3}{5}}}{2},-\frac{\sqrt{\frac{3}{5}}}{2},
-\frac{\sqrt{\frac{3}{5}}}{4},0,-\frac{\sqrt{5}}{4},0,
   \frac{\sqrt{5}}{4}\right\},
\end{split}
\label{eigvecU0st6}
\end{align} 
\begin{align}
\begin{split}
& \phi_{7r(8l)}^{b,\cu=0}= \\
& \{0.553362,-0.189903,-0.419079,0.142399,0.142399,\\
& -0.419079,-0.189903,0.232583,-0.348804,0.232583\},
\end{split}
\label{eigvecU0st7}
\end{align}
\begin{align}
\begin{split}
& \phi_{8r(7l)}^{b,\cu=0}= \\
& \{0.079316,-0.346679,0.165823,-0.205481,-0.205481,\\
& 0.165823,-0.346679,0.424594,0.503325,0.424594\},
\end{split}
\label{eigvecU0st8}
\end{align} 

Finally, the remaining two eigenvectors No. 9 and No. 10 are given by,
\begin{align}
\begin{split}
& \phi_9^{b,\cu=0}= \\
&  \left\{0,\frac{1}{2},-\frac{1}{4
   \sqrt{2}},-\frac{1}{2 \sqrt{2}},\frac{1}{2
   \sqrt{2}},\frac{1}{4
   \sqrt{2}},-\frac{1}{2},-\frac{\sqrt{\frac{3}{2}}}{
   4},0,\frac{\sqrt{\frac{3}{2}}}{4}\right\},
\end{split}
\label{eigvecU0st9}
\end{align}
and 
\begin{align}
\begin{split}
& \phi_{10}^{b,\cu=0}= \\
& \left\{-\frac{\sqrt{3}}{4},-\frac{\sqrt{\frac{3}{2}}}{4},
   \frac{\sqrt{3}}{8},\frac{\sqrt{3}}{4},
   \frac{\sqrt{3}}{4},\frac{\sqrt{3}}{8},
   -\frac{\sqrt{\frac{3}{2}}}{4},\frac{1}{8},
   -\frac{1}{2\sqrt{2}},\frac{1}{8}\right\}. 
\end{split}
\label{eigvecU0st10}
\end{align}

\section{Third-order momentum correlations for 3 bosons in 3 wells:
The infinite-interaction limit ($\cu \rightarrow \pm \infty$) for the remaining eight states}
\label{a1}

This Appendix complements Sec.\ \ref{3rdcorrUpmI} by listing without commentary the momentum-space wave
functions, $\Phi^{b,\pm\infty}_i(k_1,k_2,k_3)$ (with $i=3,\ldots,10$), associated with the corresponding
Hubbard eigenvectors, $\phi^{b,\pm\infty}_i$ [see Eqs.\ (\ref{phi3})-(\ref{phi10})], at the limits of infinite 
repulsive or attractive strength (i.e., for
$\cu \rightarrow \pm \infty$). The commentary integrating these wave functions into the broader scheme of their
evolution as a function of any interaction strength $-\infty < \cu < +\infty$ is left for Appendix\ \ref{a3rd}.

\begin{widetext}
\begin{align}
\begin{split}
\Phi^{b,+\infty}_3(& k_1, k_2, k_3) =  
   \frac{ 2^{3/4} } { 5 \sqrt{3} \pi ^{3/4} } s^{3/2} e^{ -(k_1^2+k_2^2+k_3^2)s^2 } \\ 
   & \times \left[ \sqrt{5} \cos (d(-k_1+k_2+k_3))+\sqrt{5} \cos (d(k_1+k_2-k_3)) \right.
   +\sqrt{5} \cos (d(k_1-k_2+k_3)) \\
   &\;\; +5 \cos (d(k_1+k_2)) +5 \cos (d(k_1+k_3)) + 5 \cos (d(k_2+k_3)) +2 \sqrt{5} \cos (d k_1)
   \left. +2 \sqrt{5} \cos (d k_2)+2 \sqrt{5} \cos (dk_3)\right],\\
\Phi^{b,-\infty}_3(& k_1, k_2, k_3) =  
    \frac{ 2 \times 2^{1/4} } {\pi ^{3/4} } s^{3/2} e^{ -(k_1^2+k_2^2+k_3^2)s^2 }
   \cos (d(k_1+k_2+k_3)).
\end{split}
\label{phibUpmI_3}
\end{align}
\begin{align}
\begin{split}
\Phi^{b,+\infty}_4(& k_1, k_2, k_3) =  
  -\frac{ 2 \times 2^{1/4} }{\sqrt{15} \pi^{3/4} } s^{3/2} e^{ -(k_1^2+k_2^2+k_3^2)s^2 } \\
  & \times \left[ \cos(d k_1) + \cos(d k_2) + \cos(d k_3) - 2 \cos(d (-k_1 + k_2 + k_3)) 
    - 2 \cos(d (k_1 - k_2 + k_3)- 2 \cos(d (k_1 + k_2 - k_3)] \right],\\
\Phi^{b,-\infty}_4(& k_1, k_2, k_3) =  
   - \frac{ 2^{3/4} } { 5 \sqrt{3} \pi ^{3/4} } s^{3/2} e^{ -(k_1^2+k_2^2+k_3^2)s^2 } \\
   & \times \left[ \sqrt{5} \cos (d(-k_1+k_2+k_3))+\sqrt{5} \cos (d(k_1+k_2-k_3)) \right.
   +\sqrt{5} \cos (d(k_1-k_2+k_3)) \\
   &\;\; +5 \cos (d(k_1+k_2)) +5 \cos (d(k_1+k_3)) + 5 \cos (d(k_2+k_3)) +2 \sqrt{5} \cos (d k_1)
   \left. +2 \sqrt{5} \cos (d k_2)+2 \sqrt{5} \cos (dk_3)\right].
\end{split}
\label{phibUpmI_4}
\end{align}
\begin{align}
\begin{split}
\Phi^{b,+\infty}_5(& k_1, k_2, k_3) =  
  -\frac{ i 2 \times 2^{1/4} }{\sqrt{15} \pi^{3/4} } s^{3/2} e^{ -(k_1^2+k_2^2+k_3^2)s^2 } \\
  & \times \left[ \sin(d k_1) + \sin(d k_2) + \sin(d k_3) - 2 \sin(d (-k_1 + k_2 + k_3)) 
    - 2 \sin(d (k_1 - k_2 + k_3))- 2 \sin(d (k_1 + k_2 - k_3)) \right],\\
\Phi^{b,-\infty}_5(& k_1, k_2, k_3) =  
   - \frac{ i 2^{3/4} } { 5 \sqrt{3} \pi ^{3/4} } s^{3/2} e^{ -(k_1^2+k_2^2+k_3^2)s^2 } \\ 
   & \times \left[ \sqrt{5} \sin (d(-k_1+k_2+k_3))+\sqrt{5} \sin (d(k_1+k_2-k_3)) \right.
   +\sqrt{5} \sin (d(k_1-k_2+k_3)) \\
   &\;\; + 5 \sin (d(k_1+k_2)) + 5 \sin (d(k_1+k_3)) + 5 \sin (d(k_2+k_3)) +2 \sqrt{5} \sin (d k_1)
   \left. +2 \sqrt{5} \sin (d k_2)+2 \sqrt{5} \sin (dk_3)\right].
\end{split}
\label{phibUpmI_5}
\end{align}
\begin{align}
\begin{split}
\Phi^{b,+\infty}_6(& k_1, k_2, k_3) =  
   - \frac{ i 2^{3/4} } { 5 \sqrt{3} \pi ^{3/4} } s^{3/2} e^{ -(k_1^2+k_2^2+k_3^2)s^2 } \\ 
   & \times \left[ \sqrt{5} \sin (d(-k_1+k_2+k_3))+\sqrt{5} \sin (d(k_1+k_2-k_3)) \right.
   +\sqrt{5} \sin (d(k_1-k_2+k_3)) \\
   &\;\; - 5 \sin (d(k_1+k_2)) - 5 \sin (d(k_1+k_3)) - 5 \sin (d(k_2+k_3)) +2 \sqrt{5} \sin (d k_1)
   \left. +2 \sqrt{5} \sin (d k_2)+2 \sqrt{5} \sin (dk_3)\right],\\
\Phi^{b,-\infty}_6(& k_1, k_2, k_3) =
  -\frac{ i 2 \times 2^{1/4} }{\sqrt{15} \pi^{3/4} } s^{3/2} e^{ -(k_1^2+k_2^2+k_3^2)s^2 } \\
  & \times \left[ \sin(d k_1) + \sin(d k_2) + \sin(d k_3) - 2 \sin(d (-k_1 + k_2 + k_3))
    - 2 \sin(d (k_1 - k_2 + k_3)) - 2 \sin(d (k_1 + k_2 - k_3)) \right].
\end{split}
\label{phibUpmI_6}
\end{align}
\begin{align}
\begin{split}
\Phi^{b,+\infty}_7(& k_1, k_2, k_3) =  
    - \frac{ 2^{3/4} } { 5 \sqrt{3} \pi ^{3/4} } s^{3/2} e^{ -(k_1^2+k_2^2+k_3^2)s^2 } \\ 
   & \times \left[ \sqrt{5} \cos (d(-k_1+k_2+k_3))+\sqrt{5} \cos (d(k_1+k_2-k_3)) \right.
   +\sqrt{5} \cos (d(k_1-k_2+k_3)) \\
   &\;\; - 5 \cos (d(k_1+k_2)) - 5 \cos (d(k_1+k_3)) - 5 \cos (d(k_2+k_3)) +2 \sqrt{5} \cos (d k_1)
   \left. +2 \sqrt{5} \cos (d k_2)+2 \sqrt{5} \cos (dk_3)\right],\\
\Phi^{b,-\infty}_7(& k_1, k_2, k_3) =  
  -\frac{ 2 \times 2^{1/4} }{\sqrt{15} \pi^{3/4} } s^{3/2} e^{ -(k_1^2+k_2^2+k_3^2)s^2 } \\
  & \times \left[ \cos(d k_1) + \cos(d k_2) + \cos(d k_3) - 2 \cos(d (-k_1 + k_2 + k_3))
    - 2 \cos(d (k_1 - k_2 + k_3)) - 2 \cos(d (k_1 + k_2 - k_3)) \right].
\end{split}
\label{phibUpmI_7}
\end{align}
\begin{align}
\begin{split}
\Phi^{b,+\infty}_8(& k_1, k_2, k_3) =  
    \frac{ 2 \times 2^{1/4} } {\pi ^{3/4} } s^{3/2} e^{ -(k_1^2+k_2^2+k_3^2)s^2 } 
   \cos (d(k_1+k_2+k_3)),\\
\Phi^{b,-\infty}_8(& k_1, k_2, k_3) =
    \frac{ 2^{3/4} } { 5 \sqrt{3} \pi ^{3/4} } s^{3/2} e^{ -(k_1^2+k_2^2+k_3^2)s^2 } \\
   & \times \left[ \sqrt{5} \cos (d(-k_1+k_2+k_3))+\sqrt{5} \cos (d(k_1+k_2-k_3)) \right.
   +\sqrt{5} \cos (d(k_1-k_2+k_3)) \\
   &\;\; - 5 \cos (d(k_1+k_2)) - 5 \cos (d(k_1+k_3)) - 5 \cos (d(k_2+k_3)) +2 \sqrt{5} \cos (d k_1)
   \left. +2 \sqrt{5} \cos (d k_2)+2 \sqrt{5} \cos (dk_3)\right].
\end{split}
\label{phibUpmI_8}
\end{align}
\begin{align}
\begin{split}
\Phi^{b,+\infty}_9(& k_1, k_2, k_3) =  
    \frac{ 2 i 2^{1/4} } {\pi ^{3/4} } s^{3/2} e^{ -(k_1^2+k_2^2+k_3^2)s^2 } 
   \sin (d(k_1+k_2+k_3)),\\
\Phi^{b,-\infty}_9(& k_1, k_2, k_3) =
 \frac{ i 2^{3/4} } { 5 \sqrt{3} \pi ^{3/4} } s^{3/2} e^{ -(k_1^2+k_2^2+k_3^2)s^2 } \\
   & \times \left[ \sqrt{5} \sin (d(-k_1+k_2+k_3))+\sqrt{5} \sin (d(k_1+k_2-k_3)) \right.
   +\sqrt{5} \sin (d(k_1-k_2+k_3)) \\
   &\;\; -5 \sin (d(k_1+k_2)) -5 \sin (d(k_1+k_3)) - 5 \sin (d(k_2+k_3)) +2 \sqrt{5} \sin (d k_1)
   \left. +2 \sqrt{5} \sin (d k_2)+2 \sqrt{5} \sin (dk_3)\right].
\end{split}
\label{phibUpmI_9}
\end{align}
\begin{align}
\begin{split}
\Phi^{b,+\infty}_{10}(& k_1, k_2, k_3) =  
    - \left( \frac{2}{\pi} \right)^{3/4} s^{3/2} e^{ -(k_1^2+k_2^2+k_3^2)s^2 },\\
\Phi^{b,-\infty}_{10}(& k_1, k_2, k_3) =  
- \frac { 2 \times 2^{1/4} } { \sqrt{3} \pi^{3/4} } s^{3/2} e^{ -(k_1^2+k_2^2+k_3^2)s^2 }
[ \cos( d(k_1-k_2) ) + \cos( d(k_1-k_3) ) + \cos( d(k_2-k_3) ) ].
\end{split}
\label{phibUpmI_10}
\end{align}
\end{widetext}

\section{Third-order momentum correlations for 3 bosons in 3 wells:
The non-interacting limit $\cu=0$ for the remaining eight states}
\label{a2}

This Appendix complements Sec.\ \ref{3rdcorrU0} by listing without commentary the momentum-space three-body
wave functions for the remaining 8 excited states, that is:
\begin{widetext}
\begin{align}
\begin{split}
& s^{-3/2} e^{(k_1^2+k_2^2+k_3^2)s^2} \Phi_{3r(4l)}^{b,\cu=0}(k_1,k_2,k_3) = 
   0.248595 + 0.117189 \big(\cos (d k_1) + \cos (d k_2) + \cos (d k_3) \big) \\
&  - 0.322013 \big(\cos [d (k_1-k_2)] + \cos [d (k_1-k_3)] + \cos [d (k_2-k_3)] \big)\\
&  + 0.156283 \big(\cos [d (k_1+k_2)] + \cos [d (k_1+k_3)] + \cos [d (k_2+k_3)] \big) \\
&  -0.344886 \big(\cos [d (k_1-k_2+k_3)] + \cos [d (k_1-k_2-k_3)] + \cos [d (k_1+k_2-k_3)] \big)  
   +0.331527 \cos [d (k_1+k_2+k_3)],
\end{split}
\label{phibU0_3}
\end{align}
\begin{align}
\begin{split}
& s^{-3/2} e^{(k_1^2+k_2^2+k_3^2)s^2} \Phi_{4r(3l)}^{b,\cu=0}(k_1,k_2,k_3) =
   -0.358722 -0.169103 \big(\cos (d k_1) + \cos (d k_2) + \cos (d k_3) \big) \\
&  -0.0461557 \big(\cos [d (k_1-k_2)] + \cos [d (k_1-k_3)] + \cos [d (k_2-k_3)] \big)\\
&  +0.285304 \big(\cos [d (k_1+k_2)] + \cos [d (k_1+k_3)] + \cos [d (k_2+k_3)] \big) \\
&  +0.136466 \big(\cos [d (k_1-k_2+k_3)] + \cos [d (k_1-k_2-k_3)] + \cos [d (k_1+k_2-k_3)] \big)
   +0.605221 \cos [d (k_1+k_2+k_3)],
\end{split}
\label{phibU0_4}
\end{align}
\begin{align}
\begin{split}
\Phi_{5r(6l)}^{b,\cu=0}(k_1,k_2,k_3) = \Phi_5^{b,+\infty}(k_1,k_2,k_3) = \Phi_6^{b,-\infty}(k_1,k_2,k_3), 
\end{split}
\label{phibU0_5}
\end{align}
\begin{align}
\begin{split}
& \frac{-i 2^{1/4}\sqrt{5} \pi^{3/4}}{s^{3/2}} e^{(k_1^2+k_2^2+k_3^2)s^2} \Phi_{6r(5l)}^{b,\cu=0}(k_1,k_2,k_3) =
   2 \big(\sin (d k_1) + \sin (d k_2) + \sin (d k_3) \big) \\
&  + \sin [d (k_1-k_2+k_3)] + \sin [d (-k_1+k_2+k_3)] + \sin [d (k_1+k_2-k_3)]
   - 5 \sin [d (k_1+k_2+k_3)],
\end{split}
\label{phibU0_6}
\end{align}
\begin{align}
\begin{split}
& s^{-3/2} e^{(k_1^2+k_2^2+k_3^2)s^2} \Phi_{7r(8l)}^{b,\cu=0}(k_1,k_2,k_3) =
   -0.248595 + 0.117189 \big(\cos (d k_1) + \cos (d k_2) + \cos (d k_3) \big) \\
&  + 0.322013 \big(\cos [d (k_1-k_2)] + \cos [d (k_1-k_3)] + \cos [d (k_2-k_3)] \big)\\
&  - 0.156283 \big(\cos [d (k_1+k_2)] + \cos [d (k_1+k_3)] + \cos [d (k_2+k_3)] \big) \\
&  -0.344886 \big(\cos [d (k_1-k_2+k_3)] + \cos [d (k_1-k_2-k_3)] + \cos [d (k_1+k_2-k_3)] \big)
   +0.331527 \cos [d (k_1+k_2+k_3)],
\end{split}
\label{phibU0_7}
\end{align}
\begin{align}
\begin{split}
& s^{-3/2} e^{(k_1^2+k_2^2+k_3^2)s^2} \Phi_{8r(7l)}^{b,\cu=0}(k_1,k_2,k_3) =
   0.358722 -0.169103 \big(\cos (d k_1) + \cos (d k_2) + \cos (d k_3) \big) \\
&  +0.0461557 \big(\cos [d (k_1-k_2)] + \cos [d (k_1-k_3)] + \cos [d (k_2-k_3)] \big)\\
&  -0.285304 \big(\cos [d (k_1+k_2)] + \cos [d (k_1+k_3)] + \cos [d (k_2+k_3)] \big) \\
&  +0.136466 \big(\cos [d (k_1-k_2+k_3)] + \cos [d (k_1-k_2-k_3)] + \cos [d (k_1+k_2-k_3)] \big)
   +0.605221 \cos [d (k_1+k_2+k_3)],
\end{split}
\label{phibU0_8}
\end{align}
\begin{align}
\begin{split}
& \frac{i (2 \pi )^{3/4} \sqrt{3}}{s^{3/2}} e^{(k_1^2+k_2^2+k_3^2)s^2} \Phi_9^{b,\cu=0}(k_1,k_2,k_3) =
   2 \big(\sin (d k_1) + \sin (d k_2) + \sin (d k_3) \big) \\
&  - 2 \sqrt{2} \big(\sin [d (k_1+k_2)] + \sin [d (k_1+k_3)] + \sin [d (k_2+k_3)] \big) \\
&  + \sin [d (k_1-k_2+k_3)] + \sin [d (-k_1+k_2+k_3)] + \sin [d (k_1+k_2-k_3)]
   + 3 \sin [d (k_1+k_2+k_3)].
\end{split}
\label{phibU0_9}
\end{align}
\begin{align}
\begin{split}
& \frac{(2 \pi )^{3/4}}{s^{3/2}} e^{(k_1^2+k_2^2+k_3^2)s^2} \Phi_{10}^{b,\cu=0}(k_1,k_2,k_3) =
   -1 + \sqrt{2} \big( \cos (d k_1) + \cos (d k_2) + \cos (d k_3) \big) \\
&  - \cos [d (k_1-k_2)] - \cos [d (k_1-k_3)] - \cos [d (k_2-k_3)]
   - \cos [d (k_1+k_2)] - \cos [d (k_1+k_3)] - \cos [d (k_2+k_3)] \\
&  + \frac{1}{\sqrt{2}} \big(\cos [d (k_1+k_2-k_3)] + \cos [d (k_1-k_2+k_3)] + \cos [d (-k_1+k_2+k_3)]
   + \cos [d (k_1+k_2+k_3)]\big)
\end{split}
\label{phibU0_10}
\end{align}
\end{widetext}

\begin{figure*}[t]
\includegraphics[width=17.5cm]{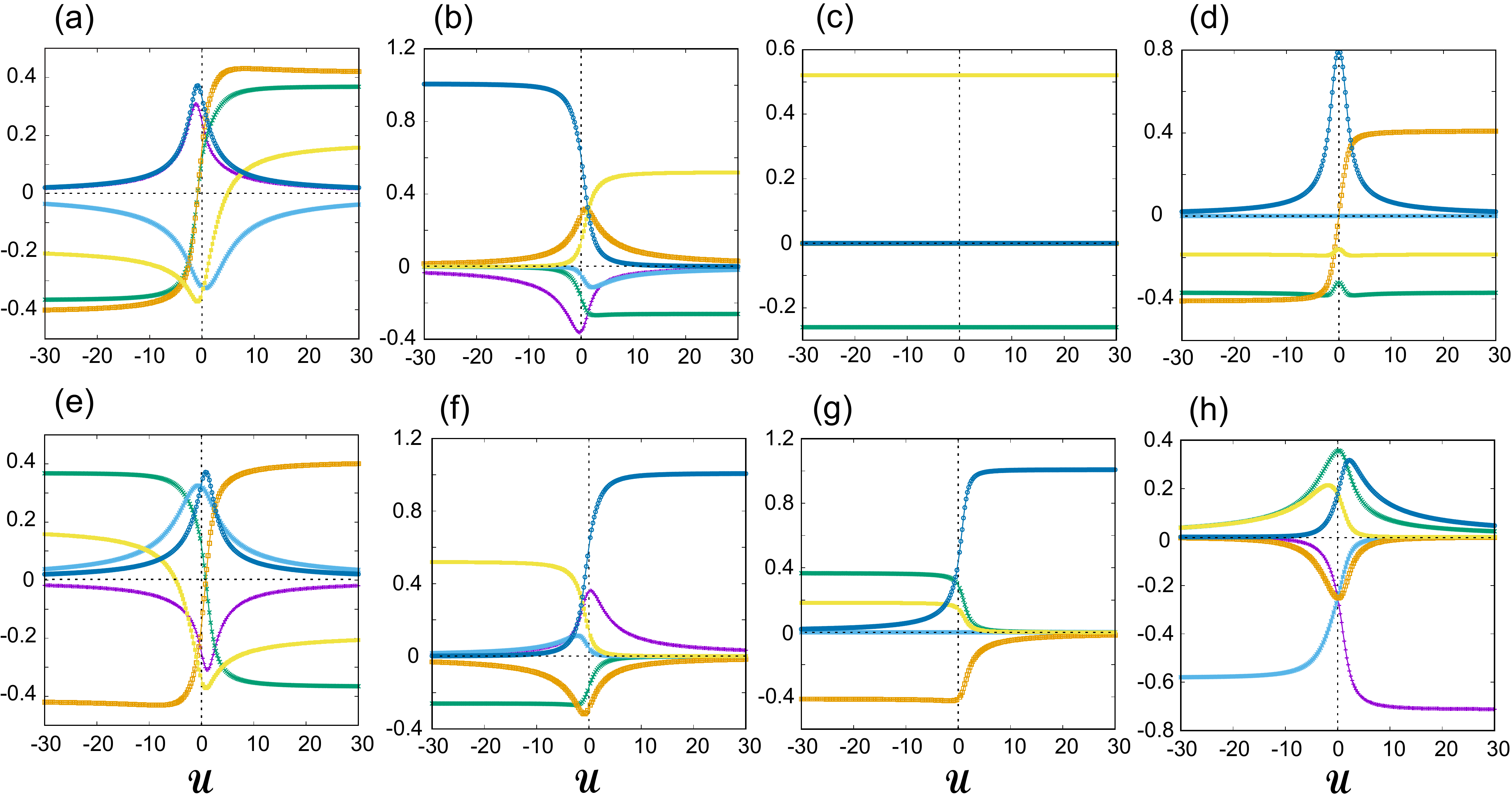}
\caption{The six different ${\cal C}$-coefficients (dimensionless) [see Eq.\ (\ref{wfbexpr})] for the eight 
remaining excited eigenstates of 3 bosons trapped in 3 linearly arranged wells as a function of $\cu$ 
(horizontal axis, dimensionless). 
This figure complements Fig.\ \ref{ccoef} in the main text. 
(a) $i=3r(4l)$. (b) $i=4r(3l)$. (c) $i=5r(6l)$. (d) $i=6r(5l)$. (e) $i=7r(8l)$. (f) $i=8r(7l)$. (g) $i=9$. 
(h) $i=10$. The choice of online colors is the same as in Fig.\ \ref{ccoef}, that is: 
$\cc_0 \rightarrow$ Violet, $\cc_1 \rightarrow$ Green, $\cc_{1-1} \rightarrow$ Light Blue,
$\cc_{1+1} \rightarrow$ Brown, $\cc_{1+1-1} \rightarrow$ Yellow, $\cc_{1+1+1} \rightarrow$ Dark Blue.
For the print grayscale version, the positioning (referred to as \#$n$, with $n=1,2,3,\dots$)
of the curves from top to bottom at the point $\cu = -30$ is as follows: 
(a) $\cc_0 \rightarrow$ \#2 (overlaps with \#1), $\cc_1 \rightarrow$ \#5, $\cc_{1-1} \rightarrow$ \#3,
$\cc_{1+1} \rightarrow$ \#6, $\cc_{1+1-1} \rightarrow$ \#4, $\cc_{1+1+1} \rightarrow$ \#1 (overlaps with \#2).
(b) $\cc_0 \rightarrow$ \#6, $\cc_1 \rightarrow$ \#5 (overlaps with \#3 and \#4), 
$\cc_{1-1} \rightarrow$ \#4,
$\cc_{1+1} \rightarrow$ \#2, $\cc_{1+1-1} \rightarrow$ \#3, $\cc_{1+1+1} \rightarrow$ \#1. 
(c) $\cc_0 = \cc_{1-1} = \cc_{1+1} = \cc_{1+1+1} =0 $, $\cc_1 \rightarrow$ lower curve, 
$\cc_{1+1-1} \rightarrow$ upper curve. 
(d) $\cc_0 \rightarrow$ \#3 (overlaps with \#2), $\cc_1 \rightarrow$ \#5, $\cc_{1-1} \rightarrow$ \#2,
$\cc_{1+1} \rightarrow$ \#6, $\cc_{1+1-1} \rightarrow$ \#4, $\cc_{1+1+1} \rightarrow$ \#1. 
(e) $\cc_0 \rightarrow$ \#5, $\cc_1 \rightarrow$ \#1, $\cc_{1-1} \rightarrow$ \#3,
$\cc_{1+1} \rightarrow$ \#6, $\cc_{1+1-1} \rightarrow$ \#2, $\cc_{1+1+1} \rightarrow$ \#4.
(f) $\cc_0 \rightarrow$ \#3 (overlaps with \#2), $\cc_1 \rightarrow$ \#6, $\cc_{1-1} \rightarrow$ \#2,
$\cc_{1+1} \rightarrow$ \#5, $\cc_{1+1-1} \rightarrow$ \#1, $\cc_{1+1+1} \rightarrow$ \#4. 
(g) $\cc_0 \rightarrow$ \#5 (overlaps with \#4), $\cc_1 \rightarrow$ \#1, $\cc_{1-1} \rightarrow$ \#4,
$\cc_{1+1} \rightarrow$ \#6, $\cc_{1+1-1} \rightarrow$ \#2, $\cc_{1+1+1} \rightarrow$ \#3. 
(h) $\cc_0 \rightarrow$ \#4 (overlaps with \#3 and \#5), $\cc_1 \rightarrow$ \#1 (overlaps with \#2), 
$\cc_{1-1} \rightarrow$ \#6,
$\cc_{1+1} \rightarrow$ \#5, $\cc_{1+1-1} \rightarrow$ \#2, $\cc_{1+1+1} \rightarrow$ \#3. 
}
\label{ccoefa}
\end{figure*}

\section{Third-order momentum correlations as a function of $\cu$ for the remaining eight excited states}
\label{a3rd}

Fig.\ \ref{ccoefa} complements Fig.\ \ref{ccoef} in that it displays the six coefficients $\cc^i(\cu)$'s for the 
remaining eight excited states (explicit numerical values can be found in the supplemental material \cite{supp}).
The dependence of these coefficients on the interaction strength $\cu$ is 
better deciphered by using as reference points the special cases at $\cu \rightarrow \pm \infty$ and $\cu=0$.
Note that in all cases the $\cc^i$ values at the end points $\cu=\pm 30$ in the figure are close to the 
corresponding limiting values at $\cu \rightarrow \pm \infty$. In particular, 

{\it The excited state denoted as $i=3r(4l)$ ($i=3$ for $0 < \cu < +\infty$ and $i=4$ for 
$-\infty < \cu < 0$):\/}
For $\cu \rightarrow -\infty$, only three coefficients, 
${\cc}^{4,-\infty}_1=-2\times2^{3/4}/(\sqrt{15}\pi^{3/4})=-0.3680$, 
${\cc}^{4,-\infty}_{1+1}=-2^{3/4}/(\sqrt{3}\pi^{3/4})=-0.4115$, and
${\cc}^{4,-\infty}_{1+1-1}=-2^{3/4}/(\sqrt{15}\pi^{3/4})=-0.1840$, 
survive in expression (\ref{wfbexpr}) [see frame (a) in Fig.\ \ref{ccoefa}]; 
the corresponding Hubbard eigenvector, $\phi^{b,-\infty}_4$ [second line in Eq.\ (\ref{phi4})] consists
of all 6 primitive kets [see Eq.\ (\ref{3b-kets})] representing exclusively doubly-occupied wells,
and the corresponding wave function in momentum space has 9 cosinusoidal terms and is given by the second 
expression in Eq.\ (\ref{phibUpmI_4}).

For $\cu =0$, all 6 coefficients, $\cc^{3r(4l),\cu=0}$'s, are present, and their numerical values from
frame (a) in Fig.\ \ref{ccoefa} agree with the numerical values for $\Phi_{3r(4l)}^{b,\cu=0}(k_1,k_2,k_3)$ in
Eq.\ (\ref{phibU0_3}).

For $\cu \rightarrow +\infty$, again only three coefficients, 
${\cc}^{3,+\infty}_1=2\times2^{3/4}/(\sqrt{15}\pi^{3/4})=0.3680$, 
${\cc}^{3,+\infty}_{1+1}=2^{3/4}/(\sqrt{3}\pi^{3/4})=0.4115$, and
${\cc}^{3,+\infty}_{1+1-1}=2^{3/4}/(\sqrt{15}\pi^{3/4})=0.1840$, 
survive in expression (\ref{wfbexpr}) [see frame (a) in Fig.\ \ref{ccoefa}]; 
the corresponding Hubbard eigenvector, $\phi^{b,+\infty}_3$ [first line in Eq.\ (\ref{phi3})] consists
of all 6 primitive kets [see Eq.\ (\ref{3b-kets})] representing exclusively doubly-occupied wells, and the 
corresponding wave function in momentum space has 9 cosinusoidal terms and is given by the second expression
in Eq.\ (\ref{phibUpmI_3}).

{\it The excited state denoted as $i=4r(3l)$ ($i=4$ for $0 < \cu < +\infty$ and $i=3$ for $-\infty < \cu < 0$):\/}
For $\cu \rightarrow -\infty$, only one coefficient, 
${\cc}^{3,-\infty}_{1+1+1}=2\times2^{1/4}/\pi^{3/4}=1.0079$, 
survives in expression (\ref{wfbexpr}) (see frame (b) in Fig.\ \ref{ccoefa}); 
the corresponding Hubbard eigenvector, $\phi^{b,-\infty}_3$ [second line in Eq.\ (\ref{phi3})] is a NOON
state of the form $(|300\rangle + |003\rangle)/\sqrt{2}$, and the corresponding wave function in momentum
space is given by the second expression in Eq.\ (\ref{phibUpmI_3}), which includes a cos term only.

For $\cu =0$, all 6 coefficients, $\cc^{4r(3l),\cu=0}$'s, are present, and their numerical values from
frame (b) in Fig.\ \ref{ccoefa} agree with the numerical values for $\Phi_{4r(3l)}^{b,\cu=0}(k_1,k_2,k_3)$ in
Eq.\ (\ref{phibU0_4}).

For $\cu \rightarrow +\infty$, only two coefficients, 
${\cc}^{4,+\infty}_1=-2\times2^{1/4}/(\sqrt{15}\pi^{3/4})=-0.2602$, and 
${\cc}^{4,+\infty}_{1+1-1}=4\times2^{1/4}/(\sqrt{15}\pi^{3/4})=0.5205$, 
survive in expression (\ref{wfbexpr}) [see frame (b) in Fig.\ \ref{ccoefa}]; 
the corresponding Hubbard eigenvector, $\phi^{b,+\infty}_4$ [first line in Eq.\ (\ref{phi4})] consists
of 4 primitive kets [see Eq.\ (\ref{3b-kets})] representing exclusively doubly-occupied wells, and the 
corresponding wave function in momentum space has 6 cosinusoidal terms and is given by the first expression
in Eq.\ (\ref{phibUpmI_4}).

{\it The excited state denoted as $i=5r(6l)$ ($i=5$ for $0 < \cu < +\infty$ and $i=6$ for $-\infty < \cu < 0$):\/}
The Hubbard eigenvector solution for this state state is $\cu$-independent; see first expression in Eq.\ 
(\ref{phi5}), second expression in Eq.\ (\ref{phi6}), or Eq.\ (\ref{eigvecU0st5}). 
In this case, 2 distinct coefficients survive in expression (\ref{wfbexpr}), that is,
${\cc}^{6,-\infty}_1={\cc}^{5,+\infty}_1={\cc}^{5r(6l),\cu=0}_1=-2\times2^{1/4}/(\sqrt{15}\pi^{3/4})=-0.2602$, 
and ${\cc}^{6,-\infty}_{1+1-1}={\cc}^{5,+\infty}_{1+1-1}={\cc}^{5r(6l),\cu=0}_{1+1-1}=
4\times2^{1/4}/(\sqrt{15}\pi^{3/4})=0.5205$, in agreement with frame (c) in Fig.\ \ref{ccoefa}. 
The corresponding Hubbard eigenvectors, $\phi^{b,+\infty}_5=\phi^{b,-\infty}_6=\phi^{b,\cu=0}_{5r(6l)}$, 
consist of 4 primitive kets [see Eq.\ (\ref{3b-kets})] representing exclusively doubly-occupied wells, and the
corresponding wave function in momentum space has 6 {\it cosinusoidal\/} terms and is given by the second 
expression in Eq.\ (\ref{phibUpmI_6}) or the first expression in Eq.\ (\ref{phibUpmI_5}).

{\it The excited state denoted as $i=6r(5l)$ ($i=6$ for $0 < \cu < +\infty$ and $i=5$ for $-\infty < \cu < 0$):\/}
For $\cu \rightarrow -\infty$, only three coefficients, 
${\cc}^{5,-\infty}_1=-2\times2^{3/4}/(\sqrt{15}\pi^{3/4})=-0.3680$, 
${\cc}^{5,-\infty}_{1+1}=-2^{3/4}/(\sqrt{3}\pi^{3/4})=-0.4115$, and
${\cc}^{5,-\infty}_{1+1-1}=-2^{3/4}/(\sqrt{15}\pi^{3/4})=-0.1840$, 
survive in expression (\ref{wfbexpr}) [see frame (d) in Fig.\ \ref{ccoefa}]; 
the corresponding Hubbard eigenvector, $\phi^{b,-\infty}_5$ [second line in Eq.\ (\ref{phi5})] consists
of all 6 primitive kets [see Eq.\ (\ref{3b-kets})] representing exclusively doubly-occupied wells,
and the corresponding wave function in momentum space has 9 {\it sinusoidal\/} terms and is given by the second 
expression in Eq.\ (\ref{phibUpmI_5}).

For $\cu=0$, three coefficients are present, namely ${\cc}^{6r(5l),\cu=0}_1$, ${\cc}^{6r(5l),\cu=0}_{1+1-1}$,
and ${\cc}^{6r(5l),\cu=0}_{1+1+1}$. Their numerical values from frame (d) in Fig.\ \ref{ccoefa} agree with the 
corresponding algebraic expressions for $\Phi_{6r(5l)}^{b,\cu=0}(k_1,k_2,k_3)$ in Eq.\ (\ref{phibU0_6}).

For $\cu \rightarrow +\infty$, again only three coefficients, 
${\cc}^{6,+\infty}_1=-2\times2^{3/4}/(\sqrt{15}\pi^{3/4})=-0.3680$, 
${\cc}^{6,+\infty}_{1+1}=2^{3/4}/(\sqrt{3}\pi^{3/4})=0.4115$, and
${\cc}^{6,+\infty}_{1+1-1}=-2^{3/4}/(\sqrt{15}\pi^{3/4})=-0.1840$, 
survive in expression (\ref{wfbexpr}) [see frame (d) in Fig.\ \ref{ccoefa}]; 
the corresponding Hubbard eigenvector, $\phi^{b,+\infty}_6$ [first line in Eq.\ (\ref{phi6})] consists
of all 6 primitive kets [see Eq.\ (\ref{3b-kets})] representing exclusively doubly-occupied wells, and the 
corresponding wave function in momentum space has 9 {\it sinusoidal\/} terms and is given by the first expression
in Eq.\ (\ref{phibUpmI_6}).

{\it The excited state denoted as $i=7r(8l)$ ($i=7$ for $0 < \cu < +\infty$ and $i=8$ for $-\infty < \cu < 0$):\/}
For $\cu \rightarrow -\infty$, only three coefficients, 
${\cc}^{8,-\infty}_1=2\times2^{3/4}/(\sqrt{15}\pi^{3/4})=0.3680$, 
${\cc}^{8,-\infty}_{1+1}=-2^{3/4}/(\sqrt{3}\pi^{3/4})=-0.4115$, and
${\cc}^{8,-\infty}_{1+1-1}=2^{3/4}/(\sqrt{15}\pi^{3/4})=0.1840$, 
survive in expression (\ref{wfbexpr}) [see sixth frame (e) in Fig.\ \ref{ccoefa}]; 
the corresponding Hubbard eigenvector, $\phi^{b,-\infty}_8$ [second line in Eq.\ (\ref{phi8})] consists
of all 6 primitive kets [see Eq.\ (\ref{3b-kets})] representing exclusively doubly-occupied wells,
and the corresponding wave function in momentum space has 9 cosinusoidal terms and is given by the second 
expression in Eq.\ (\ref{phibUpmI_8}).

For $\cu =0$, all 6 coefficients, $\cc^{7r(8l),\cu=0}$'s, are present, and their numerical values from
frame (e) in Fig.\ \ref{ccoefa} agree with the numerical values for $\Phi_{7r(8l)}^{b,\cu=0}(k_1,k_2,k_3)$ in
Eq.\ (\ref{phibU0_7}).

For $\cu \rightarrow +\infty$, again only three coefficients, 
${\cc}^{7,+\infty}_1=-2\times2^{3/4}/(\sqrt{15}\pi^{3/4})=-0.3680$, 
${\cc}^{7,+\infty}_{1+1}=2^{3/4}/(\sqrt{3}\pi^{3/4})=0.4115$, and
${\cc}^{7,+\infty}_{1+1-1}=-2^{3/4}/(\sqrt{15}\pi^{3/4})=-0.1840$, 
survive in expression (\ref{wfbexpr}) [see frame (e) in Fig.\ \ref{ccoefa}]; 
the corresponding Hubbard eigenvector, $\phi^{b,+\infty}_7$ [first line in Eq.\ (\ref{phi7})] consists
of all 6 primitive kets [see Eq.\ (\ref{3b-kets})] representing exclusively doubly-occupied wells,
and the corresponding wave function in momentum space has 9 cosinusoidal terms and is given by the first 
expression in Eq.\ (\ref{phibUpmI_7}).

{\it The excited state denoted as $i=8r(7l)$ ($i=8$ for $0 < \cu < +\infty$ and $i=7$ for $-\infty < \cu < 0$):\/}
For $\cu \rightarrow -\infty$, only two coefficients, 
${\cc}^{7,-\infty}_1=-2\times2^{1/4}/(\sqrt{15}\pi^{3/4})=-0.2602$, and 
${\cc}^{7,-\infty}_{1+1-1}=4\times2^{1/4}/(\sqrt{15}\pi^{3/4})=0.5205$, 
survive in expression (\ref{wfbexpr}) [see frame (f) in Fig.\ \ref{ccoefa}]; 
the corresponding Hubbard eigenvector, $\phi^{b,-\infty}_7$ [second line in Eq.\ (\ref{phi7})] consists
of 4 primitive kets [see Eq.\ (\ref{3b-kets})] representing exclusively doubly-occupied wells, and the 
corresponding wave function in momentum space has 6 cosinusoidal terms and is given by the second expression
in Eq.\ (\ref{phibUpmI_7}).

For $\cu =0$, all 6 coefficients, $\cc^{8r(7l),\cu=0}$'s, are present, and their numerical values from
frame (f) in Fig.\ \ref{ccoefa} agree with the numerical values for $\Phi_{8r(7l)}^{b,\cu=0}(k_1,k_2,k_3)$ in
Eq.\ (\ref{phibU0_8}).

For $\cu \rightarrow +\infty$, only one coefficient,
${\cc}^{8,+\infty}_{1+1+1}=2\times2^{1/4}/\pi^{3/4}=1.0079$,
survives in expression (\ref{wfbexpr}) [see frame (f) in Fig.\ \ref{ccoefa}];
the corresponding Hubbard eigenvector, $\phi^{b,+\infty}_8$ [first line in Eq.\ (\ref{phi8})] is a NOON
state of the form $(|300\rangle + |003\rangle)/\sqrt{2}$, and the corresponding wave function in momentum
space is given by the first expression in Eq.\ (\ref{phibUpmI_8}), which includes a cos term only.

{\it The excited state denoted as $i=9$ for $-\infty < \cu < +\infty$:\/}
For $\cu \rightarrow -\infty$, only three coefficients, 
${\cc}^{5,-\infty}_1=2\times2^{3/4}/(\sqrt{15}\pi^{3/4})=0.3680$, 
${\cc}^{5,-\infty}_{1+1}=-2^{3/4}/(\sqrt{3}\pi^{3/4})=-0.4115$, and
${\cc}^{5,-\infty}_{1+1-1}=2^{3/4}/(\sqrt{15}\pi^{3/4})=0.1840$, 
survive in expression (\ref{wfbexpr}) [see frame (g) in Fig.\ \ref{ccoefa}]; 
the corresponding Hubbard eigenvector, $\phi^{b,-\infty}_9$ [second line in Eq.\ (\ref{phi9})] consists
of all 6 primitive kets [see Eq.\ (\ref{3b-kets})] representing exclusively doubly-occupied wells,
and the corresponding wave function in momentum space has 9 {\it sinusoidal\/} terms and is given by the second 
expression in Eq.\ (\ref{phibUpmI_9}).

For $\cu=0$, four coefficients are present, namely ${\cc}^{9,\cu=0}_1$, ${\cc}^{9,\cu=0}_{1+1}$,
${\cc}^{9,\cu=0}_{1+1-1}$, and ${\cc}^{9,\cu=0}_{1+1+1}$. 
Their numerical values from frame (g) in Fig.\ \ref{ccoefa} agree with the
corresponding algebraic expressions for $\Phi_9^{b,\cu=0}(k_1,k_2,k_3)$ in Eq.\ (\ref{phibU0_9}).

For $\cu \rightarrow +\infty$, only one coefficient,
${\cc}^{9,+\infty}_{1+1+1}=2\times2^{1/4}/\pi^{3/4}=1.0079$,
survives in expression (\ref{wfbexpr}) [see frame (g) in Fig.\ \ref{ccoefa}];
the corresponding Hubbard eigenvector, $\phi^{b,+\infty}_9$ [first line in Eq.\ (\ref{phi9})] is a NOON
state of the form $(-|300\rangle + |003\rangle)/\sqrt{2}$, and the corresponding wave function in momentum
space is given by the first expression in Eq.\ (\ref{phibUpmI_9}), {\it which includes a sin term only\/}.

{\it The highest excited state denoted as $i=10$ for $-\infty < \cu < +\infty$:\/}
For $\cu \rightarrow -\infty$, only the coefficient $\cc^{10,-\infty}_{1-1}=
-2\times2^{1/4}/(\sqrt{3}\pi^{3/4}) = - 0.5819$ survives in expression (\ref{wfbexpr});
see frame (h) in Fig.\ \ref{ccoefa}. The corresponding momentum-space wave function 
comprises three cosinusoidal terms and is given by the second expression in Eq.\ (\ref{phibUpmI_10}).
The corresponding Hubbard eigenvector $\phi^{b,-\infty}_{10}$ [second line in Eq.\ (\ref{phi10})] contains
only a single component from the primitive kets listed in Eq.\ (\ref{3b-kets}), i.e., the basis ket No. 1 
$\rightarrow |111\rangle$, reflecting the fact that all three wells are singly occupied.

For $\cu =0$, all 6 coefficients, $\cc^{10,\cu=0}$'s, are present, and their numerical values from
frame (h) in Fig.\ \ref{ccoefa} agree with the numerical values for $\Phi_{10}^{b,\cu=0}(k_1,k_2,k_3)$ in
Eq.\ (\ref{phibU0_10}).

For $\cu \rightarrow +\infty$, it is seen from frame (h) in Fig.\ \ref{ccoefa} that only the constant 
coefficient $\cc^{10,-\infty}_0=-(2/\pi)^{3/4}=-0.7127$ survives in expression (\ref{wfbexpr}); the corresponding 
wave function in momentum space is given by the first expression in Eq.\ (\ref{phibUpmI_10}). 
It is a simple Gaussian distribution associated with a Bose-Einstein condensate, reflecting
the fact that all three bosons are localized in the middle well and occupy the same orbital; the corresponding
Hubbard eigenvector is given by $\phi^{b,+\infty}_{10}$ [first line in Eq.\ (\ref{phi10})] which contains only a
single component from the primitive kets listed in Eq.\ (\ref{3b-kets}), i.e., the basis ket No. 9 
$\rightarrow |030\rangle$.  

\begin{figure*}[t]
\includegraphics[width=17.5cm]{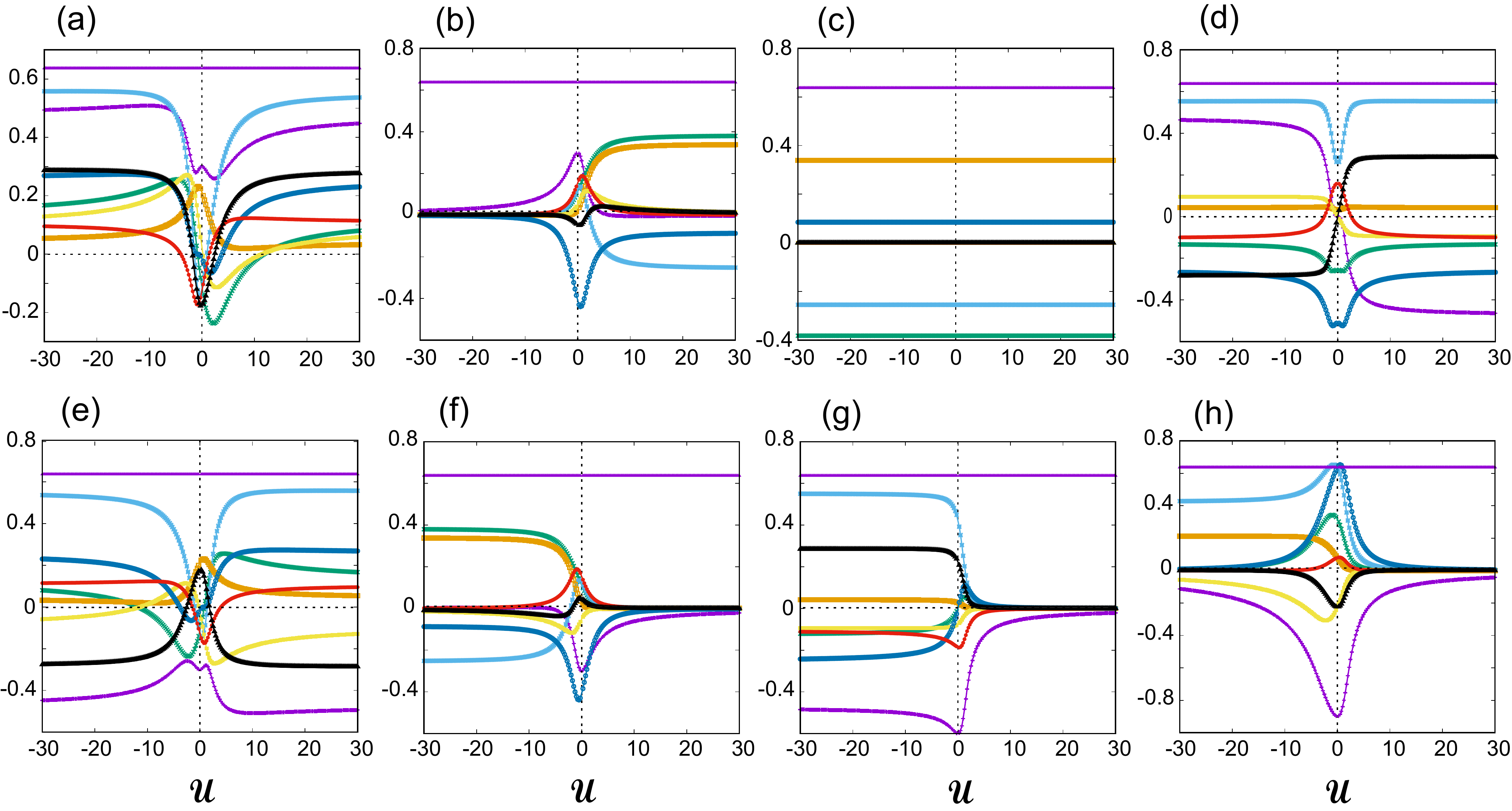}
\caption{The nine ${\cal B}$-coefficients (dimensionless) [see Eq.\ (\ref{2ndbexpr})] for the remaining 
8 excited eigenstates of 3 bosons trapped in 3 linearly arranged wells as a function of $\cu$ 
(horizontal axis, dimensionless). This figure 
complements Fig.\ \ref{bcoef} in the main text. (a) $i=3r(4l)$. (b) $i=4r(3l)$. (c) $i=5r(6l)$. 
(d) $i=6r(5l)$. (e) $i=7r(8l)$. (f) $i=8r(7l)$, (g) $i=9$, (h) $i=10$.
See text for a detailed description. The choice of colors is the same as in Fig.\ \ref{bcoef}, that is:
$\cb_0 \rightarrow$ Constant (Violet), $\cb_1 \rightarrow$ Second Violet, 
$\cb_2 \rightarrow$ Green, $\cb_{1-1} \rightarrow$ Light Blue, $\cb_{2-2} \rightarrow$ Brown,
$\cb_{2-1} \rightarrow$ Yellow, $\cb_{1+1} \rightarrow$ Dark Blue, 
$\cb_{2+2} \rightarrow$ Red, $\cb_{2+1} \rightarrow$ Black.
For the print grayscale version, the positioning (referred to as \#$n$, with $n=1,2,3,\dots$)
of the curves from top to bottom at the point $\cu = -30$ is as follows:
(a) $\cb_0 \rightarrow$ \#1, $\cb_1 \rightarrow$ \#3,
$\cb_2 \rightarrow$ \#6, $\cb_{1-1} \rightarrow$ \#2, $\cb_{2-2} \rightarrow$ \#9,
$\cb_{2-1} \rightarrow$ \#7, $\cb_{1+1} \rightarrow$ \#5,
$\cb_{2+2} \rightarrow$ \#8, $\cb_{2+1} \rightarrow$ \#4.
(b) $\cb_0 \rightarrow$ \#1, $\cb_1 \rightarrow$ \#2,
$\cb_2 \rightarrow$ \#5 (overlaps with \#3, \#4, \#6, \#7, \#8, \#9),
$\cb_{1-1} \rightarrow$ \#4, $\cb_{2-2} \rightarrow$ \#7,
$\cb_{2-1} \rightarrow$ \#6, $\cb_{1+1} \rightarrow$ \#9,
$\cb_{2+2} \rightarrow$ \#3, $\cb_{2+1} \rightarrow$ \#8.
(c) $\cb_0 \rightarrow$ \#1, $\cb_1 \rightarrow$ \#4 (overlaps with \#5, \#6, \#7),
$\cb_2 \rightarrow$ \#9, $\cb_{1-1} \rightarrow$ \#8, $\cb_{2-2} \rightarrow$ \#2,
$\cb_{2-1} \rightarrow$ \#5, $\cb_{1+1} \rightarrow$ \#3,
$\cb_{2+2} \rightarrow$ \#6, $\cb_{2+1} \rightarrow$ \#7.
(d) $\cb_0 \rightarrow$ \#1, $\cb_1 \rightarrow$ \#3,
$\cb_2 \rightarrow$ \#7, $\cb_{1-1} \rightarrow$ \#2, $\cb_{2-2} \rightarrow$ \#5,
$\cb_{2-1} \rightarrow$ \#4, $\cb_{1+1} \rightarrow$ \#8,
$\cb_{2+2} \rightarrow$ \#6, $\cb_{2+1} \rightarrow$ \#9.
(e) $\cb_0 \rightarrow$ \#1, $\cb_1 \rightarrow$ \#9,
$\cb_2 \rightarrow$ \#5, $\cb_{1-1} \rightarrow$ \#2, $\cb_{2-2} \rightarrow$ \#6,
$\cb_{2-1} \rightarrow$ \#7, $\cb_{1+1} \rightarrow$ \#3,
$\cb_{2+2} \rightarrow$ \#4, $\cb_{2+1} \rightarrow$ \#8.
(f) $\cb_0 \rightarrow$ \#1, $\cb_1 \rightarrow$ \#5 (overlaps with \#4, \#6, \#7),
$\cb_2 \rightarrow$ \#2, $\cb_{1-1} \rightarrow$ \#9, $\cb_{2-2} \rightarrow$ \#3,
$\cb_{2-1} \rightarrow$ \#7, $\cb_{1+1} \rightarrow$ \#8,
$\cb_{2+2} \rightarrow$ \#4, $\cb_{2+1} \rightarrow$ \#6.
(g) $\cb_0 \rightarrow$ \#1, $\cb_1 \rightarrow$ \#9, $\cb_2 \rightarrow$ \#7 (overlaps with \#5, \#6), 
$\cb_{1-1} \rightarrow$ \#2, $\cb_{2-2} \rightarrow$ \#4,
$\cb_{2-1} \rightarrow$ \#5, $\cb_{1+1} \rightarrow$ \#8,
$\cb_{2+2} \rightarrow$ \#6, $\cb_{2+1} \rightarrow$ \#3.
(h) $\cb_0 \rightarrow$ \#1, $\cb_1 \rightarrow$ \#9, $\cb_2 \rightarrow$ \#5 (overlaps wih \#4), 
$\cb_{1-1} \rightarrow$ \#2, $\cb_{2-2} \rightarrow$ \#3,
$\cb_{2-1} \rightarrow$ \#8, $\cb_{1+1} \rightarrow$ \#4,
$\cb_{2+2} \rightarrow$ \#6 (overlaps with \#7), $\cb_{2+1} \rightarrow$ \#7.
}
\label{bcoefa}
\end{figure*}

\section{Second-order momentum correlations as a function of $\cu$ for the remaining eight excited states}
\label{a2nd}

Fig.\ \ref{bcoefa} complements Fig.\ \ref{bcoef} in that it displays the 9 distinct coefficients $\cb^i(\cu)$'s
for the remaining eight excited states (explicit numerical values can be found in the supplemental 
material \cite{supp}). The dependence of these coefficients on the interaction strength $\cu$ is
better deciphered by using as reference points the special cases at $\cu \rightarrow \pm \infty$ and $\cu=0$.
Note that in all cases the $\cb^i$ values at the end points $\cu=\pm 30$ in the figure are close to the
corresponding limiting values at $\cu \rightarrow \pm \infty$. In particular,

{\it The excited state denoted as $i=3r(4l)$ ($i=3$ for $0 < \cu < +\infty$ and $i=4$ for $-\infty<\cu<0$)):\/}
For $\cu \rightarrow -\infty$ all 9 distinct coefficients survive [see frame (a) in Fig.\ \ref{bcoefa}]; this
state consists of only doubly-and-singly occupied sites [see second line of Eq.\ (\ref{phi4})]. In this case, 
the 9 distinct coefficients are:
$\cb^{4,-\infty}_0=2/\pi$, 
$\cb^{4,-\infty}_1=2\sqrt{5}/(3\pi)$, 
$\cb^{4,-\infty}_2=2/(5\pi)$, 
$\cb^{4,-\infty}_{1-1}=26/(15\pi)$, 
$\cb^{4,-\infty}_{2-2}=2/(15\pi)$, 
$\cb^{4,-\infty}_{2-1}=2/(3\sqrt{5}\pi)$, 
$\cb^{4,-\infty}_{1+1}=4/(5\pi)$, 
$\cb^{4,-\infty}_{2+2}=1/(3\pi)$, and
$\cb^{4,-\infty}_{2+1}=2/(\sqrt{5}\pi)$. 

In the noninteracting case ($\cu=0$), for which the Hubbard eigenvector is given by Eq.\ (\ref{eigvecU0st3}),
all 13 cosinusoidal terms and 9 distinct coefficients are present in Eq.\ (\ref{2ndbexpr}), in agreement with
the frame (a) of Fig.\ \ref{bcoefa}, that is,
$\cb^{3r(4l),\cu=0}_0=2/\pi=0.63662$,
$\cb^{3r(4l),\cu=0}_1=0.300105$,
$\cb^{3r(4l),\cu=0}_2=-0.111766$,
$\cb^{3r(4l),\cu=0}_{1-1}=-0.124185$,
$\cb^{3r(4l),\cu=0}_{2-2}=0.214057$,
$\cb^{3r(4l),\cu=0}_{2-1}=0.0243414$,
$\cb^{3r(4l),\cu=0}_{1+1}=-0.0135114$,
$\cb^{3r(4l),\cu=0}_{2+2}=-0.127997$, and
$\cb^{3r(4l),\cu=0}_{2+1}=-0.178398$.

For $\cu \rightarrow +\infty$, all 13 cosinusoidal terms survive in expression (\ref{2ndbexpr}); 
the corresponding state is 
given by the first expression in Eq.\ (\ref{phi3}). In this case, the 9 distinct coefficients are:
$\cb^{3,+\infty}_0=2/\pi$, 
$\cb^{3,+\infty}_1=2\sqrt{5}/(3\pi)$, 
$\cb^{3,+\infty}_2=2/(5\pi)$, 
$\cb^{3,+\infty}_{1-1}=26/(15\pi)$, 
$\cb^{3,+\infty}_{2-2}=2/(15\pi)$, 
$\cb^{3,+\infty}_{2-1}=2/(3\sqrt{5}\pi)$, 
$\cb^{3,+\infty}_{1+1}=4/(5\pi)$, 
$\cb^{3,+\infty}_{2+2}=1/(3\pi)$, and
$\cb^{3,+\infty}_{2+1}=2/(\sqrt{5}\pi)$. 

We note that $^2{\cal G}_4^{b,-\infty}(k_1,k_2)\;=\;^2{\cal G}_3^{b,+\infty}(k_1,k_2)$.

{\it The excited state denoted as $i=4r(3l)$ ($i=4$ for $0<\cu<+\infty$ and $i=3$ for $-\infty < \cu < 0$):\/}
For $\cu \rightarrow -\infty$ only the constant term survives in expression (\ref{2ndbexpr}); 
the corresponding state is given by the second expression in Eq.\ (\ref{phi3}) and is a NOON state 
of the form $(|300\rangle + |003\rangle)/\sqrt{2}$. In this case, the second-order correlation
function is given by:
\begin{align}
^2{\cal G}_3^{b,-\infty}(k_1,k_2)= \frac{2}{\pi}s^2 e^{-2(k_1^2+k_2^2)s^2}.
\label{3b2ndst3UmI}
\end{align}

In the noninteracting case ($\cu=0$), for which the Hubbard eigenvector is given by Eq.\ (\ref{eigvecU0st4}),
all 13 cosinusoidal terms and 9 distinct coefficients are present in Eq.\ (\ref{2ndbexpr}), in agreement with
the frame (b) of Fig.\ \ref{bcoefa}, that is,
$\cb^{4r(3l),\cu=0}_0=2/\pi=0.63662$,
$\cb^{4r(3l),\cu=0}_1=0.300105$,
$\cb^{4r(3l),\cu=0}_2=0.111766$,
$\cb^{4r(3l),\cu=0}_{1-1}=0.124185$,
$\cb^{4r(3l),\cu=0}_{2-2}=0.0246755$,
$\cb^{4r(3l),\cu=0}_{2-1}=0.0506849$,
$\cb^{4r(3l),\cu=0}_{1+1}=-0.410902$,
$\cb^{4r(3l),\cu=0}_{2+2}=0.154523$, and
$\cb^{4r(3l),\cu=0}_{2+1}=-0.0466808$.

For $\cu \rightarrow +\infty$, 6 cosinusoidal terms survive in expression (\ref{2ndbexpr}); see frame (b) in Fig.\ 
\ref{bcoefa}. The corresponding state is given by the first expression in Eq.\ (\ref{phi4}). In this case, the 
5 non-zero distinct coefficients are:
$\cb^{4,+\infty}_0=2/\pi$, 
$\cb^{4,+\infty}_1=0$, 
$\cb^{4,+\infty}_2=6/(5\pi)$, 
$\cb^{4,+\infty}_{1-1}=-4/(5\pi)$, 
$\cb^{4,+\infty}_{2-2}=16/(15\pi)$, 
$\cb^{4,+\infty}_{2-1}=0$, 
$\cb^{4,+\infty}_{1+1}=-4/(15\pi)$, 
$\cb^{4,+\infty}_{2+2}=0$, and
$\cb^{4,+\infty}_{2+1}=0$. 

{\it The excited state denoted as $i=5r(6l)$ ($i=5$ for $0<\cu<+\infty$ and $i=6$ for $-\infty < \cu < 0$):\/}
This state is $\cu$-independent; see first expression in Eq.\ (\ref{phi5}) or second expression in
Eq.\ (\ref{phi6}). In this case, 5 distinct coefficients (corresponding to 6 cosinusoidal terms) survive in 
expression (\ref{2ndbexpr}), that is,
$\cb^{5,+\infty}_0=\cb^{6,-\infty}_0=\cb^{5r(6l),\cu=0}_0=2/\pi$, 
$\cb^{5,+\infty}_1=\cb^{6,-\infty}_1=\cb^{5r(6l),\cu=0}_1=0$, 
$\cb^{5,+\infty}_2=\cb^{6,-\infty}_2=\cb^{5r(6l),\cu=0}_2=-6/(5\pi)$, 
$\cb^{5,+\infty}_{1-1}=\cb^{6,-\infty}_{1-1}=\cb^{5r(6l),\cu=0}_{1-1}=-4/(5\pi)$, 
$\cb^{5,+\infty}_{2-2}=\cb^{6,-\infty}_{2-2}=\cb^{5r(6l),\cu=0}_{2-2}=16/(15\pi)$, 
$\cb^{5,+\infty}_{2-1}=\cb^{6,-\infty}_{2-1}=\cb^{5r(6l),\cu=0}_{2-1}=0$, 
$\cb^{5,+\infty}_{1+1}=\cb^{6,-\infty}_{1+1}=\cb^{5r(6l),\cu=0}_{1+1}=4/(15\pi)$, 
$\cb^{5,+\infty}_{2+2}=\cb^{6,-\infty}_{2+2}=\cb^{5r(6l),\cu=0}_{2+2}=0$, and
$\cb^{5,+\infty}_{2+1}=\cb^{6,-\infty}_{2+1}=\cb^{5r(6l),\cu=0}_{2+1}=0$;
see frame (c) in Fig.\ \ref{bcoefa}. 

{\it The excited state denoted as $i=6r(5l)$ ($i=6$ for $0<\cu<+\infty$ and $i=5$ for $-\infty < \cu < 0$):\/}
For $\cu \rightarrow -\infty$, all 9 distinct coefficients [see frame (d) in Fig.\ \ref{bcoefa}] and 13 
cosinusoidal terms survive in expression (\ref{2ndbexpr}); 
the corresponding state is given by the second expression in Eq.\ (\ref{phi5}). In this case, the 9 distinct 
coefficients are:
$\cb^{5,-\infty}_0=2/\pi$, 
$\cb^{5,-\infty}_1=2\sqrt{5}/(3\pi)$, 
$\cb^{5,-\infty}_2=-2/(5\pi)$, 
$\cb^{5,-\infty}_{1-1}=26/(15\pi)$, 
$\cb^{5,-\infty}_{2-2}=2/(15\pi)$, 
$\cb^{5,-\infty}_{2-1}=2/(3\sqrt{5}\pi)$, 
$\cb^{5,-\infty}_{1+1}=-4/(5\pi)$, 
$\cb^{5,-\infty}_{2+2}=-1/(3\pi)$, and
$\cb^{5,-\infty}_{2+1}=-2/(\sqrt{5}\pi)$. 

In the noninteracting case ($\cu=0$), for which the Hubbard eigenvector is given by Eq.\ (\ref{eigvecU0st6}),
6 distinct coefficients (corresponding to 7 cosinusoidal terms) are present in expression (\ref{2ndbexpr}), 
in agreement with frame (d) of Fig.\ \ref{bcoefa}. That is,
$\cb^{6r(5l),\cu=0}_0=2/\pi=0.63662$,
$\cb^{6r(5l),\cu=0}_1=0$,
$\cb^{6r(5l),\cu=0}_2=-4/(5\pi)$,
$\cb^{6r(5l),\cu=0}_{1-1}=4/(5\pi)$,
$\cb^{6r(5l),\cu=0}_{2-2}=1/(10\pi)$,
$\cb^{6r(5l),\cu=0}_{2-1}=0$,
$\cb^{6r(5l),\cu=0}_{1+1}=-8/(5\pi)$,
$\cb^{6r(5l),\cu=0}_{2+2}=1/(2\pi)$, and
$\cb^{6r(5l),\cu=0}_{2+1}=0$.

For $\cu \rightarrow +\infty$, all 13 cosinusoidal terms survive in expression (\ref{2ndbexpr}); 
the corresponding state is given by the first expression in Eq.\ (\ref{phi6}). In this case, 
in agreement with the frame (d) of Fig.\ \ref{bcoefa}, the 9 distinct coefficients are:
$\cb^{6,+\infty}_0=2/\pi$, 
$\cb^{6,+\infty}_1=-2\sqrt{5}/(3\pi)$, 
$\cb^{6,+\infty}_2=-2/(5\pi)$, 
$\cb^{6,+\infty}_{1-1}=26/(15\pi)$, 
$\cb^{6,+\infty}_{2-2}=2/(15\pi)$, 
$\cb^{6,+\infty}_{2-1}=-2/(3\sqrt{5}\pi)$, 
$\cb^{6,+\infty}_{1+1}=-4/(5\pi)$, 
$\cb^{6,+\infty}_{2+2}=-1/(3\pi)$, and
$\cb^{6,+\infty}_{2+1}=2/(\sqrt{5}\pi)$. 

{\it The excited state denoted as $i=7r(8l)$ ($i=7$ for $0<\cu<+\infty$ and $i=8$ for $-\infty < \cu < 0$):\/}
For $\cu \rightarrow -\infty$, all 9 distinct coefficients [see frame (e) in Fig.\ \ref{bcoefa}] and 13 
cosinusoidal terms survive in expression (\ref{2ndbexpr}); 
the corresponding state is given by the second expression in Eq.\ (\ref{phi8}). In this case, the 9 
distinct coefficients are:
$\cb^{8,-\infty}_0=2/\pi$, 
$\cb^{8,-\infty}_1=-2\sqrt{5}/(3\pi)$, 
$\cb^{8,-\infty}_2=2/(5\pi)$, 
$\cb^{8,-\infty}_{1-1}=26/(15\pi)$, 
$\cb^{8,-\infty}_{2-2}=2/(15\pi)$, 
$\cb^{8,-\infty}_{2-1}=-2/(3\sqrt{5}\pi)$, 
$\cb^{8,-\infty}_{1+1}=4/(5\pi)$, 
$\cb^{8,-\infty}_{2+2}=1/(3\pi)$, and
$\cb^{8,-\infty}_{2+1}=-2/(\sqrt{5}\pi)$. 

In the noninteracting case ($\cu=0$), for which the Hubbard eigenvector is given by Eq.\ (\ref{eigvecU0st7}),
all 13 cosinusoidal terms and 9 distinct coefficients are present in Eq.\ (\ref{2ndbexpr}), in agreement with
frame (e) of Fig.\ \ref{bcoefa}, that is,
$\cb^{7r(8l),\cu=0}_0=2/\pi=0.63662$,
$\cb^{7r(8l),\cu=0}_1=-0.300105$,
$\cb^{7r(8l),\cu=0}_2=-0.111766$,
$\cb^{7r(8l),\cu=0}_{1-1}=-0.124185$,
$\cb^{7r(8l),\cu=0}_{2-2}=0.214057$,
$\cb^{7r(8l),\cu=0}_{2-1}=-0.0243414$
$\cb^{7r(8l),\cu=0}_{1+1}=-0.0135114$,
$\cb^{7r(8l),\cu=0}_{2+2}=-0.127997$, and
$\cb^{7r(8l),\cu=0}_{2+1}=0.178398$.

For $\cu \rightarrow +\infty$,  all 9 distinct coefficients [see frame (e) in Fig.\ \ref{bcoefa}] and 13 
cosinusoidal terms survive in expression (\ref{2ndbexpr}); 
the corresponding state is given by the first expression in Eq.\ (\ref{phi7}). In this case, the 9 distinct 
coefficients are:
$\cb^{7,+\infty}_0=2/\pi$, 
$\cb^{7,+\infty}_1=-2\sqrt{5}/(3\pi)$, 
$\cb^{7,+\infty}_2=2/(5\pi)$, 
$\cb^{7,+\infty}_{1-1}=26/(15\pi)$, 
$\cb^{7,+\infty}_{2-2}=2/(15\pi)$, 
$\cb^{7,+\infty}_{2-1}=-2/(3\sqrt{5}\pi)$, 
$\cb^{7,+\infty}_{1+1}=4/(5\pi)$, 
$\cb^{7,+\infty}_{2+2}=1/(3\pi)$, and
$\cb^{7,+\infty}_{2+1}=-2/(\sqrt{5}\pi)$. 

We note that $^2{\cal G}_8^{b,-\infty}(k_1,k_2)\;=\;^2{\cal G}_7^{b,+\infty}(k_1,k_2)$.

{\it The excited state denoted as $i=8r(7l)$ ($i=8$ for $0<\cu<+\infty$ and $i=7$ for $-\infty < \cu < 0$):\/}
For $\cu \rightarrow -\infty$, 5 distinct coefficients [see frame (f) in Fig.\ \ref{bcoefa}] and 6 
cosinusoidal terms survive in expression (\ref{2ndbexpr}); 
the corresponding state is given by the second expression in Eq.\ (\ref{phi7}). In this case, the 5 
distinct coefficients are:
$\cb^{7,-\infty}_0=2/\pi$, 
$\cb^{7,-\infty}_1=0$, 
$\cb^{7,-\infty}_2=6/(5\pi)$, 
$\cb^{7,-\infty}_{1-1}=-4/(5\pi)$, 
$\cb^{7,-\infty}_{2-2}=16/(15\pi)$, 
$\cb^{7,-\infty}_{2-1}=0$, 
$\cb^{7,-\infty}_{1+1}=-4/(15\pi)$, 
$\cb^{7,-\infty}_{2+2}=0$, and
$\cb^{7,-\infty}_{2+1}=0$. 

In the noninteracting case ($\cu=0$), for which the Hubbard eigenvector is given by Eq.\ (\ref{eigvecU0st8}),
all 13 cosinusoidal terms and 9 distinct coefficients are present in Eq.\ (\ref{2ndbexpr}), in agreement with
frame (f) of Fig.\ \ref{bcoefa}, that is,
$\cb^{8r(7l),\cu=0}_0=2/\pi=0.63662$,
$\cb^{8r(7l),\cu=0}_1=-0.300105$
$\cb^{8r(7l),\cu=0}_2=0.111766$,
$\cb^{8r(7l),\cu=0}_{1-1}=0.124185$,
$\cb^{8r(7l),\cu=0}_{2-2}=0.0246755$,
$\cb^{8r(7l),\cu=0}_{2-1}=-0.0506849$
$\cb^{8r(7l),\cu=0}_{1+1}=-0.410902$,
$\cb^{8r(7l),\cu=0}_{2+2}=0.154523$, and
$\cb^{8r(7l),\cu=0}_{2+1}=0.0466808$.

For $\cu \rightarrow +\infty$, only the constant coefficient survives [see frame (f) in Fig.\ \ref{bcoefa}].
The corresponding state is given by the first expression in Eq.\ (\ref{phi8}) and it is a NOON state of the
form $(|300\rangle + |003\rangle)/\sqrt{2}$. In this case, the second-order correlation is: 
\begin{align}
^2{\cal G}_8^{b,+\infty}(k_1,k_2)= \frac{2}{\pi}s^2 e^{-2(k_1^2+k_2^2)s^2}.
\label{3b2ndst8UI}
\end{align}

{\it The excited state denoted as $i=9$ for $-\infty < \cu < +\infty$:\/}
For $\cu \rightarrow -\infty$, all 9 distinct coefficients [see frame (g) in Fig.\ \ref{bcoefa}] and 13 
cosinusoidal terms survive in expression (\ref{2ndbexpr}); 
the corresponding state is given by the second expression in Eq.\ (\ref{phi9}). In this case, the 9 distinct 
coefficients are:
$\cb^{9,-\infty}_0=2/\pi$,
$\cb^{9,-\infty}_1=-2\sqrt{5}/(3\pi)$,
$\cb^{9,-\infty}_2=-2/(5\pi)$,
$\cb^{9,-\infty}_{1-1}=26/(15\pi)$,
$\cb^{9,-\infty}_{2-2}=2/(15\pi)$,
$\cb^{9,-\infty}_{2-1}=-2/(3\sqrt{5}\pi)$,
$\cb^{9,-\infty}_{1+1}=-4/(5\pi)$,
$\cb^{9,-\infty}_{2+2}=-1/(3\pi)$, and
$\cb^{9,-\infty}_{2+1}=2/(\sqrt{5}\pi)$.

In the noninteracting case ($\cu=0$), for which the Hubbard eigenvector is given by Eq.\ (\ref{eigvecU0st9}),
10 cosinusoidal terms and 7 distinct coefficients are present in Eq.\ (\ref{2ndbexpr}), in agreement with
frame (g) of Fig.\ \ref{bcoefa}, that is,
$\cb^{9,\cu=0}_0=2/\pi=0.63662$,
$\cb^{9,\cu=0}_1=-4\sqrt{2}/(3\pi)$,
$\cb^{9,\cu=0}_2=0$,
$\cb^{9,\cu=0}_{1-1}=4/(3\pi)$,
$\cb^{9,\cu=0}_{2-2}=1/(12\pi)$,
$\cb^{9,\cu=0}_{2-1}=-1/(3\sqrt{2}\pi)$,
$\cb^{9,\cu=0}_{1+1}=0$,
$\cb^{9,\cu=0}_{2+2}=-7/(12\pi)$, and
$\cb^{9,\cu=0}_{2+1}=1/(\sqrt{2}\pi)$.

For $\cu \rightarrow +\infty$, only the constant coefficient survives [see frame (g) in Fig.\ \ref{bcoefa}].
The corresponding state is given by the first expression in Eq.\ (\ref{phi9}) and it is a NOON state of the
form $(-|300\rangle + |003\rangle)/\sqrt{2}$. In this case, the second-order correlation is: 
\begin{align}
^2{\cal G}_9^{b,+\infty}(k_1,k_2)= \frac{2}{\pi}s^2 e^{-2(k_1^2+k_2^2)s^2}.
\label{3b2ndst9UI}
\end{align}

{\it The excited state denoted as $i=10$ for $-\infty < \cu < +\infty$:\/} 
For $\cu \rightarrow -\infty$, three terms survive, including the constant one [see frame (h) in Fig.\ 
\ref{bcoefa}]. The corresponding state is 
that of all three wells being singly occupied. In this case, the second-order correlation function is given by
\begin{align}
\begin{split}
^2{\cal G}_{10}^{b,-\infty}& (k_1,k_2) = \frac{2}{3\pi} s^2 e^{-2(k_1^2+k_2^2)s^2} \{ 3 \\
& + 2 \cos[d(k_1-k_2)] + \cos[2d(k_1-k_2)] \}.
\end{split}
\label{3b2ndst10UI}
\end{align}

In the noninteracting case ($\cu=0$), for which the Hubbard eigenvector is given by Eq.\ (\ref{eigvecU0st10}),
all 13 cosinusoidal terms and 9 distinct coefficients are present in Eq.\ (\ref{2ndbexpr}), in agreement with
frame (h) of Fig.\ \ref{bcoefa}, that is,
$\cb^{10,\cu=0}_0=2/\pi=0.63662$,
$\cb^{10,\cu=0}_1=-2\sqrt{2}/\pi$,
$\cb^{10,\cu=0}_2=1/\pi$,
$\cb^{10,\cu=0}_{1-1}=2/\pi$,
$\cb^{10,\cu=0}_{2-2}=1/(4\pi)$,
$\cb^{10,\cu=0}_{2-1}=-1/(\sqrt{2}\pi)$,
$\cb^{10,\cu=0}_{1+1}=2/\pi$,
$\cb^{10,\cu=0}_{2+2}=1/(4\pi)$, and
$\cb^{10,\cu=0}_{2+1}=-1/(\sqrt{2}\pi)$.

For $\cu \rightarrow +\infty$ only the constant term, ${\cal B}^{10}_0=2/\pi$, survives [see frame (h) in 
Fig.\ \ref{bcoefa}]. The corresponding state is the triply occupied middle well. In this case, the second-order 
correlation function is
\begin{align}
^2{\cal G}_{10}^{b,+\infty}(k_1,k_2)= \frac{2}{\pi}s^2 e^{-2(k_1^2+k_2^2)s^2}.
\label{3b2ndst10UmI}
\end{align}

\begin{figure*}[t]
\includegraphics[width=17.5cm]{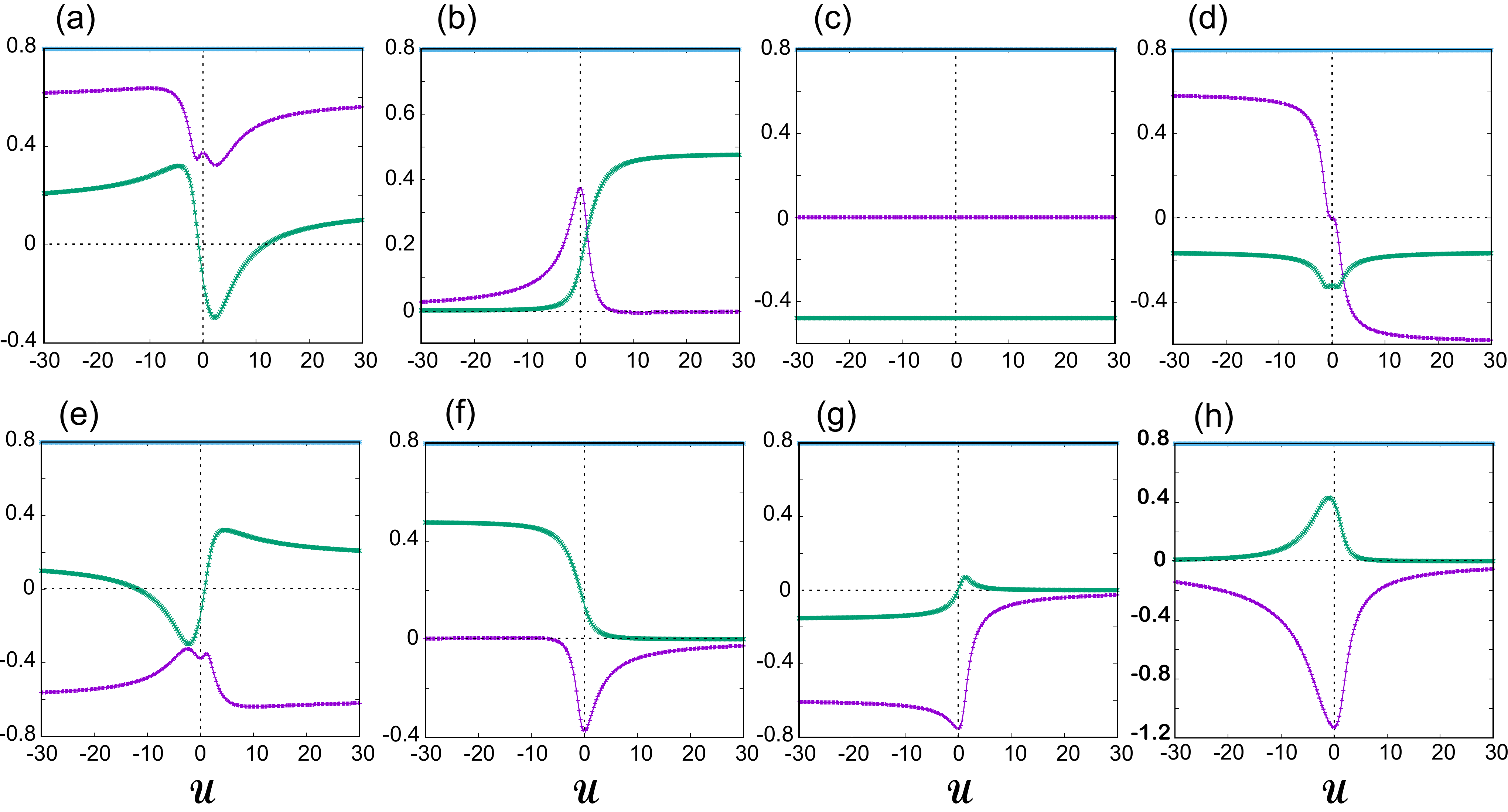}
\caption{The ${\cal A}$-coefficients (dimensionless) [see Eq.\ (\ref{frstbexpr})] for the remaining 
8 excited eigenstates of 3 bosons trapped in 3 linearly arranged wells as a function of $\cu$ 
(horizontal axis, dimensionless). This figure complements Fig.\ \ref{acoef} in the main text. 
(a) $i=3r(4l)$. (b) $i=4r(3l)$. (c) $i=5r(6l)$.
(d) $i=6r(5l)$. (e) $i=7r(8l)$. (f) $i=8r(7l)$, (g) $i=9$, (h) $i=10$.
See text for a detailed description. The choice of online colors is the same as in Fig.\ \ref{acoef}, that is:
$\ca_0 \rightarrow$ Constant (Light Blue), $\ca_1 \rightarrow$ Violet, $\ca_2 \rightarrow$ Green.
For the print grayscale version, excluding the top constant $\ca_0$ horizontal line, the positioning of the two
remaining curves at $\cu=-30$ is as follows: (a,b,c,d) $\ca_1 \rightarrow$ upper curve, $\ca_2 \rightarrow$
lower curve. (e,f,g,h)  $\ca_1 \rightarrow$ lower curve, $\ca_2 \rightarrow$ upper curve.
}
\label{acoefa}
\end{figure*}

\section{First-order momentum correlations as a function of $\cu$ for the remaining eight excited states}
\label{a1st}

Fig.\ \ref{acoefa} complements Fig.\ \ref{acoef} in that it displays the 3 distinct coefficients $\ca^i(\cu)$'s
for the remaining eight excited states (explicit numerical values can be found in the supplemental 
material \cite{supp}). The dependence of these coefficients on the interaction strength $\cu$ is
better deciphered by using as reference points the special cases at $\cu \rightarrow \pm \infty$ and $\cu=0$.
Note that in all cases the $\cb^i$ values at the end points $\cu=\pm 30$ in the figure are close to the
corresponding limiting values at $\cu \rightarrow \pm \infty$. In particular,

{\it The excited state denoted as $i=3r(4l)$ ($i=3$ for $0<\cu<+\infty$ and $i=4$ for $-\infty < \cu < 0$):\/}
For $\cu \rightarrow -\infty$, all 3 cosinusoidal terms survive in expression (\ref{frstbexpr}) [see frame (a)
in Fig.\ \ref{acoefa}]; specifically one has:
$\ca^{4,-\infty}_0=0.797885=\sqrt{2/\pi}$,
$\ca^{4,-\infty}_1=\sqrt{10}/(3\sqrt{\pi})$, and
$\ca^{4,-\infty}_2=\sqrt{2}/(5\sqrt{\pi})$.

For the non-interacting case ($\cu=0$), all 3 coefficients survive in expression (\ref{frstbexpr}) [see frame (a) 
in Fig.\ \ref{acoefa}]; specifically one has:
$\ca^{3r(4l),\cu=0}_0=0.797885=\sqrt{2/\pi}$,
$\ca^{3r(4l),\cu=0}_1=0.376126$, and
$\ca^{3r(4l),\cu=0}_2=-0.140078$.

For $\cu \rightarrow +\infty$, all 3 cosinusoidal terms survive in expression (\ref{frstbexpr}) [see frame (a)
in Fig.\ \ref{acoefa}]; specifically one has:
$\ca^{3,+\infty}_0=0.797885=\sqrt{2/\pi}$,
$\ca^{3,+\infty}_1=\sqrt{10}/(3\sqrt{\pi})$, and
$\ca^{3,+\infty}_2=\sqrt{2}/(5\sqrt{\pi})$.

{\it The excited state denoted as $i=4r(3l)$ ($i=4$ for $0<\cu<+\infty$ and $i=3$ for $-\infty < \cu < 0$):\/}
For $\cu \rightarrow -\infty$, only 1 cosinusoidal term survives in expression (\ref{frstbexpr}) [see frame (b)
in Fig.\ \ref{acoefa}]; specifically one has:
$\ca^{3,-\infty}_0=0.797885=\sqrt{2/\pi}$,
$\ca^{3,-\infty}_1=0$, and
$\ca^{3,-\infty}_2=0$.

For the non-interacting case ($\cu=0$), all 3 coefficients survive in expression (\ref{frstbexpr}) [see frame (b) 
in Fig.\ \ref{acoefa}]; specifically one has:
$\ca^{4r(3l),\cu=0}_0=0.797885=\sqrt{2/\pi}$,
$\ca^{4r(3l),\cu=0}_1=0.376126$, and
$\ca^{4r(3l),\cu=0}_2=0.140078$.

For $\cu \rightarrow +\infty$, 2 cosinusoidal terms survive in expression (\ref{frstbexpr}) [see frame (b)
in Fig.\ \ref{acoefa}]; specifically one has:
$\ca^{4,+\infty}_0=0.797885=\sqrt{2/\pi}$,
$\ca^{4,+\infty}_1=0$, and
$\ca^{4,+\infty}_2=3\sqrt{2}/(5\sqrt{\pi})$.

{\it The excited state denoted as $i=5r(6l)$ ($i=5$ for $0<\cu<+\infty$ and $i=6$ for $-\infty < \cu < 0$):\/}
This state is $\cu$-independent; see first expression in Eq.\ (\ref{phi5}) or second expression in
Eq.\ (\ref{phi6}). In this case, 2 distinct coefficients (corresponding to 2 cosinusoidal terms) survive in
expression (\ref{frstbexpr}), that is,
$\ca^{5,+\infty}_0=\ca^{6,-\infty}_0=\sqrt{2/\pi}=\ca^{5r(6l),\cu=0}_0=0.797885$,
$\ca^{5,+\infty}_1=\ca^{6,-\infty}_1=\ca^{5r(6l),\cu=0}_1=0$, and
$\ca^{5,+\infty}_2=\ca^{6,-\infty}_2=\ca^{5r(6l),\cu=0}_2=-3\sqrt{2}/(5\sqrt{\pi})$;
see frame (c) in Fig.\ \ref{acoefa}.

{\it The excited state denoted as $i=6r(5l)$ ($i=6$ for $0<\cu<+\infty$ and $i=5$ for $-\infty < \cu < 0$):\/}
For $\cu \rightarrow -\infty$, all 3 cosinusoidal terms survive in expression (\ref{frstbexpr}) [see frame (d)
in Fig.\ \ref{acoefa}]; specifically one has:
$\ca^{5,-\infty}_0=0.797885=\sqrt{2/\pi}$,
$\ca^{5,-\infty}_1=\sqrt{10}/(3\sqrt{\pi})$, and
$\ca^{5,-\infty}_2=-\sqrt{2}/(5\sqrt{\pi})$.

For the non-interacting case ($\cu=0$), 2 coefficients are present in expression (\ref{frstbexpr}) [see frame (d)
in Fig.\ \ref{acoefa}]; specifically one has:
$\ca^{6r(5l),\cu=0}_0=0.797885=\sqrt{2/\pi}$,
$\ca^{6r(5l),\cu=0}_1=0$, and
$\ca^{6r(5l),\cu=0}_2=-2\sqrt{2}/(5\sqrt{\pi})$.

For $\cu \rightarrow +\infty$, all 3 cosinusoidal terms survive in expression (\ref{frstbexpr}) [see frame (d)
in Fig.\ \ref{acoefa}]; specifically one has:
$\ca^{6,+\infty}_0=0.797885=\sqrt{2/\pi}$,
$\ca^{6,+\infty}_1=-\sqrt{10}/(3\sqrt{\pi})$, and
$\ca^{6,+\infty}_2=-\sqrt{2}/(5\sqrt{\pi})$.

{\it The excited state denoted as $i=7r(8l)$ ($i=7$ for $0<\cu<+\infty$ and $i=8$ for $-\infty < \cu < 0$):\/}
For $\cu \rightarrow -\infty$, all 3 cosinusoidal terms survive in expression (\ref{frstbexpr}) [see frame (e)
in Fig.\ \ref{acoefa}]; specifically one has:
$\ca^{8,-\infty}_0=0.797885=\sqrt{2/\pi}$,
$\ca^{8,-\infty}_1=-\sqrt{10}/(3\sqrt{\pi})$, and
$\ca^{8,-\infty}_2=\sqrt{2}/(5\sqrt{\pi})$.

For the non-interacting case ($\cu=0$), all 3 coefficients are present in expression (\ref{frstbexpr}) 
[see frame (e) in Fig.\ \ref{acoefa}]; specifically one has:
$\ca^{7r(8l),\cu=0}_0=0.797885=\sqrt{2/\pi}$,
$\ca^{7r(8l),\cu=0}_1=-0.376126$, and
$\ca^{7r(8l),\cu=0}_2=-0.140078$.

For $\cu \rightarrow +\infty$, all 3 cosinusoidal terms survive in expression (\ref{frstbexpr}) [see frame (e)
in Fig.\ \ref{acoefa}]; specifically one has:
$\ca^{7,+\infty}_0=0.797885=\sqrt{2/\pi}$,
$\ca^{7,+\infty}_1=-\sqrt{10}/(3\sqrt{\pi})$, and
$\ca^{7,+\infty}_2=\sqrt{2}/(5\sqrt{\pi})$.

{\it The excited state denoted as $i=8r(7l)$ ($i=8$ for $0<\cu<+\infty$ and $i=7$ for $-\infty < \cu < 0$):\/}
For $\cu \rightarrow -\infty$, 2 cosinusoidal terms are present in expression (\ref{frstbexpr}) [see frame (f)
in Fig.\ \ref{acoefa}]; specifically one has:
$\ca^{7,-\infty}_0=0.797885=\sqrt{2/\pi}$,
$\ca^{7,-\infty}_1=0$, and
$\ca^{7,-\infty}_2=3\sqrt{2}/(5\sqrt{\pi})$.

For the non-interacting case ($\cu=0$), all 3 coefficients are present in expression (\ref{frstbexpr}) 
[see frame (f) in Fig.\ \ref{acoefa}]; specifically one has:
$\ca^{8r(7l),\cu=0}_0=0.797885=\sqrt{2/\pi}$,
$\ca^{8r(7l),\cu=0}_1=-0.376126$, and
$\ca^{8r(7l),\cu=0}_2=0.140078$.

For $\cu \rightarrow +\infty$, only the $\cu$-independent term survives in expression (\ref{frstbexpr}) 
[see frame (f) in Fig.\ \ref{acoefa}]; specifically one has:
$\ca^{8,+\infty}_0=0.797885=\sqrt{2/\pi}$,
$\ca^{8,+\infty}_1=0$, and
$\ca^{8,+\infty}_2=0$.

{\it The excited state denoted as $i=9$ for $-\infty < \cu < +\infty$:\/}
For $\cu \rightarrow -\infty$, all 3 cosinusoidal terms are present in expression (\ref{frstbexpr}) [see frame (g)
in Fig.\ \ref{acoefa}]; specifically one has:
$\ca^{9,-\infty}_0=0.797885=\sqrt{2/\pi}$,
$\ca^{9,-\infty}_1=-\sqrt{10}/(3\sqrt{\pi})$, and
$\ca^{9,-\infty}_2=-\sqrt{2}/(5\sqrt{\pi})$.

For the non-interacting case ($\cu=0$), 2 coefficients are present in expression (\ref{frstbexpr}) 
[see frame (g) in Fig.\ \ref{acoefa}]; specifically one has:
$\ca^{9,\cu=0}_0=0.797885=\sqrt{2/\pi}$,
$\ca^{9,\cu=0}_1=-4/(3\sqrt{\pi})$, and
$\ca^{9,\cu=0}_2=0$.

For $\cu \rightarrow +\infty$, only the $\cu$-independent term survives in expression (\ref{frstbexpr}) 
[see frame (g) in Fig.\ \ref{acoefa}]; specifically one has:
$\ca^{9,+\infty}_0=0.797885=\sqrt{2/\pi}$,
$\ca^{9,+\infty}_1=0$, and
$\ca^{9,+\infty}_2=0$.

{\it The excited state denoted as $i=10$ for $-\infty < \cu < +\infty$:\/}
For $\cu \rightarrow -\infty$, only the $\cu$-independent term is present in expression (\ref{frstbexpr}) 
[see frame (h) in Fig.\ \ref{acoefa}]; specifically one has:
$\ca^{10,-\infty}_0=0.797885=\sqrt{2/\pi}$,
$\ca^{10,-\infty}_1=0$, and
$\ca^{10,-\infty}_2=0$.

For the non-interacting case ($\cu=0$), all 3 coefficients are present in expression (\ref{frstbexpr}) 
[see frame (h) in Fig.\ \ref{acoefa}]; specifically one has:
$\ca^{10,\cu=0}_0=0.797885=\sqrt{2/\pi}$,
$\ca^{10,\cu=0}_1=-2/\sqrt{\pi}$, and
$\ca^{10,\cu=0}_2=1/\sqrt{2\pi}$.

For $\cu \rightarrow +\infty$, only the $\cu$-independent term survives in expression (\ref{frstbexpr}) 
[see frame (h) in Fig.\ \ref{acoefa}]; specifically one has:
$\ca^{10,+\infty}_0=0.797885=\sqrt{2/\pi}$,
$\ca^{10,+\infty}_1=0$, and
$\ca^{10,+\infty}_2=0$.


\begin{thebibliography}{99}
\bibitem{note4}
It also applies for the case of a few strongly entangled fermions \cite{yann19.3}.
\end{thebibliography}
\end{document}